\newcommand{\kms}{{km~s$^{-1}$}}

\newcommand{\msol}{\mathcal{M}_\odot}

\newcommand{\lksol}{\mathcal{L}_{\odot,K}}		

\newcommand{\halp}{H$\alpha$}

\newcommand{\oiii}{[O{\scshape iii}]}
\newcommand{\hone}{H{\scshape i}}

\newcommand{\gaz}{\theta}

\newcommand{\vsys}{V_{\rm sys}}

\newcommand{\vc}{V_{\rm c}}

\newcommand{\vdm}{V_{\rm DM}}

\newcommand{\slos}{\sigma_{\rm LOS}}

\newcommand{\sinst}{\sigma_{\rm inst}}

\newcommand{\sigr}{\sigma_R}
\newcommand{\sigp}{\sigma_{\gaz}}
\newcommand{\sigz}{\sigma_z}

\newcommand{\sddisk}{\Sigma_{\rm dyn}}
\newcommand{\sds}{\Sigma_{\rm \ast}}
\newcommand{\sdhi}{\Sigma_{\mbox{\rm \footnotesize H{\scshape i}}}}

\newcommand{\sda}{\Sigma_{\rm atom}}
\newcommand{\sdm}{\Sigma_{\rm mol}}
\newcommand{\sdmh}{\Sigma_{\rm H_2}}

\newcommand{\sdd}{\Sigma_{\rm dust}}

\newcommand{\mdiskstar}{\mathcal{M}^{\rm disk}_{\rm \ast}}
\newcommand{\mbulgestar}{\mathcal{M}^{\rm bulge}_{\ast}}

\newcommand{\mhi}{\mathcal{M}_{\rm HI}}
\newcommand{\matom}{\mathcal{M}_{\rm atom}}
\newcommand{\mmol}{\mathcal{M}_{\rm mol}}

\newcommand{\mb}{\mathcal{M}_{\rm b}}

\newcommand{\mldyn}{\Upsilon_{\rm dyn}}

\newcommand{\mls}{\Upsilon_\ast}

\newcommand{\ml}{\Upsilon}

\newcommand{\hr}{h_{\rm R}}

\newcommand{\vmax}{V_{\rm max}}

\newcommand{\hz}{h_{z}}
\newcommand{\Fbary}{\mathcal{F}_{\rm b}}
\newcommand{\mbary}{\mathcal{M}_{\rm b}}
\newcommand{\vbary}{V_{\rm b}}
\newcommand{\mstar}{\mathcal{M}_{\ast}}

\newcommand{\itf}{i_{\rm TF}}

\newcommand{\arcdeg}{\mbox{$^\circ$}}

\newcommand{\rbulge}{R_{\rm bulge}}

\newcommand{\lk}{\mathcal{L}_K}

\documentclass[traditabstract]{aa} 
\usepackage{txfonts}
\usepackage{graphicx}
\usepackage{longtable}
\usepackage{supertabular}
\usepackage{lscape}
\usepackage{subfigure}
\usepackage{rotating}
\usepackage{caption}
\usepackage[title,titletoc]{appendix}       

\usepackage[spanish,dutch,english]{babel}
\usepackage{natbib}
\usepackage{hyperref}

\bibpunct{(}{)}{;}{a}{}{,}

\defcitealias{bershady2010a}{Paper~I}
\defcitealias{bershady2010b}{Paper~II}
\defcitealias{westfall2011a}{Paper~III}
\defcitealias{westfall2011b}{Paper~IV}
\defcitealias{bershady2011}{Paper~V}         
\defcitealias{martinsson2012a}{Paper~VI}     

\begin{document}

\title{The DiskMass Survey. VII. The distribution of luminous and \\dark matter in spiral 
galaxies}

\author{Thomas P. K. Martinsson\inst{1,}\inst{2}
   \and Marc A. W. Verheijen\inst{1}
   \and Kyle B. Westfall\inst{1,}\thanks{National Science Foundation (USA) International
   Research Fellow}
   \and \\Matthew A. Bershady\inst{3}
   \and David R. Andersen\inst{4}
   \and Rob A. Swaters\inst{5}}

\institute{
  Kapteyn Astronomical Institute, University of Groningen, PO Box 800, 9700 AV Groningen, 
  The Netherlands\\
  \email{verheyen@astro.rug.nl; westfall@astro.rug.nl}
\and
  Leiden Observatory, Leiden University, PO Box 9513, 2300 RA Leiden, The Netherlands\\
  \email{martinsson@strw.leidenuniv.nl}
\and
  Department of Astronomy, University of Wisconsin, 475 N. Charter St., Madison, WI 53706, 
  USA\\
  \email{mab@astro.wisc.edu}
\and
  NRC Herzberg Institute of Astrophysics, 5071 West Saanich Road, Victoria, British 
  Columbia, Canada V9E 2E7\\
  \email{david.andersen@nrc-cnrc.gc.ca}
\and
National Optical Astronomy Observatory, 950 North Cherry Ave., Tucson, AZ 85719, USA\\
  \email{swaters@noao.edu}
}

\date{Received 1 March 2013 / Accepted 12 July 2013}

\abstract{
We present dynamically-determined rotation-curve mass decompositions of 30 spiral
galaxies, which were carried out to test the maximum-disk hypothesis and to quantify
properties of their dark-matter halos. We used measured vertical velocity dispersions of
the disk stars to calculate dynamical mass surface densities ($\sddisk$). By subtracting
our observed atomic and inferred molecular gas mass surface densities from $\sddisk$, we
derived the stellar mass surface densities ($\sds$), and thus have absolute measurements
of all dominant baryonic components of the galaxies.
Using $K$-band surface brightness profiles ($I_K$), we calculated the $K$-band
mass-to-light ratio of the stellar disks ($\mls$$=$$\sds/I_K$) and adopted the radial mean
($\overline{\mls}$) for each galaxy to extrapolate $\sds$ beyond the outermost kinematic
measurement. The derived $\overline{\mls}$ of individual galaxies are consistent with all
galaxies in the sample having equal $\mls$. We find a sample average and scatter of 
$\langle$$\overline{\mls}$$\rangle$$=$0.31$\pm$0.07.
Rotation curves of the baryonic components were calculated from their deprojected mass
surface densities. These were used with circular-speed measurements to derive the
structural parameters of the dark-matter halos, modeled as either a pseudo-isothermal
sphere (pISO) or a Navarro-Frenk-White (NFW) halo. In addition to our dynamically
determined mass decompositions, we also performed alternative rotation-curve 
decompositions by adopting the traditional maximum-disk hypothesis.
However, the galaxies in our sample are submaximal, such that at 2.2 disk scale lengths 
($\hr$) the ratios between the baryonic and total rotation curves ($\Fbary^{2.2\hr}$) are
less than 0.75. We find this ratio to be nearly constant between 1--6$\hr$ within
individual galaxies. We find a sample average and scatter of 
$\langle$$\Fbary^{2.2\hr}$$\rangle$$=$0.57$\pm$0.07, with trends of larger
$\Fbary^{2.2\hr}$ for more luminous and higher-surface-brightness galaxies. To enforce 
these being maximal, we need to scale $\mls$ by a factor 3.6 on average.
In general, the dark-matter rotation curves are marginally better fit by a pISO than by an
NFW halo.
For the nominal-$\mls$ (submaximal) case, we find that the derived NFW-halo parameters
have values consistent with $\Lambda$CDM N-body simulations, suggesting that the baryonic
matter in our sample of galaxies has only had a minor effect on the dark-matter
distribution. In contrast, maximum-$\mls$ decompositions yield halo-concentration
parameters that are too low compared to the $\Lambda$CDM simulations.
}               


 \titlerunning{DMS-VII. Distribution of Luminous and Dark Matter in Spiral Galaxies}
 \authorrunning{Martinsson et al.}
 \maketitle

\section{Introduction}
\label{sec:Introduction}
For a spiral galaxy, it should be possible to derive the mass distributions of the
different components by decomposing its observed rotation curve into separate
contributions from the various baryonic components and a dark-matter halo.
However, even though it has now been shown that a more or less flat rotation curve seems
to be a general feature of spiral galaxies \citep [e.g.,][]{bosma1978, bosma1981a,
bosma1981b, begeman1987, begeman1989, sofue1999, sofue2001}, and although the concept of a
dark-matter halo is well established and widely accepted\footnote{Here and
throughout this work we assume that the Newtonian gravitational theory holds. However,
suggestions have been made that Newtonian dynamics need modification for use at low
accelerations \citep[e.g.,][]{milgrom1983, begeman1991, sanders1996, sanders1998}.},
there is still a huge uncertainty in the observationally inferred distribution of the
dark matter. The main issue is that, since the stellar mass is unknown, the technique of
decomposing the observed rotation curve does not put a strong constraint on the detailed
shape of the dark-matter-halo density profile. In many cases the observed rotation curve
can even be explained by a two-parameter dark-matter-halo model alone, with the disk 
containing no stellar mass at all. The most commonly used approach to circumvent this
problem has been to go to the other extreme by assuming a maximum contribution from the
baryons, thereby increasing the mass-to-light ratio of the stellar component ($\mls$)
until the rotation curve of the baryons approximates the amplitude of the observed
rotation curve in the inner region 
\citep[the ``maximum-disk hypothesis'';][]{AlbadaSancisi1986}.
This approach sets an upper limit on the contribution from the baryons, but without
knowledge of $\mls$ there is still a severe \textit{disk-halo degeneracy}, making it
impossible to determine the structural properties of the dark-matter halo.

Another approach is to use $\mls$ derived from stellar-population-synthesis models;
however, these models suffer from large uncertainties since they require many assumptions
regarding the star-formation and chemical-enrichment history, the initial mass function
(IMF), and accurate accounting for late phases of stellar evolution \citep{maraston2005,
conroy2009}. For example, although \cite{kauffmann2003} find random errors of only 40\%
from the models, the choice of IMF alone results in a factor of two systematic
uncertainty in the stellar mass.

While the maximum-disk hypothesis has been a commonly used refuge in the literature
\citep[e.g.,][]{albada1985,kent1986,BroeilsCourteau1997}, there is evidence that at least
some galaxy disks are in fact submaximal, e.g., based on the lack of a surface-brightness
dependency in the Tully-Fisher relation \citep[TF;][]{tullyfisher1977} for a wide range
of spirals \citep{zwaan1995,courteau1999,courteau2003}. However, observations have
so far not lead to a consensus, and the maximality may depend on the galaxy type
\citep{bottema1993, weiner2001, kranz2003, kregel2005, byrd2006, herrmann2009, dutton2011,
dutton2013,barnabe2012}.

One of the main goals of the DiskMass Survey \citep[DMS;][hereafter 
\citetalias{bershady2010a}]{bershady2010a} is to break the disk-halo degeneracy using
stellar and gas kinematics to determine the dynamical mass-to-light ratio of the
galaxy disk ($\mldyn$). For a locally isothermal disk 
\begin{equation}
\mldyn = \frac{\sddisk}{I} = \frac{\sigz^2}{\pi G k \hz I} ,
\label{eq:mldyn}
\end{equation}
where $\sddisk$ is the dynamical mass surface density of the disk, $I$ the surface
brightness, $\sigz$ the vertical component of the stellar velocity dispersion, $G$ the 
gravitational constant, $k$ a parameter dependent on the vertical mass distribution, and 
$\hz$ the disk scale height \citep{kruit1981,bahcall1984}. Since $I$ is well known from 
photometry, and the relation between $\hz$ and the disk scale length ($\hr$) has
been statistically determined from studies of edge-on spiral galaxies \citep[e.g.,][see
also the compilation in Fig.~1 of \citealt{bershady2010b}, hereafter 
\citetalias{bershady2010b}]{GrijsKruit1996,kregel2002}, observations of $\sigz$ give us a
direct estimate of $\mldyn$. The value of $k$ is expected to range between 1.5 to 2
\citep[exponential to isothermal distribution;][]{kruit1988}. In this paper, we will
assume an exponential distribution ($k$=1.5) as a reasonable approximation for the
composite (gas+stars) density distribution \citepalias{bershady2010b}. In the expected
range of density distributions discussed by \cite{kruit1988}, our adopted value of $k$
will effectively maximize the measurement of $\sddisk$ and $\mldyn$.

The strategy of measuring the stellar velocity dispersion to obtain the dynamical mass 
surface density has been attempted before \citep{kruit1984, kruit1986, bottema1993,
kregel2005}. These studies showed that the ratio of the maximum amplitude of the disk's
rotation curve (calculated from the observed velocity dispersion) and the maximum of the
observed rotation speed ($\vmax$) was much lower than expected for a maximum disk.
\cite{bottema1993} found that disks contribute only $63\pm10$\% to the observed rotation
speed. \cite{kregel2005} found an even smaller average disk contribution of $58\pm5$\%,
with a $1\sigma$ scatter of 18\% when including two outliers. Excluding the outliers, they
find an average disk contribution of $53\pm4$\%, with a $1\sigma$ scatter of 15\%. These
results are significantly lower than the 85$\pm$10\% which is typical for a maximum-disk
case \citep{sackett1997}.
In \citet[][hereafter \citetalias{bershady2011}]{bershady2011}, we followed the approach
of using the relation between the central $\sigz$ of the disk and $\vmax$, and found that
disks contribute only $47\pm8$\% of the observed rotation speed; even lower than what was
found by the earlier studies, but consistent within the errors. 
We also found that the disk contribution depends on color, absolute $K$-band magnitude,
and $\vmax$, such that redder, more luminous, and faster-rotating galaxies have baryonic
disks that make a relatively larger contribution to their observed rotation speeds.

With a measured distribution of the baryonic mass, together with the observed rotation
speed, it is possible to derive the density distribution of the dark matter. While the
flatness of the outer part of rotation curves suggests a halo with a dark-matter density
distribution declining as $\rho \propto R^{-2}$, the inner slope is still debated
\citep[see, e.g.,][]{deBlok2010}. The difficulty in determining the inner density
distribution of the dark matter arises mainly due the uncertainty in the baryonic mass
distribution. From numerical N-body simulations \citep[e.g.,][]{NFW1997} the inner density
profiles and concentrations of the dark-matter halos are predicted in the absence of
baryons. It is argued that the baryons in the disk will tend to contract the halo while it
is forming \citep[e.g.,][]{blumenthal1986, gnedin2004}. However, several processes may
occur that could also expand the halo, such as dynamical friction between the halo and
infalling galaxies \citep[e.g.,][]{el-zant2001}, or mass outflows from central starbursts
and active galactic nuclei \citep{read2005,governato2012,pontzen2012}.

In this paper, we decompose the rotation curves of 30 spiral galaxies, 
using stellar kinematics from PPak \citep{verheyen2004, kelz2006} and ionized gas 
kinematics from SparsePak \citep{bershady2004, bershady2005}, as well as 21-cm radio 
synthesis data from the WSRT, GMRT and VLA, near-infrared (NIR) photometry from the
Two-Micron All-Sky Survey \citep[2MASS;][]{skrutski2006} and MIPS 24-$\mu$m imaging from
the Spitzer Space Telescope (hereafter {\it Spitzer}). The paper is organized in
the following way:
Section~\ref{sec:DataSummary} describes the sample and summarizes the used data. In
Sect.~\ref{sec:Sigma}, we derive mass surface densities of the atomic and molecular gas,
and show how our stellar kinematic measurements from PPak \citep[][hereafter 
\citetalias{martinsson2012a}]{martinsson2012a} provide mass surface densities of the
stars. We further investigate relative mass fractions of the baryons. 
In Sect.~\ref{sec:RotationCurves} the observed \hone+\halp\ rotation curves are combined.
In Sect.~\ref{sec:RCDecomp} we calculate the baryonic rotation curves, derive measured
dark-matter rotation curves and fit a pseudo-isothermal sphere or a NFW halo to these
rotation curves. We also perform alternative rotation-curve decompositions, using
maximum-$\mls$ solutions. From the observed \hone+\halp\ rotation curves and the
calculated baryonic rotation curves, we derive the baryonic mass fractions as a function
of radius and quantify the baryonic maximality in Sect.~\ref{sec:DiskMaximality}. In
Sect.~\ref{sec:DMRCshape} we investigate how well the measured dark-matter rotation curves
are fitted, and compare the NFW parameters with results from numerical N-body simulations.
Section~\ref{sec:Discussion} contains a discussion on what we have found in this work,
which is finally summarized in Sect.~\ref{sec:Summary_DeComp}. Throughout this paper we
adopt the Hubble parameter $H_0$$=$73~km~s$^{-1}$~Mpc$^{-1}$.

\section{Observational Data}
\label{sec:DataSummary}
The data used in this paper are presented in more detail elsewhere. The following
subsections briefly summarize these data.

\subsection{Galaxy sample}
The complete DMS sample is described in \citetalias{bershady2010a}. Here, we use the 
subsample of 30 galaxies observed with the PPak IFU
\citepalias{martinsson2012a}.\footnote{Based on observations collected at the
Centro Astron\'omico Hispano Alem\'an (CAHA) at Calar Alto, operated jointly by the 
Max-Planck Institut f\"ur Astronomie and the Instituto de Astrof\'isica de Andaluc\'ia
(CSIC).}
We refer to these galaxies as the PPak sample. Observed properties of the galaxies in the
PPak sample, such as distances, colors, coordinates and disk orientations, can be found in
Table~1 and Table~5 of \citetalias{martinsson2012a}.
In summary, these galaxies cover a range in morphological type from Sa to Im with 22
galaxies being Sc or later; absolute $K$-magnitudes ($M_K$) range from -21.0 to -25.4;
$B-K$ colors are between 2.7 and 4.2; and the central disk face-on $K$-band surface
brightness ($\mu_{0,K}^{i}$) ranges from 19.9 to 16.0 mag~arcsec$^{-2}$. The galaxies have
been selected to be close to face-on, and cover a range in inclination from $6\arcdeg$ to
$45\arcdeg$.

\subsection{Near-infrared photometry}
\label{sec:Photometry}
A description of our re-analysis of 2MASS NIR photometry of the galaxies in our sample
is presented in \citetalias{martinsson2012a}. These data were used to produce 
pseudo-$K$-band surface brightness profiles, $\mu_{K}(R)$, derived by combining 
$J$-, $H$- and $K$-band images.
The bulge/disk decomposition and the corrections to the bulge and disk surface brightness
(face-on, Galactic-extinction, and k-corrections) resulted in corrected bulge
($\mu_{K,\rm bulge}$) and disk ($\mu_{K,\rm disk}^{i}$) radial surface brightness profiles.
We defined the bulge radius ($\rbulge$) to be the radius where the light from the bulge
contributes 10\% to the light at that radius.
The NIR surface brightness will be used to trace the stellar mass of the galaxy
(Sect.~\ref{sec:Sigma}).

As described in \citetalias{martinsson2012a}, we have derived $\hr$ in an iterative way,
fitting an exponential function to the observed $K$-band surface brightness profile 
between 1--4$\hr$. Using Eq.~1 in \citetalias{bershady2010b}, we calculate $\hz$ from
$\hr$ with a systematic error of 25\%. The derived $\hz$ is used for our conversion from
$\sigz$ to $\sddisk$ (Eq.~\ref{eq:mldyn}). Table~\ref{tab:hRhz} contains the derived
$\hr$ and $\hz$ for each target galaxy. Other derived parameters, such as the fitted scale
length in arcsec, central face-on-corrected surface brightness of the disk 
($\mu_{0,K}^{i}$), bulge-to-disk ratio ($B/D$), and the absolute $K$-band magnitude 
($M_K$), are tabulated in Table~1 and Table~5 of \citetalias{martinsson2012a}.

%
\begin{table}[t]
\caption{\label{tab:hRhz}
Scale lengths and scale heights}
\centering
\begin{tabular}{|r r r |r r r|}
\hline
\multicolumn{1}{|c}{UGC}         & 
\multicolumn{1}{c}{$\hr$}        & 
\multicolumn{1}{c}{$\hz$}        &
\multicolumn{1}{|c}{UGC}         & 
\multicolumn{1}{c}{$\hr$}        & 
\multicolumn{1}{c|}{$\hz$}       \\
\multicolumn{1}{|c}{}            &
\multicolumn{1}{c}{(kpc)}        &
\multicolumn{1}{c}{(kpc)}        &
\multicolumn{1}{|c}{}            &
\multicolumn{1}{c}{(kpc)}        &
\multicolumn{1}{c|}{(kpc)}       \\
\hline
  448 & 3.9 $\pm$ 0.2 & 0.46 $\pm$ 0.10 &  4368 & 3.2 $\pm$ 0.3 & 0.41 $\pm$ 0.10 \\
  463 & 3.8 $\pm$ 0.2 & 0.45 $\pm$ 0.10 &  4380 & 5.0 $\pm$ 0.2 & 0.54 $\pm$ 0.12 \\
 1081 & 3.1 $\pm$ 0.2 & 0.40 $\pm$ 0.09 &  4458 & 9.0 $\pm$ 0.4 & 0.79 $\pm$ 0.17 \\
 1087 & 3.2 $\pm$ 0.2 & 0.41 $\pm$ 0.09 &  4555 & 4.1 $\pm$ 0.2 & 0.48 $\pm$ 0.11 \\
 1529 & 3.6 $\pm$ 0.1 & 0.44 $\pm$ 0.10 &  4622 & 7.6 $\pm$ 0.5 & 0.70 $\pm$ 0.16 \\
 1635 & 2.9 $\pm$ 0.2 & 0.39 $\pm$ 0.09 &  6903 & 4.2 $\pm$ 0.4 & 0.49 $\pm$ 0.11 \\
 1862 & 1.4 $\pm$ 0.2 & 0.24 $\pm$ 0.06 &  6918 & 1.2 $\pm$ 0.1 & 0.21 $\pm$ 0.05 \\
 1908 & 4.9 $\pm$ 0.2 & 0.53 $\pm$ 0.12 &  7244 & 3.9 $\pm$ 0.4 & 0.46 $\pm$ 0.11 \\
 3091 & 3.6 $\pm$ 0.2 & 0.44 $\pm$ 0.10 &  7917 & 8.5 $\pm$ 0.4 & 0.76 $\pm$ 0.17 \\
 3140 & 3.5 $\pm$ 0.2 & 0.43 $\pm$ 0.10 &  8196 & 4.9 $\pm$ 0.1 & 0.53 $\pm$ 0.12 \\
 3701 & 3.6 $\pm$ 0.4 & 0.44 $\pm$ 0.11 &  9177 & 7.0 $\pm$ 0.3 & 0.67 $\pm$ 0.15 \\
 3997 & 5.5 $\pm$ 0.5 & 0.58 $\pm$ 0.14 &  9837 & 5.8 $\pm$ 0.4 & 0.60 $\pm$ 0.14 \\
 4036 & 4.3 $\pm$ 0.4 & 0.49 $\pm$ 0.12 &  9965 & 3.5 $\pm$ 0.2 & 0.44 $\pm$ 0.10 \\
 4107 & 3.2 $\pm$ 0.2 & 0.41 $\pm$ 0.09 & 11318 & 4.5 $\pm$ 0.2 & 0.51 $\pm$ 0.11 \\
 4256 & 4.7 $\pm$ 0.2 & 0.52 $\pm$ 0.12 & 12391 & 3.9 $\pm$ 0.2 & 0.46 $\pm$ 0.10 \\
\hline
\end{tabular}
\tablefoot{
Table containing measured stellar-disk scale lengths ($\hr$) and inferred stellar-disk
scale heights ($\hz$). The conversion from arcsec to kpc is done using distances tabulated
in \citetalias{martinsson2012a}.
}
\end{table}


\subsection{Stellar and ionized-gas kinematics}
\label{sec:Kinematics}
\citetalias{martinsson2012a} also presented the reduction and analysis of the stellar and 
ionized-gas (\oiii$\lambda$5007\AA) kinematics from optical spectroscopy taken with PPak,
where the stellar kinematics were derived as described in \cite{westfall2011a}.
In the current paper, we will not use any of the derived \oiii\ kinematics; instead we
will rely on our \halp\ data taken with SparsePak on the WIYN 3.5m 
telescope.\footnote{The WIYN Observatory, a joint facility of the University of
Wisconsin-Madison, Indiana University, Yale University, and the National Optical Astronomy
Observatories.}
The reason why we exclude the \oiii\ kinematics is that, empirically, it show
large scatter in the velocity fields, likely due to astrophysical properties of the gas
such as local outflows associated with star-forming regions, and deviations from circular
orbits. Generally, the \oiii\ data also have a lower signal-to-noise ratio (S/N) in the
line compared to the \halp\ data. Thus, the \oiii\ data suffer more severely from both
systematic and random errors in the characterization of the gas rotation curve.

The SparsePak integral-field spectroscopy was obtained in the \halp\ region for all
galaxies in the PPak Sample using the setup as described in Table~1 of 
\citetalias{bershady2010a}. Typically, a three-pointing dither pattern was followed,
designed to fully sample the $72\arcsec \times 71\arcsec$ field-of-view of SparsePak. The
reduction of these data, such as basic data reduction, spectral extraction and wavelength
calibration, follows methods described in \cite{andersen2006}, and will be described in a
forthcoming paper.

The \halp\ kinematics are measured in a similar way as the \oiii\ kinematics 
\citepalias{martinsson2012a}, following \cite{andersen2006, andersen2008}, with both
single and double Gaussian line profiles fitted in a 20\AA\ window centered around each
line.
Velocities are calculated using the wavelengths of the Gaussian centroids. Of all fitted 
line profiles, 27\% are better fit by a double Gaussian profile (\citealt{andersen2008});
in these cases, however, a single component is used to measure the line-of-sight velocity.
The measured velocities are used in Sect.~\ref{sec:RC} to derive \halp\ rotation curves by
fitting tilted rings to the data.

\subsection{24-$\mu$m {\it Spitzer} photometry}
\label{sec:Spitzer}
For the characterization of the molecular-gas content, we use 24-$\mu$m photometry
obtained with {\it Spitzer}. The general motivation and survey strategy for our 
{\it Spitzer} observations is provided in \citetalias{bershady2010a}. Images at 4.5, 8,
24, and 70 $\mu$m were obtained for all galaxies in the PPak Sample. Here, we will only 
use the 24-$\mu$m MIPS observations to infer the CO surface-brightness distribution from 
the  24-$\mu$m flux, motivated by the well-correlated relations between CO and infrared
emission \citep[e.g.,][]{young1991, paladino2006, regan2006, bendo2010}.
The inferred CO surface brightness follows from the empirical relation derived by
\citet[][hereafter \citetalias{westfall2011b}]{westfall2011b} based on the CO and 
24-$\mu$m data provided by \cite{leroy2008}.

We obtained reliable 24-$\mu$m surface photometry and reached S/N$=$3 per spatial
resolution element at $3\hr$, matching the extent of the kinematic IFU measurements.
Images of the galaxies at 24 $\mu$m can be found in Fig.~9 of
\citetalias{bershady2010a}. More details on the reduction will be provided in a
forthcoming paper.

\subsection{21-cm radio synthesis imaging}
\label{sec:HIdata}
In \cite{martinsson2011}, we presented the reduction and results from our 21-cm radio 
synthesis observations of 28 galaxies, of which 24 are part of the PPak sample. 
The observations were obtained using the VLA, WSRT and GMRT arrays, with three galaxies
observed by both the VLA and WSRT. The data have been smoothed to $\sim$15$\arcsec$
angular resolution and $\sim$10 km s$^{-1}$ velocity resolution, giving typical 
column-density sensitivities of 2--5 $\times 10^{20}$ atoms cm$^{-2}$.
Here, we will use the derived \hone\ mass surface densities to estimate the amount of
atomic gas in the disk (Sect.~\ref{sec:SigmaHImod}), and the measured rotation speed of
the \hone\ gas to obtain extended rotation curves (Sect.~\ref{sec:RC}).

\section{Baryonic Mass Distributions}
\label{sec:Sigma}
This section describes how the mass surface densities of the baryonic components are 
derived, closely following the analysis as outlined in \citetalias{westfall2011b}.
The mass surface densities will be used in Sect.~\ref{sec:RCmod} to calculate the baryonic 
rotation curves. While the atomic gas mass surface density ($\sda$) is observed directly
from our \hone\ observations, the molecular gas mass surface density ($\sdm$)
is derived indirectly from 24-$\mu$m {\it Spitzer} observations. The stellar-kinematic 
observations of $\sigz$ are used to calculate dynamical mass surface densities ($\sddisk$),
from which we obtain the stellar mass surface densities ($\sds$) by subtracting the gas
mass contributions.
From the derived $\sds$ and $\mu_{K,\rm disk}^{i}$, we calculate the average stellar
mass-to-light ratio of the galaxy disk ($\overline{\mls}$). We assign the same $\mls$ to
the bulge and the disk, and calculate the radial mass surface density profiles of the
stellar disk ($\sds^{\rm disk}$) and bulge ($\sds^{\rm bulge}$) separately, as well as
their total masses, using radial surface brightness profiles (Sect.~\ref{sec:Photometry})
of the disk and bulge, respectively. The total masses of the various baryonic components
are tabulated in Table~\ref{tab:Masses}.
At the end of this section, we include an investigation of the masses of the various 
dynamical components in relation to each other and to global photometric and kinematic
properties of the galaxies.

\subsection{Atomic gas mass surface density}
\label{sec:SigmaHImod}
From our 21-cm radio synthesis observations
\citep[Sect.~\ref{sec:HIdata};][]{martinsson2011}, we have measured the \hone\ mass
surface densities ($\sdhi$) for 28 galaxies, of which 24 galaxies are also part of the
PPak sample. We found that the radial $\sdhi$ profiles of these galaxies are well fit with
a Gaussian function,
\begin{equation}
\Sigma_{\rm HI}(R) = \Sigma_{\rm HI}^{\rm max}
\exp \left[-\frac{\left(R-R_{\rm \Sigma,max}\right)^2}{2\sigma_{\Sigma}^2}\right] ,
\label{eq:Sigma_HI}
\end{equation}
with $R_{\rm \Sigma,max}$ being the radius at which the profile peaks, $\sigma_{\Sigma}$ 
the width of the profile, and $\Sigma_{\rm HI}^{\rm max}$ the peak density. The tightest 
fit is found when normalizing the radius with the \hone\ radius, $R_{\rm HI}$, defined as 
the radius where $\sdhi$$=$$1\msol \rm pc^{-2}$.
In \cite{martinsson2011}, we also found another tight relation between the total 
\hone\ mass ($\mhi$) and the \hone\ diameter ($D_{\rm HI}$$=$$2 R_{\rm HI}$), 
\begin{equation}
\log(\mhi) = 1.72 \log(D_{\rm HI}) + 6.92,
\label{eq:MHIDHI}
\end{equation}
where $\mhi$ is the mass in units of solar masses and $D_{\rm HI}$ is measured in kpc.

We have direct measurements of $\sdhi$ for 24 out of 30 galaxies in this paper. For the
remaining six galaxies lacking $\sdhi$ measurements, we use our results above to model 
their $\sdhi$ profiles.
We calculate $\mhi$ from literature values of the flux from single dish \hone\
observations taken from the NASA/IPAC Extragalactic Database\footnote{Operated by the Jet
Propulsion Laboratory, California Institute of Technology, under contract with the
National Aeronautics and Space Administration.} (NED). These $\mhi$ are used in
Eq.~\ref{eq:MHIDHI} to calculate $R_{\rm HI}$.
We then use Eq.~\ref{eq:Sigma_HI}, with parameter values found from averaging all 28
galaxies ($R_{\rm \Sigma,max}$$=$$0.39R_{\rm HI}$, $\sigma_{\Sigma}$$=$$0.35R_{\rm HI}$),
to calculate $\Sigma_{\rm HI}(R)$, where the normalization constant
$\Sigma_{\rm HI}^{\rm max}$ is found by calibrating the $\Sigma_{\rm HI}$ 
profile to our estimated $\mhi$. The calculated $\Sigma_{\rm HI}^{\rm max}$
(Table~\ref{tab:ModelSigmaHI}) are typical of the values found for other galaxies in our 
sample.

We test how the use of Gaussian fits instead of actual $\sdhi$ measurements on
these 6 galaxies affects our derived results in this paper by recalculating
$\overline{\mls}$ (Sect.~\ref{sec:ML}) and the baryonic maximality at $2.2\hr$
($\Fbary^{2.2\hr}$; Sect.~\ref{sec:DiskMaximality}) for the other 24 galaxies, using
Gaussian fits based on their total \hone\ fluxes \citep[taken from][]{martinsson2011}.
The effects appear to be small, with average absolute differences on $\overline{\mls}$
and $\Fbary^{2.2\hr}$ from using the measured $\sdhi$ profiles of 4\% and 2\%,
respectively. The differences on individual galaxies are always well within the errors.

To calculate the atomic gas mass surface density, we follow earlier papers in this series
\citepalias[][]{bershady2010a,bershady2010b,westfall2011b}, and multiply $\sdhi$ by a
factor 1.4 to account for the helium and metal fraction; $\sda$$=$$1.4\sdhi$.

%
\begin{table}[t]
\caption{\label{tab:ModelSigmaHI}
Modeled $\Sigma_{\rm HI}$}
\centering
\begin{tabular}{|r c c c c c|}
\hline
\multicolumn{1}{|c}{UGC}                         & 
\multicolumn{1}{c}{Distance}                     & 
\multicolumn{1}{c}{$\int S_{\rm HI}{\rm d}V$}    & 
\multicolumn{1}{c}{$\mhi$}                       & 
\multicolumn{1}{c}{$D_{\rm HI}$}                 & 
\multicolumn{1}{c|}{$\Sigma_{\rm HI}^{\rm max}$} \\
\multicolumn{1}{|c}{}                            &
\multicolumn{1}{c}{(Mpc)}                        &
\multicolumn{1}{c}{(Jy \kms)}                    &
\multicolumn{1}{c}{($10^9 \msol$)}               &
\multicolumn{1}{c}{(kpc)}                        &
\multicolumn{1}{c|}{($\msol {\rm pc^{-2}}$)}     \\
\hline
$1081^1$  &  41.8 & 6.7  & 2.8  & 29.5 & 5.9 \\
$1529^1$  &  61.6 & 4.4  & 3.9  & 36.1 & 5.6 \\
$1862^2$  &  18.4 & 3.4  & 0.3  &  7.8 & 8.2 \\
$1908^1$  & 110.0 & 5.6  & 16.0 & 80.4 & 4.6 \\
$3091^3$  &  73.8 & 4.2  & 5.4  & 43.2 & 5.3 \\
$12391^1$ &  66.8 & 16.4 & 17.3 & 82.0 & 4.5 \\
\hline
\end{tabular}
\tablefoot{Distances are taken from \citetalias{bershady2010a}. Integrated \hone\ fluxes
($\int S_{\rm HI}{\rm d}V$) are from tabulated values in NED, with three different 
sources; 
$^1$\citep[RC3;][]{deVaucouleurs1991}, 
$^2$\citep[HIPASS;][]{doyle2005}, 
$^3$\citep[][]{andersen2006}.
The total \hone\ mass ($\mhi$), \hone\ diameter ($D_{\rm HI}$) and maximum \hone\ 
mass surface density ($\Sigma_{\rm HI}^{\rm max}$) are calculated using relations derived 
in \cite{martinsson2011}.
}
\end{table}


\subsection{Molecular gas mass surface density}
The mass surface density of the molecular gas ($\sdm$) is inferred from our 24-$\mu$m
{\it Spitzer} imaging (Sect.~\ref{sec:Spitzer}), as described in
\citetalias{westfall2011b}. This is done in three steps: First, we derive the 
$^{12}{\rm CO}$($J$$=$1$\rightarrow$0) column density ($I_{\rm CO} \Delta V$) from the 
24-$\mu$m surface brightness using Eq.~1 in \citetalias{westfall2011b}, converting the
sky-subtracted 24-$\mu$m image to a CO column-density map. This conversion is expected to
provide an estimate for $I_{\rm CO} \Delta V$ to within $\sim$30\% 
\citepalias{westfall2011b}. Subsequently, we calculate the molecular hydrogen ($\rm H_2$)
mass surface density ($\sdmh$) from Eq.~2 in \citetalias{westfall2011b} adopting the same
conversion factor,
$X_{\rm CO}$$=$(2.7$\pm$0.9)$\times$$\rm10^{20}~cm^{-2}~(K~km~s^{-1})^{-1}$, calculated
from  combining the Galactic measurement of $X_{\rm CO}$ from \cite{dame2001} with the 
measurements for M31 and M33 from \cite{bolatto2008}. Finally, we multiply by a factor
1.4 to add helium and metals to the molecular gas density; $\sdm$$=$1.4$\sdmh$.

The limitations in estimating the molecular-gas content from the observed 24-$\mu$m 
emission is discussed in more detail in \citetalias{westfall2011b}. Here we note that the 
estimated systematic errors are fairly large ($\delta\sdmh/\sdmh$$=$42\%) and often an 
important error contributor to the stellar mass-to-light ratios calculated below.

\subsection{Dynamical and stellar disk mass surface densities}
As mentioned in Sect.~\ref{sec:Introduction}, the stellar velocity dispersion is a direct
indicator of the local mass surface density. In particular, Eq.~\ref{eq:mldyn} directly
relates $\sigz$ and $\sddisk$. The observed line-of-sight velocity dispersions ($\slos$)
were presented in \citetalias{martinsson2012a}. These were deprojected into the three
components ($\sigr,\sigp, \sigz$) of the stellar velocity ellipsoid (SVE), using the 
derived disk orientations from \citetalias{martinsson2012a} and an SVE shape as justified
in \citetalias{bershady2010b} with $\alpha$$\equiv$$\sigz/\sigr$$=$0.6$\pm$0.15 and 
$\beta$$\equiv$$\sigp/\sigr$$=$0.7$\pm$0.04 at all radii for all galaxies.

From Eq.~\ref{eq:mldyn}, we calculate the total dynamical mass surface density of the 
disk. In this paper, we assume a disk with a single scale height and an exponential
vertical density distribution \citep{kruit1988}. These assumptions on the structure of the
disk have their limitations, especially in the very center of the galaxy
where a non-negligible bulge or a bar may be present, and in the outer part of the disk,
which could be affected by the dark-matter halo or may be flared
\citep[e.g.,][]{GrijsPeletier1997}.
However, in our analysis we exclude any kinematic measurements inside the bulge region,
and measurements in the outer disk will be heavily down-weighted due to larger measurement
errors. More discussion on the effects of possible systematic errors is presented in 
Sect.~\ref{sec:Discussion}.

The measured stellar mass surface density ($\sds$) is derived by subtracting the atomic 
and molecular gas mass surface densities from $\sddisk$,
\begin{equation}
\sds = \sddisk - \sda - \sdm.
\label{eq:Sigma_star}
\end{equation}
We assume that any dark matter in the disk is distributed in the same way as the stars,
and is effectively incorporated into $\sds$.

In the following subsection, we calculate the average stellar mass-to-light ratios of the
disks and use these together with the surface brightness to calculate $\sds$. We do this
to be able to separate the bulge and disk, which have different density distributions and
therefore need to be treated differently when modeling their rotation curves, and to 
calculate $\sds$ profiles that reach further out in radius, beyond our stellar-kinematic
measurements. The measured and calculated $\sds$ profiles are provided in the Atlas
(Appendix~\ref{sec:Atlas}).

\subsection{Mass-to-light ratios}
\label{sec:ML}
%
\begin{figure}
\centering
\includegraphics[width=0.5\textwidth]{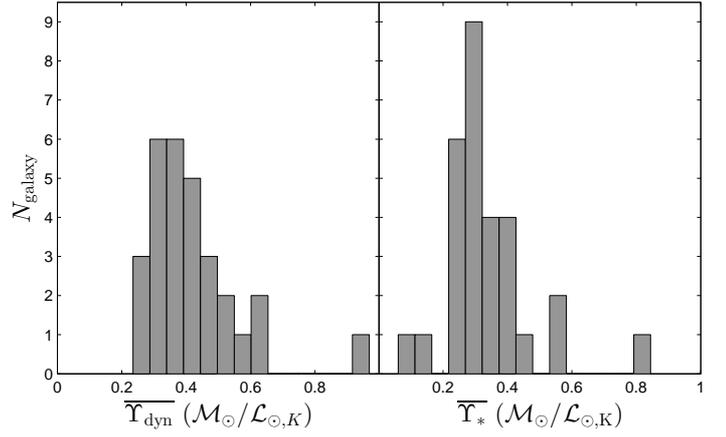}
\caption[]{{\small 
Distributions of radially-averaged dynamical (left) and stellar (right) mass-to-light
ratios.
}}
\label{fig:MLhist}
\end{figure}
%
We calculate dynamical ($\mldyn$) and stellar ($\mls$) mass-to-light ratios using our 
measurements of $\sddisk$ and $\sds$, respectively; both are calculated in the $K$-band 
using $\mu_{K,\rm disk}^{i}$. Error-weighted radial averages ($\overline{\mldyn}$ and 
$\overline{\mls}$) are calculated after excluding radial regions where our dynamical
and/or structural assumptions are less robust. In particular, we exclude data within
$R$$=$$\rbulge$, and $R$$=$$2\farcs5$ if no bulge is apparent.
For UGC~7917, we extend the excluded region to $1\hr$ due to the presence of a bar.
For UGC~4458, $\rbulge$ is larger than the field-of-view of the PPak IFU, and we shorten
the radius to be able to include the outermost measurement. At this radius, the bulge
contributes 21\% to the light. For each galaxy, the excluded regions are indicated by
gray shaded areas in the Atlas. For UGC~8196, our analysis gives us non-physical results
when calculating the baryon maximality (see Sect.~\ref{sec:DiskMaximality}), possibly due
to an erroneous measurement of $\mldyn$. We therefore exclude this galaxy from all our
results in this paper related to $\mldyn$.

Figure~\ref{fig:MLhist} shows the distributions of $\overline{\mldyn}$ and 
$\overline{\mls}$. We find the weighted radial averages and scatter of the sample to be
$\langle$$\overline{\mldyn}$$\rangle$$=$$0.39$$\pm$$0.08 \msol/\lksol$ and 
$\langle$$\overline{\mls}$$\rangle$$=$$0.31$$\pm$$0.07 \msol/\lksol$.
The effective radius at which $\overline{\mls}$ is measured ($R_{\Upsilon}$; calculated
using the same weights as used for $\overline{\mls}$) has a weighted sample average and
scatter of $\langle$$R_{\Upsilon}$$\rangle$$=$1.0$\pm$0.3$\hr$. The calculated 
$\overline{\mls}$ and $R_{\Upsilon}$ of the individual galaxies are tabulated in 
Table~\ref{tab:DMfit}.

\begin{figure}
\centering
\includegraphics[width=0.5\textwidth]{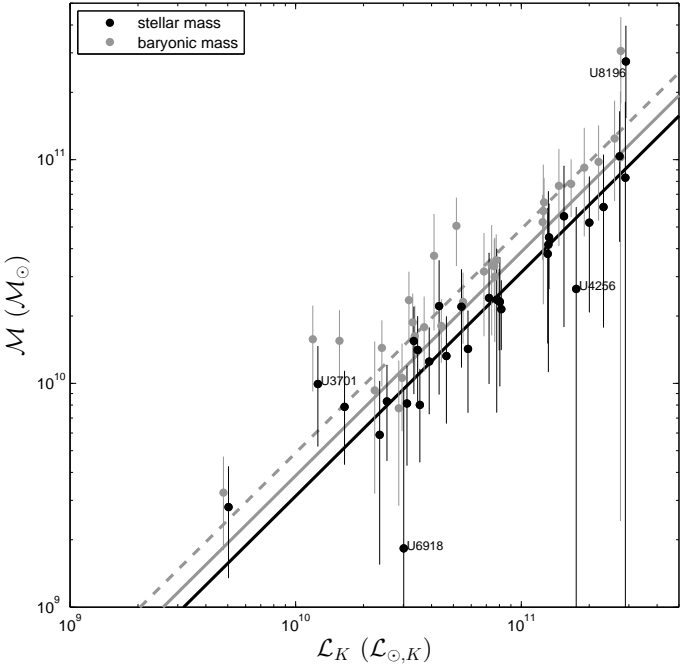}
\caption[]{{\small 
Mass ($\mathcal{M}$) versus luminosity ($\lk$). Total baryonic mass (stars+gas; gray 
symbols, slightly offset in $\lk$ for clarity) and stellar mass (disk+bulge; black 
symbols) as a function of luminosity. The solid lines indicate the average measured 
$\langle$$\overline{\mldyn}$$\rangle$ (gray) and $\langle$$\overline{\mls}$$\rangle$
(black). The average ratio $\mb/\lk$ of the sample
($\langle$$\ml_{\rm b}$$\rangle$; see text) is indicated with a gray dashed line.
}}
\label{fig:MvsL}
\end{figure}

The radially-averaged mass-to-light ratios are weighted more towards measurements in the
inner region of the galaxy due to increasing errors with radius. Even if the galaxies have
constant $\mls$ with radius, $\mldyn$ will still vary if the radial distribution of the
gas is different from that of the stars. If there is relatively more gas further out in
the disk than in the inner region (as seen in Fig.~\ref{fig:Sigma_bar}), the 
error-weighted mean $\overline{\mldyn}$ will be lower than the {\it total} baryonic 
mass-to-light ratio $\ml_{\rm b}$$=$$\mb/\mathcal{L}_{K}$, which is calculated from the 
total integrated baryonic mass ($\mb$; Table~\ref{tab:Masses}) and the total $K$-band
luminosity ($\mathcal{L}_{K}$).
Figure~\ref{fig:MvsL} shows the baryonic and stellar mass as a function of 
$\mathcal{L}_{K}$ and demonstrates that $\mb$ tends to be larger than what is expected
from $\langle$$\overline{\mldyn}$$\rangle$. In detail, we find 
$\langle$$\ml_{\rm b}$$\rangle$$=$$0.49 \msol/\lksol $$=$$
1.3\times$$\langle$$\overline{\mldyn}$$\rangle$
as expected given the generally larger radial extent of the \hone\ disks 
\citep{martinsson2011}.

%
\begin{table}
\caption{\label{tab:Masses}
Total masses of the baryonic components}
\centering
\renewcommand{\tabcolsep}{1.4mm}
\renewcommand{\arraystretch}{1.5}
\begin{tabular}{|r r r r r r|}
\hline
\multicolumn{1}{|c}{UGC}                  &    
\multicolumn{1}{c}{$\log(\mdiskstar)$}    &   
\multicolumn{1}{c}{$\log(\mbulgestar)$}   &    
\multicolumn{1}{c}{$\log(\matom)$}        &   
\multicolumn{1}{c}{$\log(\mmol)$}         &   
\multicolumn{1}{c|}{$\log(\mbary)$}        \\
\multicolumn{1}{|c}{}              &    
\multicolumn{1}{c}{$(\msol)$}      &   
\multicolumn{1}{c}{$(\msol)$}      &    
\multicolumn{1}{c}{$(\msol)$}      &  
\multicolumn{1}{c}{$(\msol)$}      &    
\multicolumn{1}{c|}{$(\msol)$}     \\
\hline
  448 & $10.25_{-0.51}^{+0.23}$ & $ 9.76_{-0.51}^{+0.23}$ & $ 9.88_{-0.05}^{+0.04}$ & $ 9.39_{-0.23}^{+0.15}$ & $10.53_{-0.31}^{+0.18}$ \\
  463 & $10.54_{-0.41}^{+0.21}$ & $ 9.35_{-0.41}^{+0.21}$ & $ 9.57_{-0.05}^{+0.04}$ & $10.04_{-0.23}^{+0.15}$ & $10.72_{-0.17}^{+0.12}$ \\
 1081 & $10.12_{-0.24}^{+0.16}$ & $ 8.81_{-0.24}^{+0.16}$ & $ 9.59_{-0.05}^{+0.04}$ & $ 8.89_{-0.23}^{+0.15}$ & $10.27_{-0.18}^{+0.13}$ \\
 1087 & $10.08_{-0.25}^{+0.16}$ & $ 8.64_{-0.25}^{+0.16}$ & $ 9.67_{-0.05}^{+0.04}$ & $ 8.84_{-0.23}^{+0.15}$ & $10.25_{-0.20}^{+0.14}$ \\
 1529 & $10.34_{-0.39}^{+0.20}$ & $ 8.98_{-0.39}^{+0.20}$ & $ 9.74_{-0.05}^{+0.04}$ & $ 9.12_{-0.23}^{+0.15}$ & $10.47_{-0.29}^{+0.17}$ \\
 1635 & $ 9.88_{-0.29}^{+0.17}$ & $ 8.74_{-0.29}^{+0.17}$ & $ 9.32_{-0.05}^{+0.04}$ & $ 8.57_{-0.23}^{+0.15}$ & $10.03_{-0.24}^{+0.15}$ \\
 1862 & $ 9.45_{-0.37}^{+0.20}$ & $    -                $ & $ 8.58_{-0.05}^{+0.04}$ & $ 7.85_{-0.24}^{+0.15}$ & $ 9.51_{-0.26}^{+0.16}$ \\
 1908 & $10.74_{-0.54}^{+0.23}$ & $ 9.68_{-0.54}^{+0.23}$ & $10.35_{-0.05}^{+0.04}$ & $10.15_{-0.23}^{+0.15}$ & $10.99_{-0.26}^{+0.16}$ \\
 3091 & $ 9.90_{-0.28}^{+0.17}$ & $    -                $ & $ 9.88_{-0.05}^{+0.04}$ & $ 8.85_{-0.23}^{+0.15}$ & $10.21_{-0.20}^{+0.14}$ \\
 3140 & $10.59_{-0.24}^{+0.15}$ & $ 9.65_{-0.24}^{+0.15}$ & $10.11_{-0.05}^{+0.04}$ & $ 9.82_{-0.23}^{+0.15}$ & $10.80_{-0.14}^{+0.11}$ \\
 3701 & $10.00_{-0.31}^{+0.18}$ & $ 8.69_{-0.31}^{+0.18}$ & $ 9.75_{-0.05}^{+0.04}$ & $ 8.28_{-0.24}^{+0.15}$ & $10.21_{-0.23}^{+0.15}$ \\
 3997 & $10.16_{-0.27}^{+0.17}$ & $ 8.53_{-0.27}^{+0.17}$ & $ 9.88_{-0.05}^{+0.04}$ & $ 8.74_{-0.22}^{+0.15}$ & $10.36_{-0.18}^{+0.13}$ \\
 4036 & $10.35_{-0.23}^{+0.15}$ & $ 8.92_{-0.23}^{+0.15}$ & $ 9.88_{-0.05}^{+0.04}$ & $ 9.42_{-0.23}^{+0.15}$ & $10.52_{-0.17}^{+0.12}$ \\
 4107 & $10.10_{-0.31}^{+0.18}$ & $ 8.65_{-0.31}^{+0.18}$ & $ 9.43_{-0.05}^{+0.04}$ & $ 9.32_{-0.23}^{+0.15}$ & $10.25_{-0.16}^{+0.12}$ \\
 4256 & $10.38_{-10.38}^{+0.36}$ & $ 9.29_{-9.29}^{+0.36}$ & $10.33_{-0.05}^{+0.04}$ & $10.48_{-0.23}^{+0.15}$ & $10.89_{-0.15}^{+0.11}$ \\
 4368 & $10.37_{-0.41}^{+0.21}$ & $ 9.21_{-0.41}^{+0.21}$ & $10.13_{-0.05}^{+0.04}$ & $ 9.17_{-0.23}^{+0.15}$ & $10.60_{-0.33}^{+0.19}$ \\
 4380 & $10.32_{-0.20}^{+0.13}$ & $ 8.86_{-0.20}^{+0.13}$ & $10.08_{-0.05}^{+0.04}$ & $ 9.29_{-0.23}^{+0.15}$ & $10.56_{-0.16}^{+0.12}$ \\
 4458 & $10.81_{-0.39}^{+0.20}$ & $10.67_{-0.39}^{+0.20}$ & $10.26_{-0.05}^{+0.04}$ & $ 9.45_{-0.23}^{+0.15}$ & $11.13_{-0.28}^{+0.17}$ \\
 4555 & $10.39_{-0.39}^{+0.20}$ & $ 8.96_{-0.39}^{+0.20}$ & $ 9.71_{-0.05}^{+0.04}$ & $ 9.39_{-0.23}^{+0.15}$ & $10.51_{-0.29}^{+0.17}$ \\
 4622 & $10.69_{-0.41}^{+0.21}$ & $ 9.89_{-0.41}^{+0.21}$ & $10.55_{-0.05}^{+0.04}$ & $ 9.65_{-0.23}^{+0.15}$ & $10.98_{-0.31}^{+0.18}$ \\
 6903 & $ 9.73_{-0.62}^{+0.25}$ & $ 8.03_{-0.62}^{+0.25}$ & $ 9.49_{-0.05}^{+0.04}$ & $ 8.48_{-0.22}^{+0.15}$ & $ 9.96_{-0.46}^{+0.22}$ \\
 6918 & $ 9.23_{-9.23}^{+0.70}$ & $ 8.04_{-8.04}^{+0.70}$ & $ 9.37_{-0.05}^{+0.04}$ & $ 9.55_{-0.23}^{+0.15}$ & $ 9.89_{-0.44}^{+0.21}$ \\
 7244 & $ 9.83_{-0.29}^{+0.17}$ & $    -                $ & $ 9.86_{-0.05}^{+0.04}$ & $ 8.68_{-0.22}^{+0.14}$ & $10.16_{-0.20}^{+0.14}$ \\
 7917 & $10.87_{-10.87}^{+0.34}$ & $10.01_{-10.01}^{+0.34}$ & $10.14_{-0.05}^{+0.04}$ & $ 9.73_{-0.23}^{+0.15}$ & $11.01_{-1.62}^{+0.30}$ \\
 8196 & $11.35_{-0.26}^{+0.16}$ & $10.73_{-0.26}^{+0.16}$ & $10.45_{-0.05}^{+0.04}$ & $ 9.56_{-0.23}^{+0.15}$ & $11.48_{-0.23}^{+0.15}$ \\
 9177 & $10.64_{-0.58}^{+0.24}$ & $ 9.55_{-0.58}^{+0.24}$ & $10.15_{-0.05}^{+0.04}$ & $ 9.45_{-0.23}^{+0.15}$ & $10.81_{-0.42}^{+0.21}$ \\
 9837 & $ 9.87_{-0.29}^{+0.17}$ & $ 8.35_{-0.29}^{+0.17}$ & $ 9.75_{-0.05}^{+0.04}$ & $ 8.65_{-0.24}^{+0.15}$ & $10.15_{-0.17}^{+0.12}$ \\
 9965 & $10.16_{-0.29}^{+0.17}$ & $    -                $ & $ 9.79_{-0.05}^{+0.04}$ & $ 9.44_{-0.23}^{+0.15}$ & $10.37_{-0.19}^{+0.13}$ \\
11318 & $10.60_{-0.50}^{+0.23}$ & $ 9.69_{-0.50}^{+0.23}$ & $10.05_{-0.05}^{+0.04}$ & $ 9.97_{-0.23}^{+0.15}$ & $10.81_{-0.27}^{+0.16}$ \\
12391 & $10.33_{-0.28}^{+0.17}$ & $ 8.98_{-0.28}^{+0.17}$ & $10.38_{-0.05}^{+0.04}$ & $ 9.64_{-0.23}^{+0.15}$ & $10.71_{-0.18}^{+0.13}$ \\
\hline
\end{tabular}
\tablefoot{Calculated masses of the stellar disk ($\mdiskstar$), stellar bulge
($\mbulgestar$), atomic gas ($\matom$), molecular gas ($\mmol$), and total baryonic mass
of the galaxies ($\mbary$).
}
\end{table}


In many galaxies, we find that $\mls$ increases toward larger radii (see UGC~4107,
UGC~4368 \& UGC~9965 in the Atlas). Three possible explanations for this are:
\begin{enumerate}
\item The mass-to-light ratio of the stellar population could in fact be rising at larger 
radii. Even though color gradients are often seen in spiral galaxies 
\citep[e.g.,][]{deJong1996}, these are usually small and do not sufficiently explain the 
increase in $\mls$ within the context of stellar-population variations \citepalias[see 
discussion on UGC~463 in][]{westfall2011b}.
\item There could be unknown systematic instrumental errors, such that the measurement of 
the observed velocity dispersion ``hit the floor'', giving systematically larger 
measurements of $\slos$. Indeed, the skewed (non-Gaussian) shape of the velocity-dispersion
error distribution for our data may indicate the presence of such an effect. However, with 
our relatively high velocity resolution ($\sinst$$\approx$16.6$\pm$1.4~\kms; 
\citetalias{martinsson2012a}) and a rather conservative rejection cut 
($\delta \slos$$\geq$8~\kms), we do not expect this to happen at the level where we see 
the increasing $\mls$ ($\slos$$>$20~\kms). Moreover, $\sigz$ often follows an exponential 
decline well, while changes in the light profile are the primary reason for the 
increased $\mls$ (e.g.,\ UGC~1862).
\item Our assumptions for calculating $\sddisk$ using Eq.~\ref{eq:mldyn} may not be 
valid at larger radii. There are several reasons why we could get larger $\sigz$ values 
than what would be expected from the local mass surface density of the disk. If the 
stellar disk is flaring at larger radii, our assumption of a constant scale height no 
longer holds, resulting in an overestimate of $\sddisk$. The dark-matter halo could also
have an effect on the stellar velocity dispersions, which might become more important at
larger radii.
\end{enumerate}

\noindent In this paper, we have simply assumed a constant $\mls$, calculated using all
measurements outside the bulge region. However, since the errors increase with radius, our
error-weighted mean $\overline{\mls}$ hinges on measurements in the inner, high S/N region
of each galaxy.
The variation of $\mls$ (within each galaxy and among different galaxies) are interesting 
in the context of star formation and the stellar initial mass function. Here, measurements
of $\mls$ often have large systematic errors, partly due to the uncertain subtraction of 
the molecular gas from $\sddisk$. However, even though we see some scatter in 
Fig.~\ref{fig:MvsL}, the measured $\overline{\mls}$ are in general consistent with being
equal for all galaxies.
A detailed discussion on mass-to-light ratios in the context of stellar populations and
star formation, including an investigation of trends with other global properties of the
disk such as luminosity, surface brightness, color, or scale length, will be presented in 
a forthcoming paper.

For all subsequent analysis, we calculate the radial mass surface density profiles of the 
stellar disk ($\sds^{\rm disk}$) and bulge ($\sds^{\rm bulge}$) based on the disk and 
bulge surface brightness profiles (Sect.~\ref{sec:Photometry}) and a single 
$\overline{\mls}$ calculated for each galaxy.

\subsection{Baryonic mass surface density profiles and total mass fractions}
The mass surface densities of the individual baryonic components ($\sda$, $\sdm$, 
$\sds^{\rm disk}$, $\sds^{\rm bulge}$) are plotted out to 50$\arcsec$ for every galaxy
in the Atlas. For a majority of the galaxies in our sample, the total baryonic mass
surface density is dominated by the stellar component over most of this radial range.
However, some galaxies have a relatively large molecular-gas content distributed in a way 
similar to the stars; 12 galaxies have a molecular gas mass larger than 10\% of the
stellar mass, of which 8 galaxies have a molecular-gas content larger than 10\% of the
total baryonic mass. Two galaxies (UGC~4256 and UGC~6918) even have a larger molecular gas
mass than stellar mass, although it should be noted that these two galaxies have the
lowest $\mls$ in the sample, with large uncertainties in both the molecular and stellar
mass.

\begin{figure}
\centering
\includegraphics[width=0.5\textwidth]{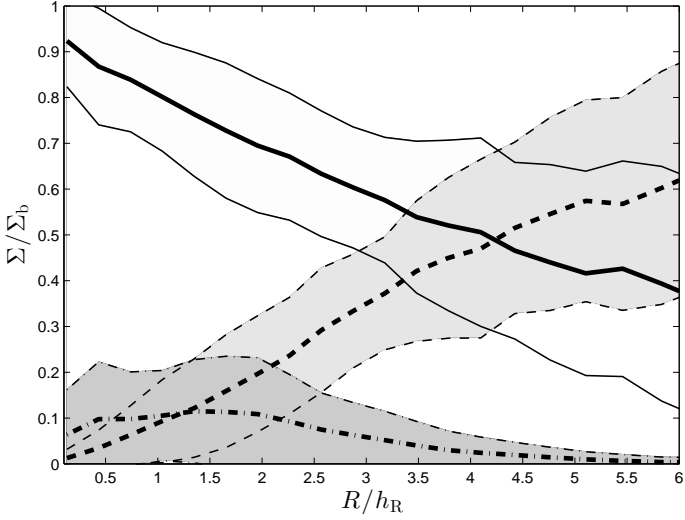}
\caption[]{{\small 
Relative fraction of stellar, atomic-gas, and molecular-gas mass surface densities as a
function of radius. The solid thick and thin lines show the average $\sds/\Sigma_{\rm b}$
and $1 \sigma$ scatter, respectively. The dashed thick and thin lines (enclosing the light
gray area) and the dashed-dotted thick and thin lines (enclosing the dark gray area) show
the average $\sda/\Sigma_{\rm b}$ and $\sdm/\Sigma_{\rm b}$, respectively, again with the
thin lines indicating $1 \sigma$ scatter.
}}
\label{fig:Sigma_bar}
\end{figure}

$\sdm$ is often higher than $\sda$ out to $\sim$1--2$\hr$. In the outer parts of
the disk, the atomic gas starts to dominate the mass surface density of the baryonic
matter. Figure~\ref{fig:Sigma_bar} shows the average fraction in mass surface density of
the stars, atomic gas, and molecular gas. In the center, typically more than 90\% of the
baryonic matter is in stars, while the relative amount of atomic gas is steadily
increasing with radius. On average, at about $4\hr$, the relative amount of gas and stars
are comparable. The atomic gas then typically dominates the baryonic mass surface density
at $R$$\gtrsim$$4\hr$.

\begin{figure}
\centering
\includegraphics[width=0.5\textwidth]{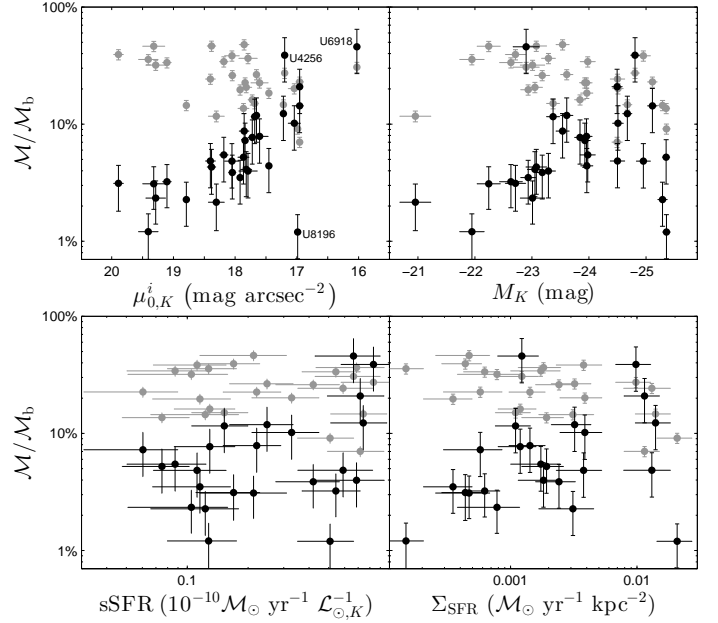}
\caption[]{{\small 
Molecular ($\mmol/\mbary$; black symbols) and atomic ($\matom/\mbary$; gray symbols)
gas-mass fractions as a function of $\mu_{0,K}^{i}$, $M_K$, sSFR and $\Sigma_{\rm SFR}$.
}}
\label{fig:molgasfrac}
\end{figure}

Figure~\ref{fig:molgasfrac} demonstrates that the molecular gas mass fraction (indicated
with black symbols) is rather well correlated with the central surface brightness of the 
disk; the galaxy disks in our sample with higher surface brightness also have higher
molecular gas mass fraction. More luminous galaxies also tend to have larger molecular gas
mass fractions, but with large scatter, especially for high-luminosity galaxies. We see no
correlation with the specific star-formation rate \citep[sSFR; calculated from the 21-cm
continuum flux in ][]{martinsson2011}, but note that the two galaxies with the largest
molecular-gas fraction (UGC~4256 and UGC~6918) are among the galaxies with the largest
sSFR. There is also a weak trend of increased SFR density 
\citep[$\Sigma_{\rm SFR}$;][]{martinsson2011} for galaxies with larger molecular-gas 
fractions. However, it should be noted again that the molecular-gas content is an indirect
estimate from the 24-$\mu$m emission.

The atomic gas mass fractions (indicated in Fig.~\ref{fig:molgasfrac} with gray symbols)
are in general larger than the molecular gas mass fractions. Correlations with 
$\mu_{0,K}^{i}$, $M_K$, sSFR and $\Sigma_{\rm SFR}$ are weak and have opposite sign to
correlations found for the molecular gas mass fraction, resulting in total gas mass
fractions that are consistent with having no correlations with $\mu_{0,K}^{i}$, $M_K$,
sSFR or $\Sigma_{\rm SFR}$.

\subsubsection{Comparison to literature}
To be able to compare our gas-mass fractions with the literature, we will below compare
the atomic and molecular gas masses with the derived stellar masses instead of the total
baryonic masses. It should however be noted that these measurements have larger
uncertainties due to the additional uncertainties on the stellar masses.

Recent studies of the gas content in galaxies using CO and \hone\ observations 
\citep{leroy2009,saintonge2011} have found a molecular gas mass content that is typically
3--10\% of the stellar mass. For the galaxies in our sample, we find that more than one
third have a molecular gas mass larger than 10\% of the stellar mass. This is larger than
found by \cite{leroy2009} or \cite{saintonge2011}, probably because of differences in the
way the stellar mass is determined (see below). However, this could also be due to
differences in the samples, as the molecular gas-mass fraction seems to correlate with
some global galaxy parameters (Fig.~\ref{fig:molgasfrac}).

\begin{figure}
\centering
\includegraphics[width=0.5\textwidth]{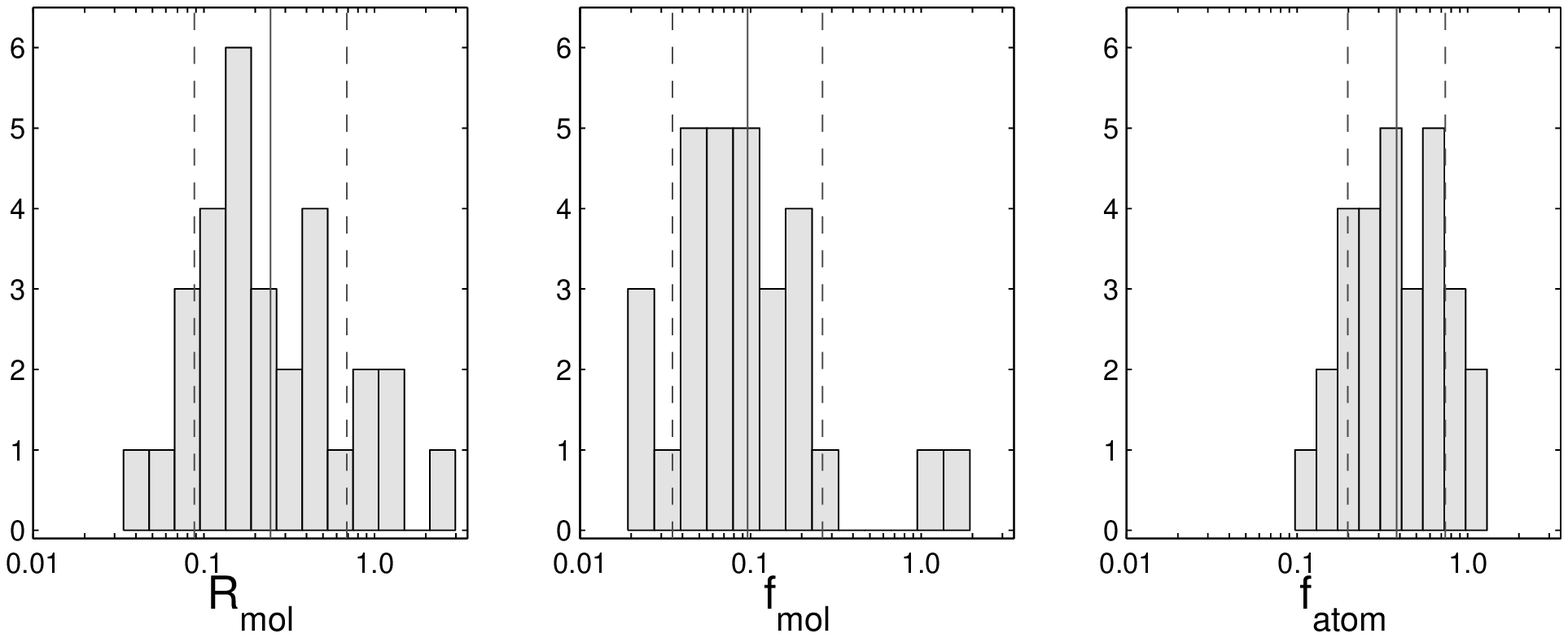}
\caption[]{{\small 
Gas-mass fractions. The vertical solid and dashed lines indicate the mean and one standard
deviation from the mean, respectively.
}}
\label{fig:gasfrac_hist}
%
\includegraphics[width=0.5\textwidth]{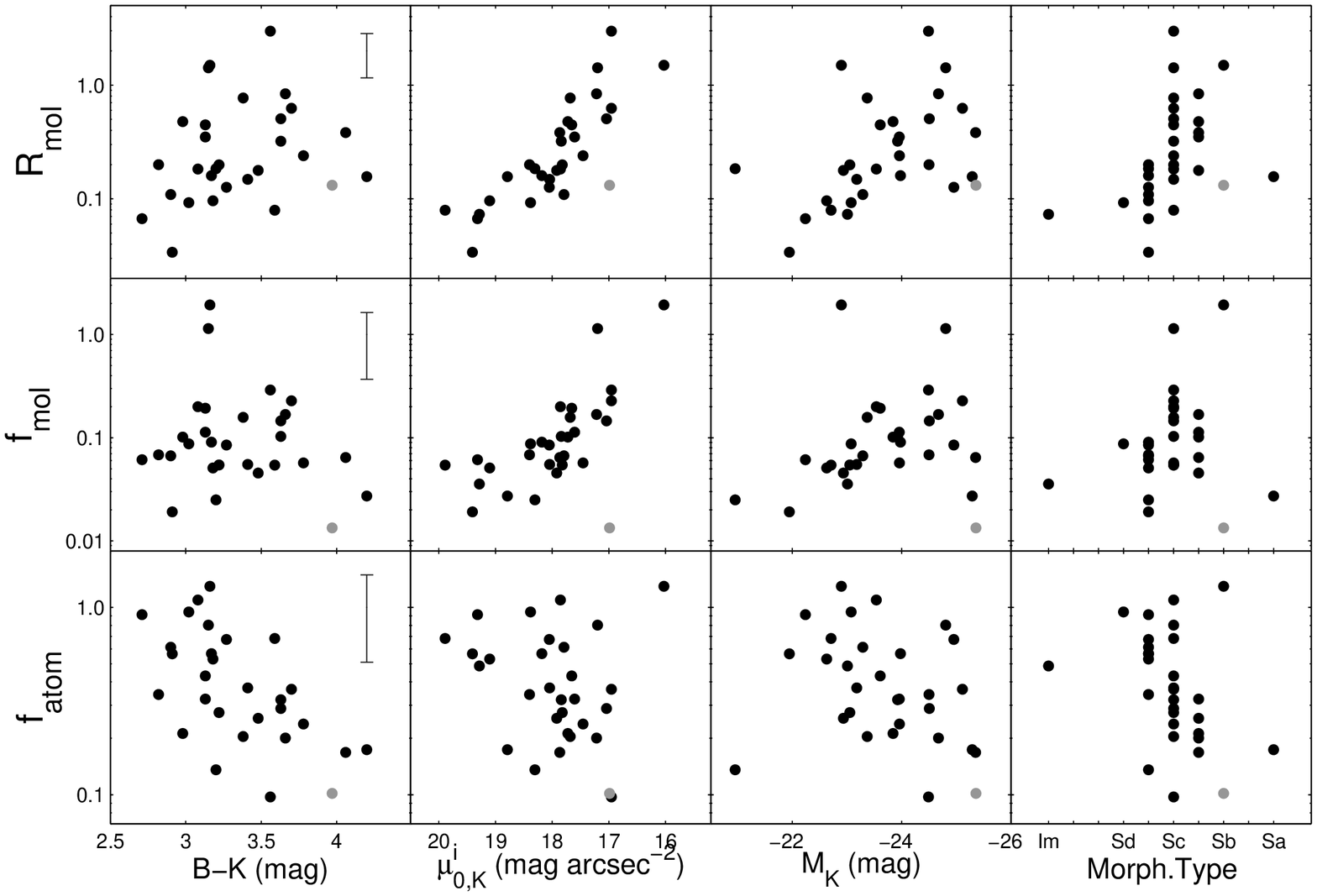}
\caption[]{{\small 
Relations between gas-mass fractions and other global properties of the galaxies.
The errorbars in the upper-right corners of the panels in the left column indicate
typical (mostly systematic) errors. UGC~8196 is marked with a gray symbol.
}}
\label{fig:gasfrac_rel}
\end{figure}

Figure~\ref{fig:gasfrac_hist} shows the distributions of $\rm R_{mol}$$=$$\mmol/\matom$, 
$\rm f_{mol}$$=$$\mmol/\mstar$ and $\rm f_{atom}$$=$$\matom/\mstar$. We calculate the 
geometric means and scatter to be $\langle$$\rm R_{mol}$$\rangle$$=$$0.25^{+0.44}_{-0.16}$,
$\langle$$\rm f_{mol}$$\rangle$$=$$0.10^{+0.17}_{-0.06}$, and 
$\langle$$\rm f_{atom}$$\rangle$$=$$0.38^{+0.36}_{-0.18}$. The values for individual
galaxies cover two orders of magnitudes for ratios including the molecular gas 
(0.03$<$$\rm R_{mol}$$<$2.98, 0.02$<$$\rm f_{mol}$$<$1.93) and one order of magnitude for
the atomic gas mass fraction (0.10$<$$\rm f_{atom}$$<$1.30). These averages
and scatters are similar to what was found by \cite{leroy2009} and \cite{saintonge2011}.
After correcting for differences in adopted $X_{\rm CO}$ values, these studies find
average $\rm R_{mol}$ that are comparable to what we find, with similar scatter. They find
gas-mass fractions ($\rm f_{atom}$ and $\rm f_{mol}$) that are $\sim$30-60\%
lower than what we report. The reason for this systematically lower gas-mass fractions are
likely partly due to the differences in how the stellar masses are determined.
While we measure the stellar masses dynamically, \cite{leroy2009} assume a
constant $\mls$$=$0.5, which is $\sim$60\% larger than our mean $\mls$
(Sec.~\ref{sec:ML}), and \cite{saintonge2011} calculate their stellar masses from
spectral energy distribution (SED) fitting techniques. These could have factors of 2--3 
systematic errors (e.g., \citealt{kauffmann2003}; \citetalias{bershady2011}); 
however, the discrepancies we find here are smaller than that.

The equivalents to our two extreme cases UGC~4256 and UGC~6918 (with $\rm f_{mol}$ equal
to 1.1 and 1.9, both with low $\mls$) cannot be found in these two studies; 
\cite{leroy2009} and \cite{saintonge2011} find a highest measurement of 0.25 and 0.20,
respectively. However, our extreme case UGC~463 with a $\rm R_{mol}$$=$2.98, can be 
compared to the extreme cases of 1.13 in \cite{leroy2009} and 4.09 in
\cite{saintonge2011}.

Figure~\ref{fig:gasfrac_rel} shows relations between the mass fractions and color, surface 
brightness, absolute magnitude and morphological type. For $\rm R_{mol}$ we see trends 
with all four properties; redder, more luminous, higher surface brightness, and 
earlier-type galaxies have a larger fraction of molecular gas compared to the amount of 
atomic gas. \cite{saintonge2011} find similar, though weaker, trends with $\rm R_{mol}$, 
but the correlations with $\rm f_{mol}$ are somewhat different, especially compared to the 
surface brightness. While they find a weak negative trend with effective stellar surface
density, we find a rather strong positive trend with the disk central surface brightness.
However, note that the stellar surface density in \cite{saintonge2011} and our 
$\mu_{0,K}^{i}$ are defined and measured in very different ways. In accordance with
\cite{catinella2010}, $\rm f_{atom}$ is correlated with color, surface brightness and 
luminosity. We also find that the late-type galaxies in our sample tend to have larger
$\rm f_{atom}$.
We note that $\rm f_{mol}$ and $\rm f_{atom}$ have consistently opposite trends with
$\mu_{0,K}^{i}$ and $M_K$; higher-surface-brightness and more luminous galaxies have
larger $\rm f_{mol}$, while the opposite is true for $\rm f_{atom}$. This implies a trend
in conversion rate of \hone\ to $\rm H_2$ and/or in star-formation efficiency.

\begin{figure}
\centering
\includegraphics[width=0.5\textwidth]{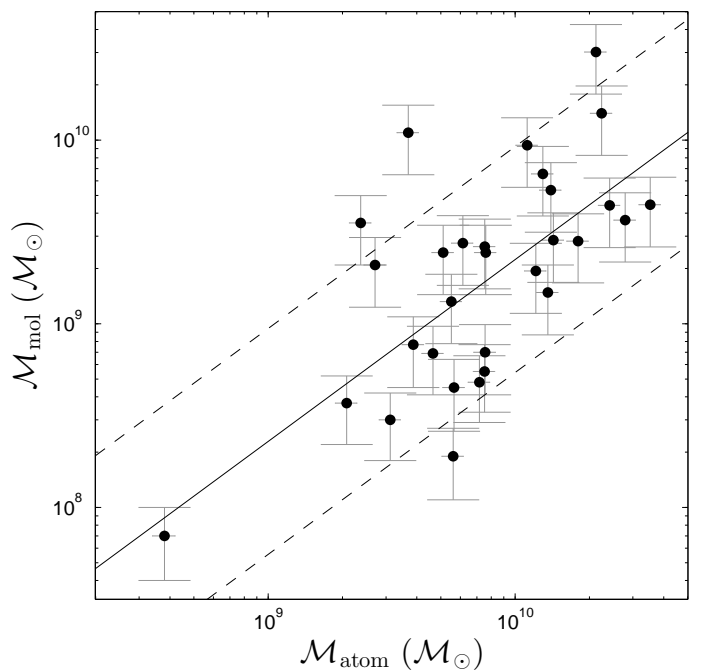}
\caption[]{{\small 
$\mmol$ versus $\matom$. The solid and two dashed lines show the fit from
\cite{saintonge2011}; $\log \mmol$$=$$0.99(\log \matom)-0.42$$\pm$$1.5\sigma$, where
$\sigma=0.41$~dex and $\mmol$ has been corrected for a 35\% difference in the
adopted $X_{\rm CO}$ value.
}}
\label{fig:Matom_mol}
\end{figure}

Assuming that molecular gas forms out of clouds of atomic gas, one might expect a tight 
correlation between $\matom$ and $\mmol$. However, as shown by \cite{saintonge2011} and 
from our own measured $\rm R_{mol}$, there is a large scatter in this relation. 
Figure~\ref{fig:Matom_mol} shows $\mmol$ versus $\matom$.
After accounting for the offset due to our adopted $X_{\rm CO}$ value, we find that our
galaxies follow the relation from \cite{saintonge2011} rather well and with similar
scatter.

\section{Observed Rotation Curves And Their Shapes}
\label{sec:RotationCurves}

\subsection{Combined \halp\ and HI rotation curves}
\label{sec:RC}
The rotational velocities of the galaxy disks were measured in
\citetalias{martinsson2012a} (stellar and \oiii\ gas) and in \cite{martinsson2011} (\hone\
gas). Here, we exclude the stellar rotation curves, which are affected by asymmetric
drift, as well as the \oiii\ rotation curves (see Sect.~\ref{sec:Kinematics}).
Instead, we combine our \hone\ rotation curves with \halp\ rotation curves, derived from
data taken with the SparsePak IFU (Sect.~\ref{sec:Kinematics}).
We use the available \halp\ kinematics to create \halp\ rotation curves by azimuthally
averaging velocity measurements in 5$\arcsec$ broad concentric tilted rings, sampled at
radii $R_{j}$$=$$2.5\arcsec$+$j$$\times$$5\arcsec$ (where $j$$=$0,1,2,...), complementing
the \hone\ velocities sampled at radii $R_{j}$$=$$5\arcsec$+$j$$\times$$10\arcsec$.
The \halp\ rotation curves are derived in the same way as the stellar and \oiii\ rotation
curves, as described in \citetalias{martinsson2012a}. 
Due to uncertainties in the instrumental position angle of the SparsePak instrument, we
let the position angle be a free parameter while fitting the \halp\ rotation curve. The
differences in the derived position angles between the \halp\ and \hone\ velocity fields
are however small \citep[see Fig.~5.8 in][]{martinsson2011} and generally consistent with
no offset, with an average absolute measured difference of $1\fdg8$ and a maximum
difference of $5\arcdeg$.

Also, the adopted center of the galaxy might be slightly different between the two data 
sets. While the \hone\ rotation velocities are measured using the morphological center 
obtained from fitting reconstructed continuum images of the PPak data to optical DSS and 
SDSS images \citepalias{martinsson2012a}, the \halp\ rotation curves are derived using the
fitted kinematic center.
We expect the differences between the morphological and best-fitted kinematic centers to 
be less than $2\arcsec$ \citepalias{martinsson2012a}, and mostly dependent on the
regularity of the \halp\ velocity fields. The inclinations of the disks have been fixed to
the inverse Tully-Fisher inclinations \citep[$\itf$;][]{RixZaritsky1995,verheyen2001b} as
derived in \citetalias{martinsson2012a}. The combined \hone\ and \halp\ rotation curves
are shown for each galaxy in the Atlas with open and solid symbols, respectively.

The velocities of the \halp\ and \hone\ gas are expected to be nearly identical with only
minimal variations due to small differences in asymmetric drift. Indeed, the derived
\halp\ and \hone\ rotation curves are typically in excellent agreement. However, for some
galaxies (UGC~3140, UGC~3997, UGC~4256, UGC~7244; see Atlas), we note that the derived 
\halp\ and \hone\ rotation curves have slightly different shapes. We do not believe these
differences to be physical, but likely the result of systematic errors in deriving the two
rotation curves independently, e.g., due to the small differences in center and position 
angle mentioned above.
In one extreme case (UGC~7244), the difference arises due to a large kinematic asymmetry
between the receding and approaching side of the galaxy, with the rotation curve rising 
much steeper on the receding side than on the approaching side. This is reflected in the
large errorbars on the \hone\ velocities, where the errors are dominated by the difference
between the receding and approaching side of the rotation curve 
\citep[see][]{martinsson2011}. Curiously, we do not find the same asymmetry in the \halp\
kinematics, which follow a more shallow shape on both the receding and approaching side.

\subsection{The shapes of the observed rotation curves}
\begin{figure}
\centering
\includegraphics[width=0.5\textwidth]{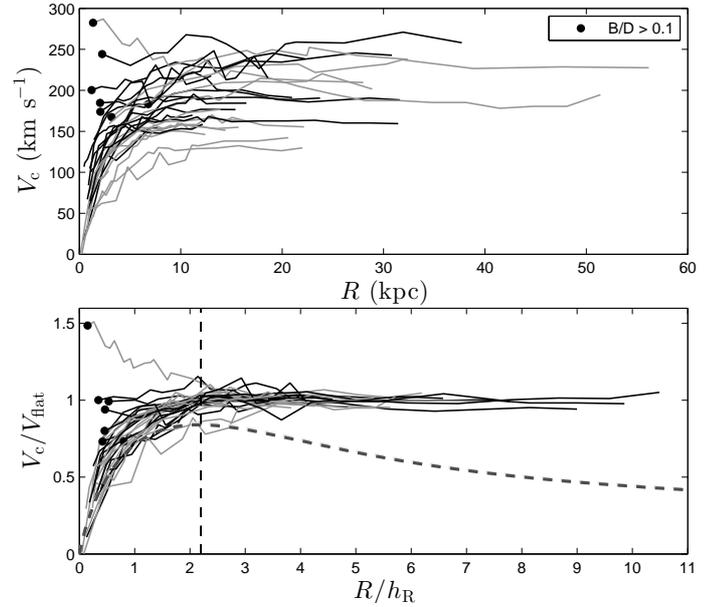}
\caption[]{{\small 
Observed rotation curves. Top: Radius in kpc and amplitude in \kms. Bottom: Radius scaled
by $\hr$ and amplitude scaled by the rotational velocity of the flat part of the rotation
curve ($V_{\rm flat}$). Galaxies with $B/D$$>$0.1 have their first velocity measurement
marked by a filled circle. In both panels, black and gray lines indicate high and low 
surface brightness, respectively. The thick, dashed line in the bottom panel indicates the
rotation curve of a thin exponential disk with $\vmax$$=$$0.85\vc$. $R$$=$$2.2\hr$ is
indicated with a vertical dashed line.
}}
\label{fig:Vobs}
\end{figure}

\begin{figure}
\centering
\includegraphics[width=0.5\textwidth]{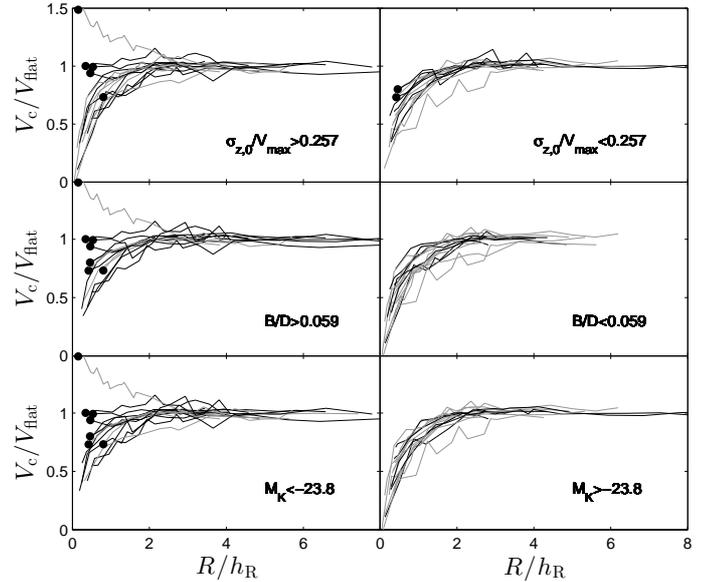}
\caption[]{{\small 
Observed rotation curves, divided into bins of high and low 
$\sigma_{z,0}/\vmax^{\rm OIII}$ (top row), $B/D$  (middle row), and $M_K$ (bottom row).
Lines are coded as in Fig.~\ref{fig:Vobs}.
}}
\label{fig:Vobs_rel}
\end{figure}

The combined \hone+\halp\ rotation curves are shown together in Fig.~\ref{fig:Vobs}. As
seen in this figure, galaxies with a higher bulge-to-disk ratio ($B/D$$>$0.1) have a
steeper rise in their rotation curves, indicating a higher central concentration of mass.
In the bottom panel we see that all rotation curves are fairly self-similar, and that
they have reached the flat part by $3\hr$. No difference can be seen in the shape of 
the scaled rotation curves between low (17.84$<$$\mu_{0,K}^{i}$$<$19.89) and high
(16.03$<$$\mu_{0,K}^{i}$$<$17.84) surface brightness galaxies.
The dashed dark gray line in the bottom panel shows the rotation curve of a thin
exponential disk which has a maximum rotation speed that is 85\% of the observed circular
speed at its maximum\footnote{In this paper we use 
$R_{\rm max}^{\rm disk}$$\equiv$$2.2\hr$, the radius at which the circular speed peaks for
a razor-thin, radially exponential disk.}, the fraction a maximum disk would have
\citep{sackett1997}. As expected, even though the stellar disk often displays a 
close-to-exponential decline, we find that the observed rotation curves depart from the
exponential-disk case, both at small and large radii. This is due to the presence of 
additional mass components, such as the bulge and the dark-matter halo.

In Fig.~\ref{fig:Vobs_rel}, we divide the rotation curves into bins of high and low 
dispersion/velocity ratio ($\sigma_{z,0}/\vmax$; an indicator of the disk maximality 
\citepalias{bershady2011}), bulge/disk ratio ($B/D$) and absolute magnitude ($M_K$).
In the upper row of Fig.~\ref{fig:Vobs_rel} we see that galaxies with the steepest inner
rise of the rotation curve tend to have a higher $\sigma_{z,0}/\vmax$ ratio.
The separation in $B/D$ (middle row) shows the same, but even clearer, and
in the bottom row we see it again (low-luminosity galaxies have often smaller bulges and
tend to have a smaller $\sigma_{z,0}/\vmax$ ratio).

\section{Rotation-Curve Mass Decompositions}
\label{sec:RCDecomp}
This section describes how we decompose the observed rotation curves presented in the
previous section into their baryonic components (Sect.~\ref{sec:RCmod}) and a parametrized
dark-matter-halo component (Sect.~\ref{sec:DM-RCs}).
We define two cases with different scaling of the stellar rotation curve;
Case~I (Sect.~\ref{sec:UniqueRCDecomp}) assumes our nominal $\overline{\mls}$
measurements, while Case~II (Sect.~\ref{sec:maxdisk}) scales $\overline{\mls}$ by a
galaxy-dependent factor $f_{\ast}$ to satisfy the maximum-disk hypothesis. These cases are
divided into Cases Ia and IIa, which have a parameterized dark-matter halo modeled as a
spherical pseudo-isothermal sphere \citep[pISO; e.g.,][]{ostriker1979,schmidt1985,kent1986},
and Cases Ib and IIb which instead include a Navarro-Frenk-White 
\citep[NFW;][]{NFW1997} dark-matter-halo model.
In Sect.~\ref{sec:compDMmod} we compare how well the different cases fit the observed
rotation curves, with the results discussed further in Sect.~\ref{sec:Discussion}.

\subsection{Rotation curves of the baryonic components}
\label{sec:RCmod}
The rotation curves of the various baryonic components are calculated from their radial 
mass surface density profiles and an assigned three-dimensional axisymmetric density 
distribution using the task {\tt ROTMOD} in {\it GIPSY}\footnote{Groningen Image
Processing SYstem} \citep{hulst1992,vogelaar2001}.
The shape of the stellar rotation curve follows from the light profile of the galaxy,
while the amplitude is set by $\sqrt{\mls}$.
We adopt an exponential vertical distribution for the stellar disk with a constant scale
height $\hz$, as calculated in Sect.~\ref{sec:Photometry} and presented in
Table~\ref{tab:hRhz}. For the stellar bulge, we assume a spherical density distribution.
The atomic and molecular gas are assumed to be distributed in razor-thin disks.

Non-exponential features in the disk light profile result in ``bumps'' in the rotation
curve and a peak that does not necessarily occur at $2.2\hr$. Due to the ring-like
distribution of the \hone\ gas, a ``negative'' rotation speed\footnote{In fact, it is the
{\it square} of the rotation speed that is negative. The rotation velocity itself is
imaginary, but is by convention denoted as negative.} of the atomic-gas rotation curve is
often found in the inner part of the galaxy. This arises due to the outward force from the
atomic gas; the \hone\ disk is not counter-rotating in the center. The derived rotation
curves are shown for individual galaxies in the Atlas.

The total gravitational potential can be considered as being composed from independent and
separable density distributions. This allows us to add the calculated circular velocities
of the stellar disk ($V_{\rm disk,\ast}$), stellar bulge ($V_{\rm bulge,\ast}$),
atomic gas ($V_{\rm atom}$) and molecular gas ($V_{\rm mol}$) in quadrature to obtain the
total baryonic rotation curve
\begin{equation}
\vbary = \left[ V_{\rm disk,\ast}^2 + V_{\rm bulge,\ast}^2 + 
V_{\rm atom}^2 + V_{\rm mol}^2\right]^{1/2}.
\label{eq:Vbary}
\end{equation}

\subsection{Parameterized rotation curves of the dark-matter halos}
\label{sec:DM-RCs}
Although we have now calculated the rotation curves of all dynamically important
baryonic components, and we can therefore, together with our observed rotation curves,
directly determine the rotation curves of the dark-matter halos, we will also perform
rotation-curve mass decompositions using two different parameterizations of the
dark-matter halo; either using a pISO or a NFW halo model.

The pISO-halo is parameterized by its central density ($\rho_0$) and its core radius 
($R_C$), with a rotation curve following from
\begin{equation}
\vdm^{\rm pISO}(R) = \sqrt{4\pi G \rho_0 R_C^2\left[1-\frac{R_C}{R} 
\arctan\left(\frac{R}{R_C}\right)\right]}.
\label{eq:pISO}
\end{equation}

The NFW halo is parameterized by its mass ($\mathcal{M}_{200}^{\rm halo}$) within the 
virial radius ($R_{200}$; defined as the radius of a sphere of mean interior density 200
times the critical density $\rho_{\rm crit}$$=$$3H_{0}^{2}/8 \pi G$) and its concentration
($C$) as defined in \cite{NFW1997}. The rotation curve is given by
\begin{equation}
\frac{\vdm^{\rm NFW}(R)}{V_{200}} = \sqrt{\frac{1}{X} 
\frac{\ln(1+CX)-(CX)/(1+CX)}{\ln(1+C)-C/(1+C)}},
\label{eq:NFW}
\end{equation}
where $V_{200}$ is the circular velocity at $R_{200}$ and $X$$=$$R/R_{200}$.

The pISO is an ad-hoc form of the density distribution, fitted to the observed rotation
curves to describe the dark-matter distribution. The NFW halo, on the other hand, is a
prediction from dark-matter-only simulations, and therefore predicts how the halo looked
like prior to the galaxy forming in it. In principal, the baryons in the forming galaxy
could affect the distribution of the dark matter.
However, as discussed in Sect.~\ref{sec:ISOvsNFW}, several processes may occur that could
both contract and expand the halo, and even effectively leaving it unmodified.

\subsection{Unique rotation-curve mass decompositions (Case~I)}
\label{sec:UniqueRCDecomp}
%
%
\begin{figure*}
\centering
\includegraphics[width=1.0\textwidth]{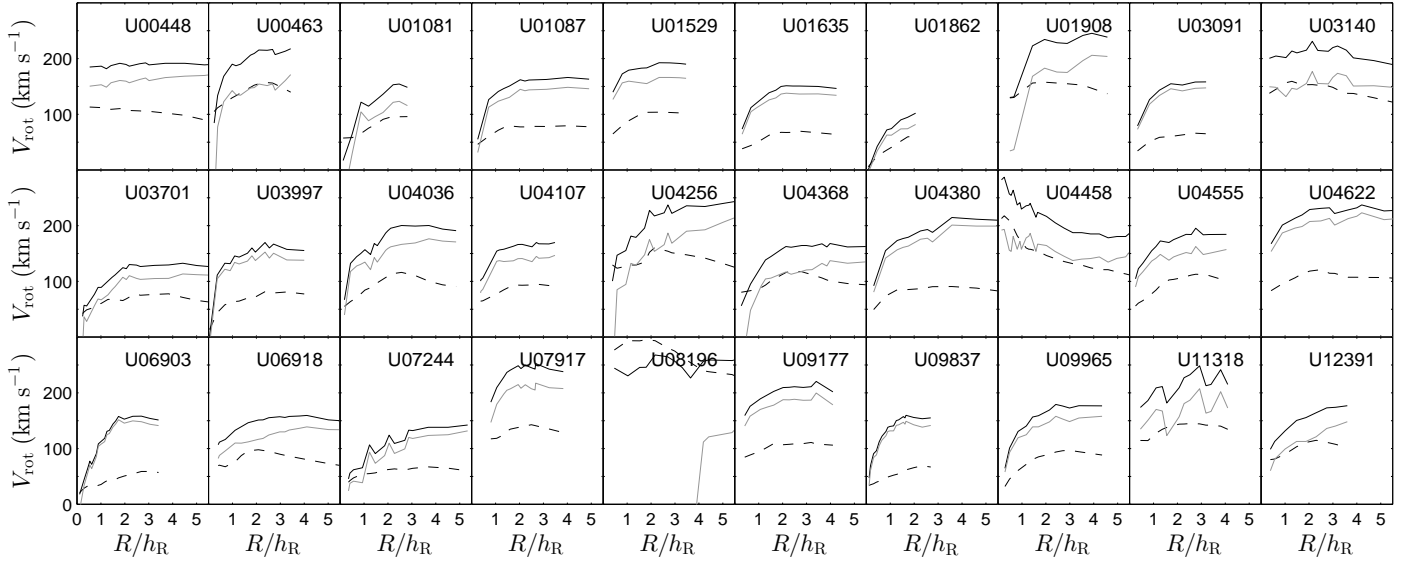}
\caption[]{{\small 
Rotation curves: Observed (solid black), baryonic (dashed black) and dark matter (solid
gray).
}}
\label{fig:RCs_all}
\end{figure*}
%
%
Using the baryonic rotation curves derived in Sect.~\ref{sec:RCmod}, we determine the 
dark-matter distribution in our galaxy sample taking an approach similar to traditional 
rotation-curve mass decompositions \citep[e.g.,][]{albada1985, carignan1985, begeman1991}. 
Assuming the \halp\ and \hone\ rotation curves trace the circular speed ($\vc$) of the 
composite potential, the circular speed of the dark-matter component ($\vdm$) is
calculated from
\begin{equation}
\vdm^2 = \vc^2 - \vbary^2.
\label{eq:RCdecomp}
\end{equation}
Therefore, our data provide $\vdm$ directly such that we calculate a unique, dynamically
constrained, non-parametric dark-matter rotation curve for each galaxy in our sample.

The derived $\vdm$, calculated from Eq.~\ref{eq:RCdecomp}, are plotted in the Atlas 
figures. In Fig.~\ref{fig:RCs_all} we plot the baryonic ($\vbary$), dark matter ($\vdm$) 
and observed total rotation curves ($\vc$) in the same figure. What can already be noted 
from this figure is that the contribution from the dark-matter rotation curve to the 
observed total rotation curve is in general dominant at most radii. This is not 
unexpected given the result in \citetalias{bershady2011} (and
Sect.~\ref{sec:DiskMaximality} of this paper), where we found that these galaxies are
submaximal.

We use the {\it GIPSY} task {\tt ROTMAS} to model the dark-matter rotation curves via a
pISO or NFW-halo parameterization. The baryonic and dark-matter rotation curves are added
together in quadrature into a total rotation curve, matching the observed rotation curve 
using a minimum $\chi^2$-fitting routine. Since all baryonic components are fixed, this is
basically only fitting the two-parameter dark-matter-halo models to the measured 
dark-matter rotation curve. Each fitted halo is plotted in the Atlas and the fitted 
parameters are provided in Table~\ref{tab:DMfit}.

\subsection{Maximum-$\mls$ rotation-curve decompositions (Case~II)}
\label{sec:maxdisk}
Traditionally, without an independent measurement of $\mls$, rotation-curve mass 
decompositions have often been performed with the assumption of a maximum contribution 
from the stellar disk and bulge to the total potential
\citep[e.g.,][]{albada1985,kent1986,BroeilsCourteau1997}. This approach sets a lower
limit to the contribution of the dark matter, often with the result that no dark matter is 
needed in the inner part of the galaxy. For comparison, we consider a maximum-$\mls$ case
here as well. The fitting is performed using a pISO or NFW halo with ${\overline{\mls}}$
scaled up by a factor $f_\ast$ (listed in Table~\ref{tab:DMfit}). This
factor, which is usually (but not always exactly) the same for the NFW and pISO halo
models, is forced to be as high as allowed by the observed rotation curve, which
effectively sets $\vbary$ close to $\vc$ in the inner region of the galaxies. As for
Case~I, we enforce ${\overline{\mls}}$ of the stellar bulge and disk to be equal.

To enforce these galaxies to be maximal, $\overline{\mls}$ must on average be 3.6
times what was measured in Sect.~\ref{sec:ML}, with a range in $f_\ast$ from 1.6 
(UGC~7244)\footnote{When excluding UGC~8196; see Sect.~\ref{sec:DiskMaximality}.} to 8.1
(UGC~6903), and with 70\% of the galaxies having values of $f_\ast$ between 2.1 and 4.6.
A detailed discussion on this difference between our measured $\mls$ and what is found for
the maximum-$\mls$ cases, and the effects of our results on stellar-population
models, will be the subject of a forthcoming paper.
We find the ratio between the maximum-$\mls$ baryonic rotation curve and the
measured total rotation curve at $R$$=$$2.2\hr$ ($\mathcal{F}_{\rm b,max}^{2.2\hr}$) to
range between 0.66 (UGC~7244) and 1.03 (UGC~4622), with an average value of 0.92$\pm$0.08
(std)\footnote{As in Sect.~\ref{sec:DiskMaximality}, we exclude UGC~1862.}.
As expected (see Sect.~\ref{sec:Disc-maxi}), this is somewhat higher than the
0.85$\pm$0.10 found by \cite{sackett1997}. Table~\ref{tab:F} includes
$\mathcal{F}_{\rm b,max}^{2.2\hr}$ for the individual galaxies.

\subsection{Comparison of $\chi^2$ for the various cases}
\label{sec:compDMmod}
%
%
\begin{figure}
\centering
\includegraphics[width=0.5\textwidth]{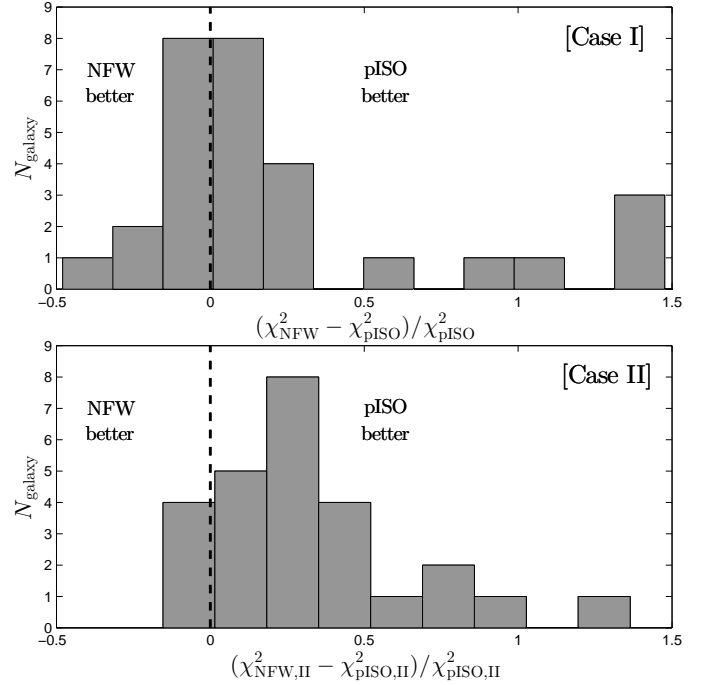}
\caption[]{{\small 
Relative difference in calculated $\chi^2$ between the fits using NFW and pISO 
dark-matter halos. Upper and lower row show the differences in the nominal-$\mls$ and
maximum-$\mls$ case, respectively.
}}
\label{fig:Chi2}
\end{figure}
%
%
Here we make a relative comparison between the quality of the decompositions of the
modeled dark-matter+baryonic rotation curves to the observed rotation curves when using
either our nominal-$\mls$ (submaximal) case (Case~I; Sect.~\ref{sec:UniqueRCDecomp}) or
the maximal-$\mls$ case (Case~II; Sect.~\ref{sec:maxdisk}). For both cases, we have used
two different models (pISO and NFW) for the dark-matter halo. We calculate the $\chi^2$
values from a weighted quadrature sum of the differences between the observed and modeled
rotation velocities. Due to complex and non-Gaussian errors we will not consider the
reduced $\chi^2$ values, but instead look at relative values.

We find that the nominal-$\mls$ fits produce significantly better $\chi^2$ results than
the maximum-$\mls$ fits. The nominal-$\mls$ case with a pISO dark-matter halo
(Case~Ia) gives on average the best fit with the lowest $\chi^2$. The average relative
difference in $\chi^2$ between Case~Ia ($\chi^2_{\rm pISO}$) and Case~Ib 
($\chi^2_{\rm NFW}$), normalized to $\chi^2_{\rm pISO}$, is 
$(\chi^2_{\rm NFW}$$-$$\chi^2_{\rm pISO})/\chi^2_{\rm pISO}$$=$0.46. For the
maximum-$\mls$ cases (Case~IIa,b) we find the relative differences in $\chi^2$
compared to Case~Ia to be
$(\chi^2_{\rm pISO,II}$$-$$\chi^2_{\rm pISO})/\chi^2_{\rm pISO}$$=$1.42 and
$(\chi^2_{\rm NFW,II}$$-$$\chi^2_{\rm pISO})/\chi^2_{\rm pISO}$$=$2.36.

Figure~\ref{fig:Chi2} shows the distribution of relative differences between the pISO and
NFW-halo models (nominal-$\mls$ and maximum-$\mls$ case separately). Even though
Case~Ia on average results in a slightly better fit than Case~Ib, and Case~IIa is a better
fit than Case~IIb, the differences are not significant as indicated by the relative
$\chi^2$-values.
This inability to differentiate between a pISO and a NFW-halo parameterization of the
observed data was also seen for luminous galaxies in THINGS \citep{deBlok2008}. However,
for low-luminosity galaxies, they find that a pISO parameterization better fits the data.

We find no strong correlations between how well the different models fit
the data and global properties of the galaxies \citep[see][]{martinsson2011}.

\section{The Dynamical Importance of the Baryons}
\label{sec:DiskMaximality}
\begin{figure}
\centering
\includegraphics[width=0.5\textwidth]{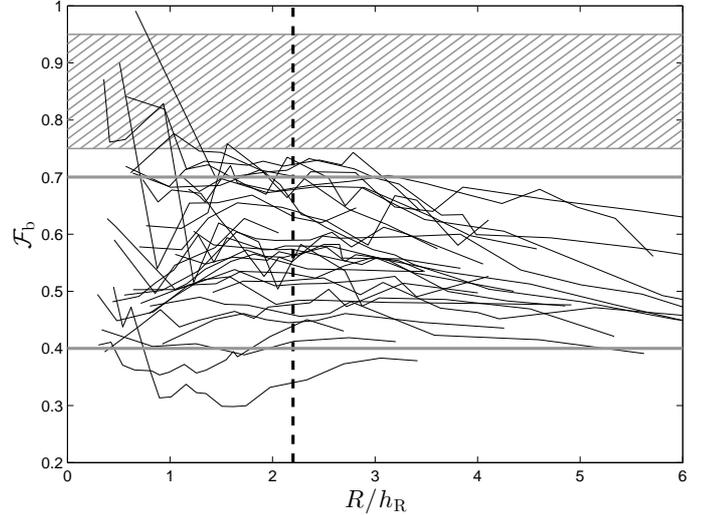}
\caption[]{{\small 
Baryonic mass fraction of all galaxies except UGC~8196 as a function of radius. As in 
Fig.~\ref{fig:F_all}, the shaded area shows 0.75$\leq$$\Fbary$$\leq$ 0.95, the vertical
dashed line $R$$=$$2.2\hr$, and we exclude measurements within $R$$<$$\rbulge$. The two
thick gray horizontal lines indicate the range $\Fbary$$=$0.4--0.7.
}}
\label{fig:F}
\end{figure}

\begin{figure*}
\centering
\includegraphics[width=1.0\textwidth]{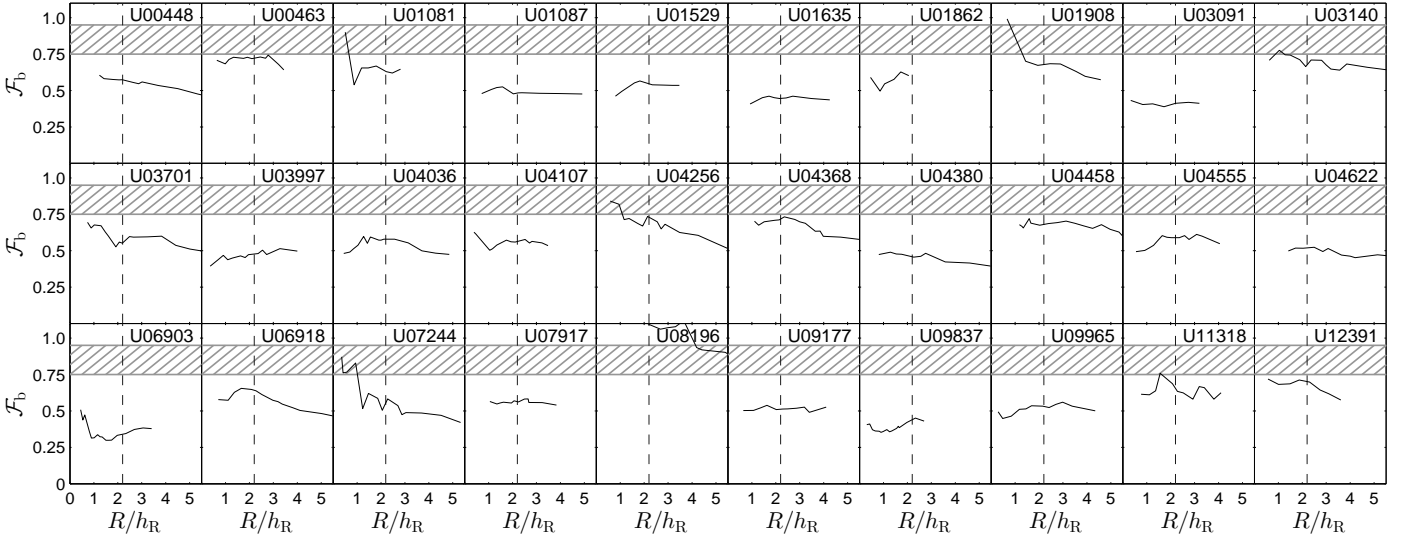}
\caption[]{{\small 
Baryonic maximality of individual galaxies as a function of radius. Measurements within
$R$$=$$\rbulge$ have been omitted. The shaded area shows 0.75$\leq$$\Fbary$$\leq$0.95, the
defined fraction \citep{sackett1997} a maximum disk should have at $R$$=$$2.2\hr$ 
(vertical dashed line).
}}
\label{fig:F_all}
\end{figure*}

In this section we investigate the ratio $\Fbary$$=$$\vbary/\vc$, or the ``baryonic
maximality'', for the galaxies in our sample. This ratio is a measure of the dynamical
importance of the baryons in the galaxy. We calculate $\Fbary$ as a function of radius for
all galaxies. These are plotted together in Fig.~\ref{fig:F} and separately in
Fig.~\ref{fig:F_all}.

%

%

We find that the galaxies have a rather constant value of $\Fbary$ with radius, ranging in
amplitude between $\sim$0.4--0.7 for the sample of galaxies (Fig.~\ref{fig:F}). The
constancy of $\Fbary$ arises due to an increasing contribution from the atomic gas at
larger radii where the contribution from the stellar disk is declining
(Fig.~\ref{fig:Sigma_bar}). This result has a consequence for the measurement of the
maximality: with the assessment of the maximality of a galaxy using $\Fbary$, it is of
less importance to measure exactly at the radius $R_{\rm max}^{\rm disk}$$\approx$$2.2\hr$
where the rotation curve of the stellar disk peaks.

\begin{figure*}
\centering
\includegraphics[width=0.95\textwidth]{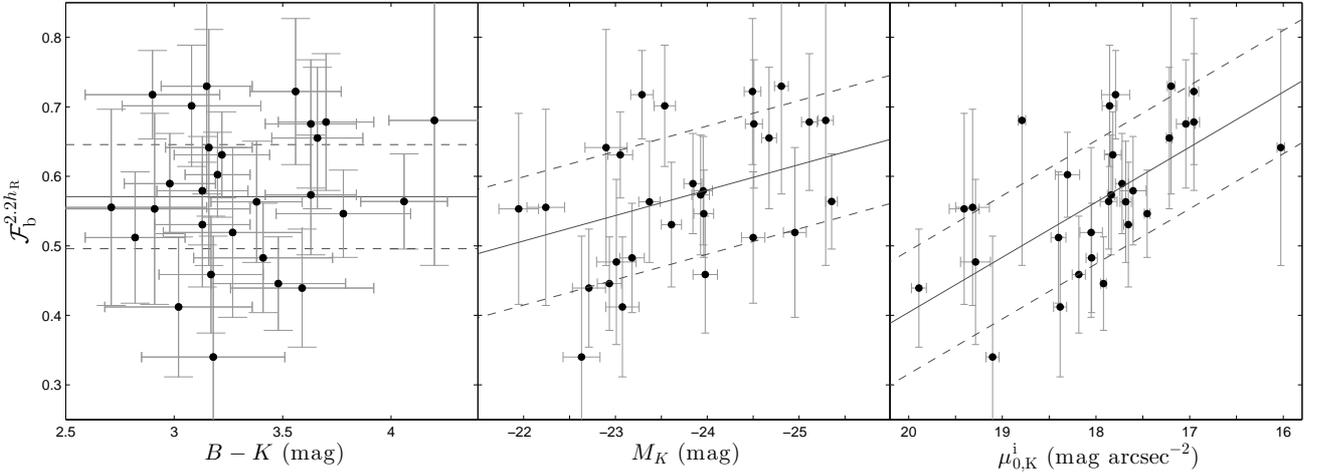}
\caption[]{{\small 
Baryonic mass fraction measured at $2.2\hr$ versus $B-K$ color (left), absolute $K$-band
magnitude (middle), and central surface brightness of the disk (right). Black dots with
gray errorbars indicate measurements from individual galaxies. Black lines are linear fits
to the data, showing that more luminous and higher-surface brightness galaxies have larger
$\Fbary^{2.2\hr}$. The lines in the left panel show the weighted average and scatter.
}}
\label{fig:F2.2hR}
\end{figure*}

In Fig.~\ref{fig:F2.2hR}, we plot the measured $\Fbary$ at $R$$=$$2.2\hr$ 
($\Fbary^{2.2\hr}$; tabulated in Table~\ref{tab:F}) versus $B-K$ color, absolute $K$-band
magnitude, and central disk surface brightness. We confirm the result in 
\citetalias{bershady2011} that all our galaxies are submaximal, with a range in
$\Fbary^{2.2\hr}$ of 0.34--0.73, equivalent to $\sim$10--50\% of the total mass within
$2.2\hr$ being baryonic. On average, after rejecting UGC~1862 which has no measurement
of $\vc$ at $2.2\hr$, $\langle$$\Fbary^{2.2\hr}$$\rangle$$=$0.57$\pm$0.07~(std).
This result is consistent with what was found by \cite{bottema1993} and
\cite{kregel2005}, and based on the correlations with $M_K$ and $\mu_{0,K}^{i}$ seen in
Fig.~\ref{fig:F2.2hR}, we conclude that small differences in the mean may be due to
differences in the galaxy samples. However, also note that both \cite{bottema1993} and
\cite{kregel2005} measured the {\it disk} mass fraction, excluding the contribution from
the stellar bulge (see discussion in Sect.~\ref{sec:Disc-maxi}).

In \citetalias{bershady2011} we found that more luminous and redder galaxies are closer to
be maximal, but with little dependence on both central surface brightness and morphology.
Here, we again find a weak trend of larger $\Fbary^{2.2\hr}$ for more luminous galaxies,
but also for galaxies with higher central disk surface brightness.
Weighted linear fits give 
$\Fbary^{2.2\hr}$$=$(0.58$\pm$0.09)$-$(0.04$\pm$0.02)($M_K+24$) and 
$\Fbary^{2.2\hr}$$=$(0.56$\pm$0.09)$-$(0.08$\pm$0.03)($\mu_{0,K}^{i}-18$).
Unlike the result in \citetalias{bershady2011} we find no correlation with color, and in
contrast to \cite{herrmann2009}, who had a similar range of morphologies, we find little
dependence on morphological type.

%
%
%
\begin{table}
\caption{\label{tab:F}
Baryonic mass fraction at $R$$=$$2.2\hr$}
\centering
\begin{tabular}{|c c c| c c c|}
\hline
\multicolumn{1}{|c}{UGC}                                &
\multicolumn{1}{c}{$\mathcal{F}_{\rm b}^{2.2\hr}$}      &
\multicolumn{1}{c|}{$\mathcal{F}_{\rm b,max}^{2.2\hr}$} &
\multicolumn{1}{c}{UGC}                                 &
\multicolumn{1}{c}{$\mathcal{F}_{\rm b}^{2.2\hr}$}      &
\multicolumn{1}{c|}{$\mathcal{F}_{\rm b,max}^{2.2\hr}$} \\
\hline
  448 &  0.57 $\pm$ 0.09 & 0.94 &  4368 &  0.72 $\pm$ 0.06 & 1.01 \\
  463 &  0.72 $\pm$ 0.11 & 0.96 &  4380 &  0.46 $\pm$ 0.08 & 0.88 \\
 1081 &  0.63 $\pm$ 0.06 & 0.89 &  4458 &  0.68 $\pm$ 0.21 & 0.90 \\
 1087 &  0.48 $\pm$ 0.08 & 0.85 &  4555 &  0.59 $\pm$ 0.07 & 0.97 \\
 1529 &  0.55 $\pm$ 0.06 & 0.98 &  4622 &  0.52 $\pm$ 0.12 & 1.03 \\
 1635 &  0.45 $\pm$ 0.07 & 0.88 &  6903 &  0.34 $\pm$ 0.17 & 0.80 \\
 1862 &  0.60 $\pm$ 0.06 & 0.91 &  6918 &  0.64 $\pm$ 0.17 & 0.97 \\
 1908 &  0.68 $\pm$ 0.10 & 0.88 &  7244 &  0.56 $\pm$ 0.14 & 0.66 \\
 3091 &  0.41 $\pm$ 0.10 & 0.93 &  7917 &  0.56 $\pm$ 0.07 & 0.97 \\
 3140 &  0.68 $\pm$ 0.09 & 0.92 &  8196 &  1.09 $\pm$ 0.05 & 1.09 \\
 3701 &  0.55 $\pm$ 0.14 & 0.79 &  9177 &  0.51 $\pm$ 0.09 & 1.03 \\
 3997 &  0.48 $\pm$ 0.12 & 0.95 &  9837 &  0.44 $\pm$ 0.08 & 0.89 \\
 4036 &  0.58 $\pm$ 0.08 & 0.96 &  9965 &  0.53 $\pm$ 0.09 & 0.94 \\
 4107 &  0.56 $\pm$ 0.09 & 1.00 & 11318 &  0.66 $\pm$ 0.10 & 1.03 \\
 4256 &  0.73 $\pm$ 0.16 & 0.93 & 12391 &  0.70 $\pm$ 0.09 & 0.94 \\
\hline
\end{tabular}
%
\end{table}


As can be seen in Fig.~\ref{fig:F_all}, UGC~8196 appears to be ``supermaximal'',
$\vbary$$>$$\vc$, which is a non-physical result. This might be either due to an 
overestimation of $\mls$ or an underestimation of $\vc$.
The derived $\mls$ is indeed the highest in the sample. An overestimation of $\mls$ could
be due to an incorrect assumption about the SVE ($\alpha$), the thickness of the disk
($\hz$), or its vertical mass distribution ($k$).
As seen in the Atlas, the measured $\sigz$ of UGC~8196 is extremely high for a disk in the
inner region compared to the other disks in our sample, and instead more comparable to 
other bulges. This could be an indication that we are not measuring the velocity
dispersion of the disk and that Eq.~\ref{eq:mldyn} is therefore invalid. If we extend
$\rbulge$ out to $15\arcsec$, excluding the measurement at this radius, we find a
$\mls$ that is only 80\% of the adopted value, resulting in $\Fbary$$=$0.98 at $2.2\hr$;
much higher than all other galaxies, but physically possible.
UGC~8196 has a rather low inclination of $16\fdg1$. If this inclination is instead
$2\fdg7$ lower ($3\sigma$), this would result in a rotation curve with a larger
amplitude, making the $\mls$ and $\Fbary$ estimates physically possible, and resulting in
$\Fbary$$=$0.91$\pm$0.05 at $2.2\hr$.
However, due to the uncertainties in what is causing the non-physical result, we have
decided to exclude UGC~8196 from our analysis and results.

\section{The Dark-Matter-Halo Rotation Curve}
\label{sec:DMRCshape}
From our unique rotation-curve mass decompositions in Sect.~\ref{sec:UniqueRCDecomp}, we
have {\it measured} the dark-matter rotation curves for all galaxies. In this section, we
will investigate the shape of these dark-matter rotation curves, and look at how well our
parameterized halo models fit the data. We compare the resulting parameters in relation to
one another and with respect to results from $\Lambda$CDM simulations.

\begin{figure*}
\includegraphics[width=1.0\textwidth]{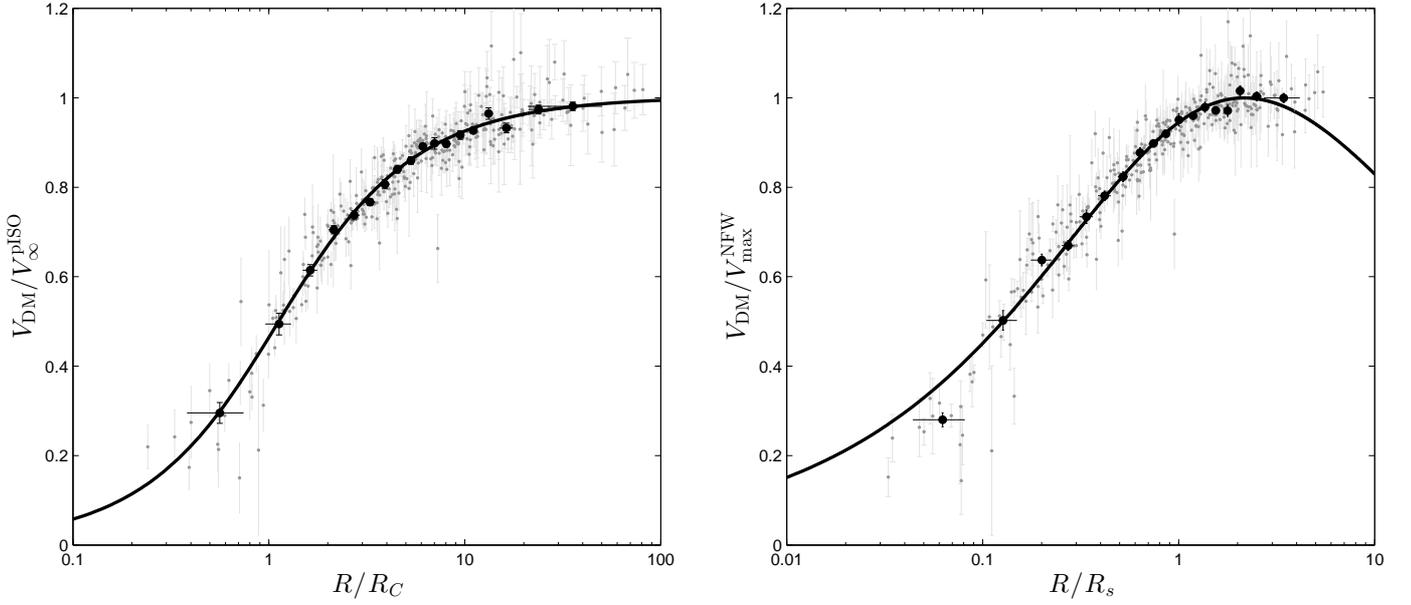}
\caption[]{{\small 
All measured dark-matter rotation velocities plotted together and compared to the fitted
models.
{\bf Left panel:} Comparison to the pISO model. $R$ and $\vdm$ have been normalized
with $R_C$ and $V_{\infty}^{\rm pISO}$, respectively. 
{\bf Right panel:} Comparison to the NFW-halo model. Here, $R$ and $\vdm$ have been
normalized with $R_s$ and $V_{\max}^{\rm NFW}$, respectively.
Gray points with light gray errorbars indicate individual $\vdm$ measurements. The black
filled circles show averages in radial bins containing 20 measurements each.
}}
\label{fig:DMmeasured}
\end{figure*}

For individual galaxies, we find in general no significant difference between the quality 
of the pISO and NFW fits to our measurements of $\vdm$; however, for our entire sample, 
the pISO model tends to result in a lower $\chi^2$ (Sect.~\ref{sec:compDMmod}). To 
further investigate any general differences between the shape of the pISO and NFW 
parameterization with respect to our data, we consider the $\vdm$ measurements of all 
galaxies simultaneously.
In Fig.~\ref{fig:DMmeasured}, we plot all the measured dark-matter rotation curves in
the same panel with the radius normalized to either $R_C$ or $R_s$$=$$R_{200}/C$, and the
velocity normalized to either $V^{\rm pISO}_{\infty}$ or $\vmax^{\rm NFW}$ (the velocity
of the NFW rotation curve at $\sim$$2.2R_s$). Measurements of $\vdm$ inside of 
$R$$=$$5\arcsec$ have been excluded. On top of the measured points, we plot the
parameterized rotation curves, which were fitted to each individual galaxy separately.

The left panel of Figure~\ref{fig:DMmeasured} shows all measured points compared to the
pISO case. In general, the data follow the fits well, with a weighted scatter of 0.024 for
all data points. The right panel shows the same as the left panel, except it adopts the
NFW-halo parameterization. The weighted scatter of all data points compared to the NFW
model is 0.027; only slightly larger than for the pISO case. In \cite{martinsson2011} we
investigated how well the dark-matter rotation curves are fitted for galaxies with low or
high $\mu_{0,K}^{i}$ and $\Fbary^{2.2\hr}$. For both the pISO and the NFW cases, we find
no dependencies on these parameters.
In a few galaxies with a steep rise in the inner region of the dark-matter rotation curve
(e.g., UGC~448 and UGC~3140; see Atlas), the NFW-halo model fails to follow the inner
steep rise. For these galaxies, the pISO model is a better fit. The inner regions exhibit
in general the largest residuals, with a scatter of 0.056 and 0.068 within $R$$=$$0.9R_C$
and $R$$=$$0.1R_s$ for the pISO and NFW fits, respectively. The measured $\vdm$ are here
systematically lower than the fitted NFW rotation curve.
In the outer regions the NFW rotation curve begins to decline at $R/R_s$$\approx$2, which
is reached at a smaller radius in galaxies with a steep inner rise in the dark-matter 
rotation curve, whereas the observed dark-matter rotation curve stays flat. However, note
that our measured dark-matter rotation curves are rather uncertain at both small and large
$R$; uncertainties in the dynamical mass of the bulge and the possible effect of adiabatic
contraction may affect the inner measurements of $\vdm$, while the outer measurements may
be systematically in error for individual galaxies because we intentionally keep the 
observed rotation curves flat by introducing a linear inclination warp of the \hone\ disk 
to ``force'' these galaxies onto the Tully-Fisher relation \citep{martinsson2011}.

\subsection{Derived parameters}
\label{sec:DMRCshape_comp}
As a consequence of the submaximality of our galaxies, the dark-matter rotation curves
rise quickly at small radii, with parameter values that differ significantly from what is
found for the maximum-$\mls$ cases. This behavior, and the inferred dark-matter
distributions, can be compared to what has been found from numerical simulations.

\begin{figure}
\centering
\includegraphics[width=0.5\textwidth]{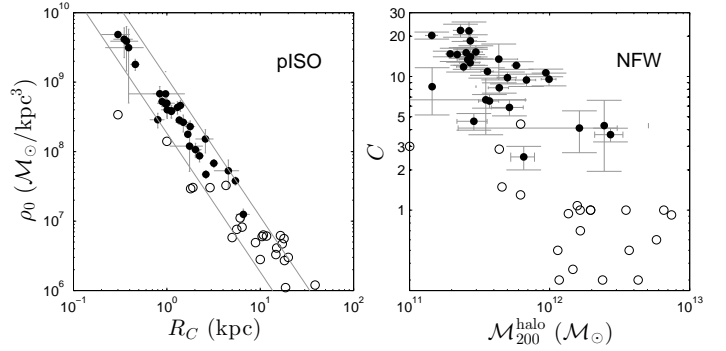}
\caption[]{{\small 
{\bf Left:} $\rho_0$ versus $R_C$. The two gray lines indicate
$V^{\rm pISO}_{\infty}$$=$100~\kms\ and 250~\kms; the range of $V^{\rm pISO}_{\infty}$ for
the nominal-$\mls$ case. {\bf Right:} Relation between the concentration ($C$) and mass
($\mathcal{M}_{200}^{\rm halo}$) of the NFW-halo profile. Filled and open circles indicate
nominal-$\mls$ and maximum-disk fits, respectively.
}}
\label{fig:CvsM200}
\end{figure}

In Fig.~\ref{fig:CvsM200}, we plot the two halo parameters from the dark-matter-halo fits
against each other, for both the pISO and the NFW-halo parameterization. In addition to
the nominal-$\mls$ case, we also include the results from the maximum-$\mls$ case for
comparison.
For the pISO fits, we find a strong correlation between $\rho_0$ and $R_C$. This is
expected given the strong covariance between the two parameters as seen in the formula 
$V^{\rm pISO}_{\infty}$$=$$\sqrt{4\pi G \rho_0 R_C^2}$ and the relatively small range in
dynamical masses of our galaxy sample, as indicated with the two lines in the figure 
showing constant $V^{\rm pISO}_{\infty}$. The spread along these lines is not explained
by the measurement errors. We find that the maximum-$\mls$ fits have larger $R_C$
and smaller $\rho_0$. This is expected since the maximum-$\mls$ decompositions
result in a much shallower rise of the dark-matter rotation curve.
For the NFW halo, we again find a correlation between the two parameters with a lower
concentration for larger halo masses, and with the maximum-$\mls$ case having
much lower concentration than the nominal-$\mls$ case.
The median values and ranges of the fitted parameters are presented in
Table~\ref{tab:DMpars}.

We have compared our results with parameter values found by \cite{deBlok2008}.
Excluding the four least massive galaxies in their sample (leaving 13 galaxies similar to
those in the PPak sample), and looking at the models with fixed stellar mass-to-light
ratios derived using a Kroupa IMF, we find rather similar parameter values.
However, there are some differences. In the case of fitting a pISO model, the core radius
of our fitted halos are on average $\sim$50\% smaller, but with a $\sim$65\% larger
central density. When fitting a NFW-halo model, the median value of the concentration
parameter in the 13 galaxies from \cite{deBlok2008} is similar to what we find ($C=9$
and $C=11$, respectively), but with a much larger scatter.

%
%
%
\begin{table}
\caption{\label{tab:DMpars}
Median dark-matter-halo parameters} 
\centering
\renewcommand{\tabcolsep}{1.0mm}
{\tiny
\begin{tabular}{|l |c c |c c |c c |c c|}
\hline
\multicolumn{1}{|c}{}                               &
\multicolumn{4}{|c}{pISO}                           &
\multicolumn{4}{|c|}{NFW}                           \\
\hline
\multicolumn{1}{|c}{}                               &
\multicolumn{2}{|c}{$\rho_0$}                       &
\multicolumn{2}{|c}{$R_C$}                          &
\multicolumn{2}{|c|}{$C$}                           &
\multicolumn{2}{c|}{$\mathcal{M}_{200}^{\rm halo}$} \\
\multicolumn{1}{|c}{}                               &
\multicolumn{2}{|c}{($10^{8}\msol/{\rm kpc^3}$)}    &
\multicolumn{2}{|c}{(kpc)}                          &
\multicolumn{2}{|c|}{}                              &
\multicolumn{2}{c|}{($10^{11}\msol$)}               \\
\hline
Case-I  & \bf{3.3}  & (0.47--32)   & \bf{1.3} & (0.39--4.6) & \bf{11}  & (4.1--18)   & \bf{1.0} & (0.59--4.4) \\
Case-II & \bf{0.06} & (0.03--0.30) & \bf{10}  & (1.9--18)   & \bf{1.0} & (0.30--3.0) & \bf{4.1} & (1.2--16)  \\
\hline
\end{tabular}
}
\tablefoot{Parentheses give ranges of parameter values for 80\% of the galaxies.
}
\end{table}


\subsection{Comparison to cosmological simulations}
\cite{bullock2001} studied dark-matter-halo density profiles in a high-resolution N-body
simulation with $\Lambda$CDM cosmology. They found similar values and scatter compared 
to what we find here for the nominal-$\mls$ case, with $C$$\approx$10--20 in our mass
range. The negative slope in the $\mathcal{M}_{200}^{\rm halo}$-$C$ relation can be seen 
clearly in their Fig.~4 (distinct halos) and, even stronger, in their Fig.~5 (subhalos).
The concentrations found for our maximum-$\mls$ fits are much lower than for the 
nominal-$\mls$ case, with typical values around or lower than $C$$=$1. As discussed by
\cite{deBlok2008}, the value of $C$ indicates the amount of collapse the halo has
undergone, where $C$$=$1 indicates no collapse and a value of $C$$<$1 is an unphysical
result in the context of cold dark matter. There is also a lack of correlation between
$\mathcal{M}_{200}^{\rm halo}$ and $C$ for the maximum-$\mls$ case. This result
implies that for the galaxies in our sample, a maximum-$\mls$ solution is
inconsistent with results from simulated NFW halos.
However, for the nominal-$\mls$ case, the shapes of our inferred dark-matter rotation
curves are well fit with parameters in agreement with what has been found from 
dark-matter-only simulations. Furthermore, this result suggests that the baryonic matter
in our sample of submaximal galaxies may have only a minor effect on the dark-matter
distribution.

%
\begin{table*}
\caption{\label{tab:DMfit}
Parameterized dark-matter-halo parameters}
\centering
{\footnotesize 
\renewcommand{\tabcolsep}{1.0mm}
\begin{tabular}{|r |r |r |r r |r r |c c c |c c c|}
\hline
\multicolumn{1}{|c}{}                                  &
\multicolumn{6}{|c}{Nominal-$\mls$ (Case Ia \& Ib)}    &
\multicolumn{6}{|c|}{Maximum-disk (Case IIa \& IIb)}    \\
\hline
\multicolumn{1}{|c}{}                                  &
\multicolumn{1}{|c}{}                                  &
\multicolumn{1}{|c}{}                                  &
\multicolumn{2}{|c}{pISO}                              &
\multicolumn{2}{|c}{NFW}                               &
\multicolumn{3}{|c}{pISO}                              &
\multicolumn{3}{|c|}{NFW}                              \\
\multicolumn{1}{|c}{UGC}                               & 
\multicolumn{1}{|c}{$\overline{\mls}$}                 &
\multicolumn{1}{|c}{$R_{\Upsilon}$}                    &
\multicolumn{1}{|c}{$R_C$}                              & 
\multicolumn{1}{c}{$\log(\rho_0)$}                     & 
\multicolumn{1}{|c}{$C$}                               & 
\multicolumn{1}{c}{$\log(\mathcal{M}_{200}^{\rm halo})$} &
\multicolumn{1}{|c}{$f_{\ast}^{\rm iso}$}              &
\multicolumn{1}{c}{$R_C$}                               & 
\multicolumn{1}{c}{$\log(\rho_0)$}                     &  
\multicolumn{1}{|c}{$f_{\ast}^{\rm nfw}$}              &
\multicolumn{1}{c}{$C$}                                & 
\multicolumn{1}{c|}{$\log(\mathcal{M}_{200}^{\rm halo})$}                   \\
\multicolumn{1}{|c}{}                                  & 
\multicolumn{1}{|c}{($\msol/\lksol$)}         &
\multicolumn{1}{|c}{($\hr$)}                           &
\multicolumn{1}{|c}{(kpc)}                             & 
\multicolumn{1}{c}{($\msol~{\rm kpc^{-3}}$)}       & 
\multicolumn{1}{|c}{}                                  & 
\multicolumn{1}{c}{($\msol$)}                      &
\multicolumn{1}{|c}{}                                  &
\multicolumn{1}{c}{(kpc)}                              & 
\multicolumn{1}{c}{($\msol~{\rm kpc^{-3}}$)}       &  
\multicolumn{1}{|c}{}                                  &
\multicolumn{1}{c}{}                                   & 
\multicolumn{1}{c|}{($\msol$)}                     \\
\hline
%
  448 & 0.31 $\pm$ 0.21 & 1.5 $\pm$ 0.1 &  0.35 $\pm$ 0.08 & 9.62 $\pm$ 0.20 & 18.5 $\pm$ 2.0 & 11.43 $\pm$ 0.05 & 3.1 & 10.5 & 6.78 & 3.1 &  1.1 & 12.20 \\
  463 & 0.29 $\pm$ 0.17 & 1.3 $\pm$ 0.1 &  1.00 $\pm$ 0.21 & 8.70 $\pm$ 0.16 & 14.1 $\pm$ 1.9 & 11.44 $\pm$ 0.07 & 2.2 &  1.8 & 7.47 & 2.1 &  0.4 & 12.17 \\
 1081 & 0.40 $\pm$ 0.16 & 1.2 $\pm$ 0.1 &  1.75 $\pm$ 0.67 & 8.08 $\pm$ 0.25 &  6.7 $\pm$ 3.0 & 11.54 $\pm$ 0.40 & 2.2 &  1.9 & 7.48 & 2.2 &  1.5 & 11.66 \\
 1087 & 0.32 $\pm$ 0.13 & 0.9 $\pm$ 0.1 &  1.12 $\pm$ 0.14 & 8.58 $\pm$ 0.09 & 12.7 $\pm$ 1.4 & 11.43 $\pm$ 0.07 & 3.4 &  2.9 & 7.48 & 3.4 &  2.9 & 11.64 \\
 1529 & 0.29 $\pm$ 0.16 & 1.2 $\pm$ 0.1 &  0.37 $\pm$ 0.12 & 9.59 $\pm$ 0.27 & 22.1 $\pm$ 2.5 & 11.36 $\pm$ 0.05 & 3.7 &  5.0 & 6.76 & 3.7 & 10.0 & 10.22 \\
 1635 & 0.26 $\pm$ 0.12 & 1.1 $\pm$ 0.1 &  0.89 $\pm$ 0.08 & 8.72 $\pm$ 0.07 & 14.8 $\pm$ 0.7 & 11.29 $\pm$ 0.03 & 4.2 &  1.0 & 8.15 & 4.8 &  3.0 & 11.00 \\
 1862 & 0.56 $\pm$ 0.29 & 0.9 $\pm$ 0.1 &  0.80 $\pm$ 0.14 & 8.46 $\pm$ 0.10 &  8.4 $\pm$ 3.2 & 11.16 $\pm$ 0.46 & 2.6 &  0.3 & 8.53 & 2.6 & 16.0 &  9.61 \\
 1908 & 0.26 $\pm$ 0.18 & 1.3 $\pm$ 0.2 &  4.56 $\pm$ 1.35 & 7.72 $\pm$ 0.19 &  4.3 $\pm$ 2.3 & 12.39 $\pm$ 0.46 & 2.0 & 16.5 & 6.79 & 2.0 &  1.0 & 12.82 \\
 3091 & 0.23 $\pm$ 0.10 & 0.8 $\pm$ 0.2 &  1.12 $\pm$ 0.13 & 8.59 $\pm$ 0.09 & 13.4 $\pm$ 1.2 & 11.42 $\pm$ 0.06 & 6.2 &  5.6 & 6.88 & 5.9 &  0.9 & 12.13 \\
 3140 & 0.33 $\pm$ 0.14 & 1.1 $\pm$ 0.1 &  0.39 $\pm$ 0.17 & 9.50 $\pm$ 0.36 & 15.3 $\pm$ 1.7 & 11.47 $\pm$ 0.05 & 2.1 & 18.2 & 6.43 & 2.1 &  0.3 & 12.63 \\
 3701 & 0.85 $\pm$ 0.40 & 1.0 $\pm$ 0.1 &  2.62 $\pm$ 0.24 & 7.67 $\pm$ 0.06 &  4.6 $\pm$ 0.7 & 11.46 $\pm$ 0.11 & 2.2 &  6.4 & 6.91 & 2.2 &  1.3 & 11.79 \\
 3997 & 0.45 $\pm$ 0.19 & 0.8 $\pm$ 0.1 &  1.01 $\pm$ 0.14 & 8.60 $\pm$ 0.11 & 11.8 $\pm$ 1.0 & 11.39 $\pm$ 0.05 & 5.0 &    - &    - & 5.0 &    - &     - \\
 4036 & 0.29 $\pm$ 0.11 & 1.2 $\pm$ 0.1 &  1.50 $\pm$ 0.25 & 8.42 $\pm$ 0.12 &  9.8 $\pm$ 1.0 & 11.70 $\pm$ 0.07 & 3.2 & 15.0 & 6.61 & 3.2 &  0.5 & 12.57 \\
 4107 & 0.28 $\pm$ 0.14 & 1.1 $\pm$ 0.1 &  0.97 $\pm$ 0.12 & 8.69 $\pm$ 0.09 & 14.6 $\pm$ 1.7 & 11.34 $\pm$ 0.07 & 4.1 &    - &    - & 4.1 &    - &     - \\
 4256 & 0.15 $\pm$ 0.19 & 1.0 $\pm$ 0.1 &  5.43 $\pm$ 0.49 & 7.58 $\pm$ 0.06 &  3.7 $\pm$ 0.4 & 12.43 $\pm$ 0.10 & 3.0 & 18.0 & 6.75 & 3.0 &  0.9 & 12.87 \\
 4368 & 0.57 $\pm$ 0.34 & 1.5 $\pm$ 0.1 &  2.24 $\pm$ 0.25 & 7.94 $\pm$ 0.08 &  6.6 $\pm$ 0.8 & 11.57 $\pm$ 0.07 & 2.1 & 14.7 & 6.52 & 2.1 &  0.7 & 12.22 \\
 4380 & 0.27 $\pm$ 0.09 & 0.7 $\pm$ 0.2 &  1.78 $\pm$ 0.22 & 8.36 $\pm$ 0.09 &  9.4 $\pm$ 1.0 & 11.83 $\pm$ 0.07 & 4.3 & 17.2 & 6.67 & 4.3 &  0.6 & 12.76 \\
 4458 & 0.42 $\pm$ 0.24 & 1.1 $\pm$ 0.1 &  0.30 $\pm$ 0.10 & 9.69 $\pm$ 0.02 & 22.0 $\pm$ 3.7 & 11.42 $\pm$ 0.04 & 1.9 & 10.0 & 6.45 & 1.9 &  0.1 & 12.00 \\
 4555 & 0.35 $\pm$ 0.20 & 1.1 $\pm$ 0.1 &  0.84 $\pm$ 0.13 & 8.83 $\pm$ 0.12 & 15.1 $\pm$ 1.4 & 11.40 $\pm$ 0.05 & 3.1 & 10.0 & 6.45 & 3.1 &  0.3 & 12.07 \\
 4622 & 0.28 $\pm$ 0.17 & 1.2 $\pm$ 0.2 &  1.40 $\pm$ 0.14 & 8.67 $\pm$ 0.08 & 10.7 $\pm$ 0.7 & 11.97 $\pm$ 0.03 & 4.6 & 38.7 & 6.08 & 4.6 &  0.1 & 12.76 \\
 6903 & 0.24 $\pm$ 0.17 & 0.4 $\pm$ 0.1 &  3.19 $\pm$ 0.34 & 7.83 $\pm$ 0.06 &  4.1 $\pm$ 1.4 & 12.21 $\pm$ 0.38 & 8.1 &  6.1 & 7.05 & 8.1 &  1.0 & 12.22 \\
 6918 & 0.06 $\pm$ 0.24 & 1.2 $\pm$ 0.1 &  0.46 $\pm$ 0.05 & 9.26 $\pm$ 0.08 & 20.3 $\pm$ 1.2 & 11.16 $\pm$ 0.04 & 6.2 &  4.3 & 7.51 & 5.5 &  4.4 & 11.79 \\
 7244 & 0.41 $\pm$ 0.18 & 0.5 $\pm$ 0.1 &  6.60 $\pm$ 1.09 & 7.10 $\pm$ 0.08 &  2.5 $\pm$ 0.5 & 11.82 $\pm$ 0.08 & 1.6 & 10.8 & 6.80 & 1.6 &  1.0 & 12.29 \\
 7917 & 0.29 $\pm$ 0.33 & 1.5 $\pm$ 0.1 &  2.59 $\pm$ 0.55 & 8.18 $\pm$ 0.16 &  9.6 $\pm$ 1.0 & 12.00 $\pm$ 0.05 & 3.4 & 18.7 & 6.04 & 3.4 &  0.1 & 11.97 \\
 8196 & 0.94 $\pm$ 0.41 & 2.1 $\pm$ 0.2 & 15.50 $\pm$ 1.50 & 6.64 $\pm$ 0.05 &  1.0 $\pm$ 1.0 & 12.55 $\pm$ 0.13 & 1.0 & 15.5 & 6.64 & 1.0 &  1.0 & 12.55 \\
 9177 & 0.36 $\pm$ 0.26 & 1.3 $\pm$ 0.1 &  1.30 $\pm$ 0.14 & 8.64 $\pm$ 0.09 & 12.1 $\pm$ 0.8 & 11.76 $\pm$ 0.03 & 4.6 &    - &    - & 4.6 &    - &     - \\
 9837 & 0.32 $\pm$ 0.14 & 0.5 $\pm$ 0.1 &  1.68 $\pm$ 0.11 & 8.25 $\pm$ 0.05 &  8.2 $\pm$ 0.7 & 11.64 $\pm$ 0.07 & 8.0 &  8.9 & 6.69 & 8.0 &  0.5 & 12.06 \\
 9965 & 0.25 $\pm$ 0.12 & 0.9 $\pm$ 0.2 &  1.36 $\pm$ 0.11 & 8.45 $\pm$ 0.06 & 10.9 $\pm$ 0.7 & 11.56 $\pm$ 0.04 & 4.0 & 20.0 & 6.48 & 4.0 &  0.3 & 12.38 \\
11318 & 0.29 $\pm$ 0.20 & 1.2 $\pm$ 0.1 &  0.97 $\pm$ 0.58 & 8.83 $\pm$ 0.47 & 13.5 $\pm$ 4.0 & 11.64 $\pm$ 0.16 & 3.0 &    - &    - & 3.0 &    - &     - \\
12391 & 0.41 $\pm$ 0.19 & 1.2 $\pm$ 0.1 &  2.03 $\pm$ 0.31 & 8.03 $\pm$ 0.10 &  5.9 $\pm$ 0.8 & 11.71 $\pm$ 0.11 & 2.0 & 11.7 & 6.79 & 2.0 &  1.0 & 12.29 \\
\hline
\end{tabular}
}
\tablefoot{Table listing derived stellar mass-to-light ratios ($\overline{\mls}$;
Sect.~\ref{sec:ML}), the effective radius at which $\overline{\mls}$ was measured
($R_{\Upsilon}$), fitted dark-matter-halo parameters (Sect.~\ref{sec:RCDecomp}), and the
scaling factors ($f_{\ast}^{\rm iso}$ \& $f_{\ast}^{\rm nfw}$) with which
$\overline{\mls}$ have to be multiplied to make the galaxy maximal.
}
\end{table*}



\section{Discussion}
\label{sec:Discussion}
In the previous sections, we presented results on the distribution of the baryonic matter 
and quantified the baryonic maximality. We derived the dark-matter rotation curves and 
compared results from the unique rotation-curve mass decompositions with the maximum-disk 
hypothesis. In this section, we further discuss the galaxy {\it disk} and {\it total
baryonic} maximality (Sect.~\ref{sec:Disc-maxi}) and the arguments for and against maximal
and submaximal disks (Sect.~\ref{sec:Disc-maxdisk}). We briefly discuss the difficulties
in distinguishing between the pISO and NFW parameterizations of the dark-matter
rotation curves (Sect.~\ref{sec:ISOvsNFW}). Finally, we include a discussion on our
adopted assumptions and the introduced uncertainties in our analysis
(Sect.~\ref{sec:Disc-uncert}).

\subsection{Disk and baryonic maximality}
\label{sec:Disc-maxi}
The ``maximality'' of a galaxy is often assessed on the basis of the {\it stellar-disk}
mass fraction, $\mathcal{F}_{\rm disk,\ast}$$=$$V_{\rm disk,\ast}/\vc$, traditionally
measured at $2.2\hr$, and including the total mass associated with the stellar disk
(e.g., with a molecular-gas component assumed to have the same distribution as the stars).
The most commonly adopted definition of a maximum disk was provided by \cite{sackett1997}, 
who proposed $\mathcal{F}_{\rm disk,\ast}^{2.2\hr}$$=$0.85$\pm$0.10 to be an appropriate 
definition for maximum disks in galaxies of Hubble type similar to the Milky Way (Sb and 
Sc). The Sackett definition allows for small contributions from a bulge and a dark-matter 
halo with a non-hollow core, as well as a contribution from the \hone\ gas, but the 
literature studies on which this number was based usually incorporated the molecular gas
component in the stellar-disk mass distribution implicitly.
In this paper, instead, we are considering the {\it total baryonic} mass fraction. In our
analysis of the maximum-$\mls$ case in Sect.~\ref{sec:maxdisk}, we define a {\it galaxy}
to be maximal if $\Fbary$$\sim$1 in the inner region of the galaxy, keeping the 
contribution from the dark-matter halo minimal.
We still compare our measured $\Fbary^{2.2\hr}$ values with the definition of a maximum
disk in \cite{sackett1997}, but note that in general 
$\Fbary^{2.2\hr}$$\geq$$\mathcal{F}_{\rm disk,\ast}^{2.2\hr}$. This is certainly true for 
all galaxies in our sample. Thus, for our definition of the baryonic maximality, the 
fraction should be larger than the Sackett definition and closer to one.
Indeed, in Sect.~\ref{sec:maxdisk} we found that for the maximum-$\mls$ case, calculated
by maximizing the masses of the stellar bulges and disks as much as possibly allowed by
the observed rotation curves, we measure an average and scatter of 
$\langle$$\mathcal{F}_{\rm b,max}^{2.2\hr}$$\rangle$$=$0.92$\pm$0.08.
If we instead calculate the sample average of the ratio between the scaled stellar disk
and the observed rotation curve at $2.2\hr$, we find it to be 0.84$\pm$0.09, similar to
what \cite{sackett1997} found. If we include the molecular gas into the stellar disk we 
get an average fraction of 0.87.

The sample average of $\Fbary^{2.2\hr}$ found in this paper,
$\langle$$\Fbary^{2.2\hr}$$\rangle$$=$0.57$\pm$0.07, is larger than what we found for the
same sample of galaxies using a different approach in \citetalias{bershady2011},
$\langle$$F_{\rm max}^{\rm disk}$$\rangle$$=$0.47$\pm$0.08.
Although we are using the same sample of galaxies here and in \citetalias{bershady2011},
the maximality measurements presented are different in several ways.
Values of $F_{\rm max}^{\rm disk}$ presented in \citetalias{bershady2011} were based
on the central vertical velocity dispersion of the disk stars ($\sigma_{\rm z,0}$),
excluding the bulge by extrapolating the fitted exponential increase of $\sigz$ to the
center.
However, here $\Fbary^{2.2\hr}$ is a measure of the {\it baryonic} maximality, with the 
stellar bulge included. We therefore expect $\Fbary^{2.2\hr}$$>$$F_{\rm max}^{\rm disk}$. 
The difference between the two is dependent on $B/D$; the generally small bulges in our
galaxy sample, with the average bulge contributing $\sim$2\% to the total rotation curve
at $2.2\hr$, leads to an average difference of 0.02 between $\Fbary^{2.2\hr}$ and
$F_{\rm max}^{\rm disk}$.
A fundamental procedural difference in the calculation of $\Fbary^{2.2\hr}$ and 
$F_{\rm max}^{\rm disk}$ is that the former is directly tied to the observed surface 
brightness profile of each galaxy, while in \citetalias{bershady2011} we instead used only
one extrapolated value of the central stellar velocity dispersion.
In \citetalias{martinsson2012a}, we found that many galaxies have measured velocity
dispersions in the outer disk which are larger than what is expected for an exponential
decline. Therefore, it is possible that there is a systematic error introduced, such that
the data points at larger radii lower the central velocity dispersion of the fit. However,
the inner points are in general well fit by an exponential decline, with little influence
from the outer data points which have larger errors. Thus, any systematic error should be
small.
To conclude, both random and systematic differences are expected between 
$F_{\rm max}^{\rm disk}$ and $\Fbary^{2.2\hr}$ due to differences in the methodology of
deriving these quantities, and because these are intrinsically two different quantities.

\subsection{Submaximal versus maximal disk}
\label{sec:Disc-maxdisk}
Related to the maximality discussed above, we will here discuss arguments for both 
submaximal and maximal-disk cases.
It has been pointed out \citep[e.g., by][]{AlbadaSancisi1986} that there are several 
arguments to support the idea that disks in spiral galaxies are maximal. For example, the 
simplest way to explain why fitting a maximum disk works is that the disks actually are 
close to maximum. It has frequently been shown that the dynamical mass distribution 
follows the light distribution in the inner regions of galaxies 
\citep[e.g.,][]{kent1986,sancisi2004,noordermeer2007,swaters2011,fraternali2011}, with a 
contribution from the stellar mass that can be scaled up to fully explain the shape of the
inner part of the rotation curve within the radius typically reached by optical rotation 
curves. It has also been found that features in the rotation curve can be seen in the
light profile as well. However, these arguments in favor of a maximum disk can be 
disputed.

Even though the rotation curves of the galaxies in our sample can be decomposed rather 
convincingly using a maximum $\mls$, our submaximal-$\mls$ cases result in significantly
better fits (Sect.~\ref{sec:compDMmod}). Furthermore, in the context of Newtonian gravity,
the extended, flat \hone\ rotation curves require dark-matter halos to explain the
non-declining outer part generally observed in spiral galaxies. If these observations
indeed convince us of the existence of dark-matter halos, why should this dark matter be
distributed in such a way that it is only required to explain the outer part of the
rotation curves? With an all-over dominating dark-matter halo, the ``conspiracy''
\citep{bahcall1985} between the stellar disk and dark-matter halo to produce flat rotation
curves weakens.
Even though features in the observed rotation curve are often seen in the light profile,
this only suggests that the gravitational potential of the visible stars or gas is 
non-negligible in the galaxy mid-plane; e.g., a bump in a non-dominant baryonic
rotation curve that is added in quadrature to a smooth dominant dark-matter rotation curve
may still be seen, especially in the inner regions where the baryons contribute most. We
have two examples of this in our sample: UGC~4256 and UGC~6918 (see Atlas). Even though
the rotation curves of these galaxies are dominated by $\vdm$, features in $V_{\rm mol}$
are still clearly seen in the observed rotation curves. Furthermore, the features seen in
the rotation curves are often associated with non-axisymmetric structures such as 
spiral arms, i.e., perturbations on both velocity and surface brightness. This requires 
neither maximality nor submaximality. 

A theoretical argument in favor of submaximal disks is presented in \cite{amorisco2010}. 
From constructed self-consistent models of nonspherical isothermal halos embedding a 
zero-thickness disk, they find a best-fit model that is significantly submaximal.

Observationally, one can infer that disk galaxies are submaximal given the lack of a
surface-brightness dependence in the Tully-Fisher relation for a wide range of spiral
galaxies \citep{zwaan1995,courteau1999,courteau2003}. Also barred galaxies, which are
suspected to be close to maximal, show no offset from the Tully-Fisher relation.
However, studies from kinematic observations have so far not reached a consensus. For 
example, \cite{weiner2001} used fluid-dynamical models of gas flows in a barred galaxy to
constrain its dynamical properties. They found that maximum-disk values are highly favored
for this galaxy, and argued that the luminous matter dominates inside the optical radius
of high surface brightness disk galaxies in general.
\cite{kranz2003} studied five high surface brightness late-type spiral galaxies using
hydrodynamic gas simulations and draw the conclusion that high surface brightness galaxies
possess maximal disks if their maximal rotation velocities are larger than 200 \kms.
If the maximal rotation velocity is less, the galaxies appear to have submaximal disks.
Analyses of one barred, resonance-ring system \citep{byrd2006} and one gravitational-lens
system \citep{dutton2011} indicate that these two galaxies are submaximal. However, in
another lensed galaxy, \cite{barnabe2012} found that the dark-matter fraction within
2.2 optical disk scale lengths ($\hr$) was only $\sim$28\%, implying a maximum disk.
\cite{dutton2013} further investigated five gravitationally-lensed, bulge-dominated disk
galaxies, and found that the {\it disks} of the galaxies are submaximal at 2.2$\hr$, but
due to their large bulges, these galaxies are still baryon dominated at this radius. 
Results from the kinematics of planetary nebulae in several nearby galaxies suggest
that the maximality may depend on Hubble type \citep{herrmann2009}. In conclusion, many
different methods have resulted in many different results.

In addition to our own study, similar analysis of stellar-kinematic observations of normal
spiral galaxies \citep[e.g.,][]{bottema1993, kregel2005, herrmann2009} generally show that
spiral galaxies are submaximal with disks that contribute $\sim$60$\pm$10\% to the
observed rotation speed, with possible dependencies on morphological type, surface
brightness, luminosity and color.
The result presented in this paper is another strong argument against maximal disks, and
it shows that normal, disk-dominated spiral galaxies are submaximal. To infer maximal
disks on the basis of our stellar-kinematic measurements, we would on average have to
increase the observed velocity dispersions by a factor of $\sim$2, or alternatively
decrease $\hz$ or $k$ with a factor of 3.6 for the entire sample. Based on the available
empirical evidence from edge-on galaxies \citep[e.g.,][]{kregel2002}, this does not seem
plausible.

In this paper, we compare our fitted NFW-halo parameters from both the submaximal and 
maximum-$\mls$ cases with what has been found from numerical N-body simulations 
\citep{bullock2001}. Interestingly, we find that the fitted parameters from the submaximal
case show a $C$-$\mathcal{M}_{200}^{\rm halo}$ relation that is consistent with these
simulations, with a concentration parameter typically between $C$$\sim$10--20. On the 
other hand, in the maximum-$\mls$ case the fitted concentration parameters are
much lower, ($C$$\sim$1), and show little correlation with $\mathcal{M}_{200}^{\rm halo}$.
The dark-matter-halo parameters derived from the maximum-$\mls$ decompositions are not
consistent with these simulations.

\subsection{The dark-matter distribution}
\label{sec:ISOvsNFW}
The flatness of the outer part of rotation curves suggests a halo with a dark-matter 
density distribution declining as $\rho$$\propto$$R^{-2}$. The inner slope is still
debated. The NFW and similar shapes of the dark-matter halo which have cuspy density
profiles with a central slope of $\rho$$\propto$$R^{-1}$ to $\rho$$\propto$$R^{-1.5}$,
have repeatedly been found in dark-matter-only numerical simulations 
\citep[e.g.,][]{NFW1996, NFW1997, moore1999,klypin2001,diemand2005}. Observationally,
however, a core-like density distribution is often inferred \citep{deBlok2001, 
swaters2003a, swaters2003b, KuziodeNaray2008, KuziodeNaray2009, deBlok2010, oh2011}.

Generally, it is thought \citep[e.g.,][]{blumenthal1986, gnedin2004} that the baryons in
the inner regions will tend to contract the halo while it is forming. The uncontracted 
halo in dark-matter-only simulations would then be more extended and have a lower
concentration than in observed galaxies. This contraction of the dark-matter halo could be
an issue when we fit our models to the data; however, for our light-weight disks we 
suspect this will only have a mild effect.
Several processes may also occur that would instead expand the halo (see
Sect.~\ref{sec:Introduction}). Recent results from, e.g., \cite{barnabe2012} and 
\cite{dutton2013} show that massive spiral galaxies have dark-matter distributions
consistent with an unmodified NFW profile.
In summary, there is no clear consensus on what can be expected for the inner slope of
the dark-matter density distribution due to uncertainties in the baryon physics of galaxy
formation. In this paper, we fit pISO and NFW models to our data. Fitting the
dark-matter halo with alternative parameterizations to account for the uncertainty in the
inner slope, e.g., using Einasto profiles \citep{einasto1979, chemin2011}, will be
deferred to a forthcoming paper.

In general, the differences between our fitted pISO and NFW rotation curves are miniscule;
both fits have similar $\chi^2$. In Sect.~\ref{sec:DMRCshape}, we stacked data from all
galaxies to investigate whether the pISO or NFW model fit the data better on average.
Although we find slightly less scatter for the pISO-model fit, these results have little
statistical significance, mainly due to limitations in our data close to the center and
outside the optical disk where the pISO and NFW models tend to differ most. To better
distinguish between pISO and NFW-halo distributions we recommend the following improvement
at large and small radii for future observations.

Because inner regions often display large gradients in velocity, corrections for beam
smearing are large and dependent on assumptions about the unresolved flux and velocity
distribution. Corrections can be minimized with observations taken at sufficient physical
resolution to resolve the inner rotation-curve rise. The influence of baryonic matter is
also highest in the inner regions, primarily due to the presence of bulge or inner-disk
components which may have systematically different $\mls$ than the outer disk. Direct,
dynamical assessment of these innermost mass components would reduce systematic errors in
this part of the baryonic rotation curve. Alternatively, observing pure-disk (bulgeless)
galaxies might remove this extra complexity, but would naturally lead to an increased bias
towards very late-type galaxies. However, chaotic motions in the innermost regions of such
systems \citep{swaters2003b} may nullify potential advantages of this strategy.

In the outer regions, robust rotation curves require a detailed assessment of both
inclination and position-angle warps.  Inclination warps are difficult to constrain for
nearly face-on galaxies, even with the availability of deep 21-cm aperture-synthesis
observations \citep{martinsson2011}. Galaxies at higher inclination ($\sim$45$\arcdeg$ to
60$\arcdeg$) are required to properly measure inclination warps free of assumptions about
the asymptotic shape of the rotation curve. Such high inclination would complicate the
measurement of $\sigz$, but this is mitigated by the relations found in this paper, namely
the calibration of $\mls$, and better knowledge of the SVE (presented in future papers of
this series).

\subsection{Uncertainties and assumptions}
\label{sec:Disc-uncert}
The steps in this paper, which have taken us from our observational data (stellar and gas 
line-of-sight kinematics, NIR, 24-$\mu$m and 21-cm imaging) to rotation-curve mass 
decompositions, include many systematic uncertainties as discussed in great detail in 
\citetalias{bershady2010b}. Here we revisit the most significant ones.

Throughout our analysis, we have adopted a constant shape of the stellar velocity 
ellipsoid (SVE) for all galaxies. Fortunately, as part of our survey strategy 
\citepalias{bershady2010a}, the resulting uncertainties in the conversion from $\slos$ to 
$\sigz$ are limited by the nearly face-on orientation of our galaxy sample. As discussed 
by \cite{herrmann2009}, from both observations
and theoretical works on the physics of disk scattering,
we do not expect $\alpha$, which is the dominant contributor to the uncertainties in our 
deprojection of $\slos$ to $\sigz$, to be larger than one. If we assume $\alpha$$=$1 then 
the calculated $\Fbary$ will certainly be larger; however, we would still find that a 
majority of the galaxies in our sample are submaximal, with an average 
$\langle$$\Fbary^{2.2\hr}$$\rangle$$=$0.65 when excluding UGC~8196, and with only 6 out
of 29 galaxies larger than 0.75.
In \citetalias{westfall2011b}, the shape of the SVE was derived directly from the data
for one of our galaxies (UGC~463). We found that it has $\alpha$$=$0.48$\pm$0.09 and
$\beta$$=$1.04$\pm$0.22. Even though these values differ from what we have adopted for
this paper, $\alpha$ is still well within the errors, and the measured $\beta$ differs
only $1.5\sigma$ from our assumed value. The shape of the SVE for the remaining galaxies
in the sample will be revisited in a later paper of the DMS series,
where a recalculation of $\sigz$ will be performed and the effect on earlier results will
be examined.

Another contribution to the systematic errors comes from uncertainties in the vertical
density distribution. It is possible that the distribution parameter ($k$) differs among
and within galaxies. If $k$ is systematically higher than the adopted value, e.g., with a
vertical isothermal $\rm sech^2$ distribution instead of the adopted exponential
distribution, our nominal calculations produce an overestimate of $\sddisk$. This would
imply that we overestimate $\Fbary$. If $k$ is systematically lower than the adopted
value, we would instead underestimate $\Fbary$.
Among the vertical mass distributions suggested by \cite{kruit1988}, we have chosen a 
value of $k$$=$1.5, maximizing $\sddisk$ and $\Fbary$. In some cases, the effective value
of $k$ could be lower; e.g., when the galaxy has a massive thin gas disk. However, the
effect from a dark-matter halo (see below) would instead increase $k$, canceling the 
effect of the massive gas disk \citepalias{westfall2011b}.

Since we observe nearly face-on galaxies, it is impossible to measure the vertical scale
height $\hz$. Instead, we calculate $\hz$ from $\hr$, using the relation found from
edge-on galaxies as described in \citetalias{bershady2010b}. This introduces a significant
systematic error (25\%) within a galaxy; however, these errors are expected
to be random when considering the sample as a whole.

Direct kinematic measurements of the disk inclination has been proven to be difficult for
our nearly-face-on galaxies (\citealt{martinsson2011}; \citetalias{martinsson2012a};
Andersen~\&~Bershady~2013, ApJ, {\it in press}), and we have adopted inclinations using
the inverse Tully-Fisher relation \citepalias{martinsson2012a}. The uncertainties from
these estimated inclinations, mainly introduced as systematic errors in the observed 
circular speed measurements, are again expected to be random for the sample.

The only known systematic uncertainty for the survey as a whole is the estimated 7\%
uncertainty on the Hubble constant \citepalias[$H_0$;][]{bershady2010b}. An error on 
$H_0$ results in an error on the derived distances, which would propagate to both the
derived scale ($\hr$, $\hz$) and luminosity ($M_K$) of a galaxy. This would affect our
measurement of $\Fbary$ in the sense that an error on $\hz$ propagates to $\vbary$ (via
$\sddisk$), and an error on $M_K$ propagates to $\vc$ (via $\itf$). Although $\vbary$ and
$\vc$ change in opposite direction (e.g., if $H_0$ is smaller, then $\vbary$ gets smaller
and $\vc$ gets larger), thereby maximizing the error on $\Fbary$, the systematic
uncertainty on $\Fbary$ due to the uncertainty on $H_0$ is only 5\%.

Additional uncertainties are introduced by our limited accounting of all the baryonic 
components. We have only considered the four (presumably) dominant baryonic contributors
to the total potential of the galaxy; the stellar bulge, stellar disk, atomic-gas disk, 
and molecular-gas disk.
Our somewhat simplistic one-dimensional bulge-disk decompositions
\citepalias{martinsson2012a} could introduce some errors, both due to an erroneous
assumption of the mass distribution, and a possibly different $\mls$ for the bulge 
compared to the disk. We have limited the uncertainties from the bulge by excluding 
kinematic measurements inside $\rbulge$ (Sect.~\ref{sec:Photometry}), e.g., when 
calculating $\overline{\mls}$. The understanding of the bulge is critical to the inner
shape of the dark-matter rotation curves. However, at larger radii (e.g., at 2.2$\hr$),
the uncertainties from our bulge decompositions are negligible for most of the galaxies
in our sample.

We have assumed that the stellar halo contains so little mass as to be 
dynamically negligible. This is also a valid argument for neglecting the contribution of 
the mass surface density of the dust in the disk, $\sdd$, which has been 
observed to have typical ratios $\sdd/\Sigma_{\rm HI}$$=$0.01--0.1 
\citep{magrini2011}. However, dust may affect our measurements of $\mls$ due to
extinction, but since we observe close-to face-on galaxies in the NIR, this effect should
be small. We further assume that there is no hidden $\rm H_2$ gas, but that it is traced
by the 24-$\mu$m emission. Finally, we do not include a thick disk component because, as
mentioned in \citetalias{bershady2010b}, the resulting differences in $k$ and $h_z$ tend
to cancel when calculating $\sddisk$.

When calculating $\sddisk$ using Eq.~\ref{eq:mldyn}, we include an infinite,
plan-parallel, self-gravitating, isolated disk. For a disk embedded in a dark-matter
halo, we expect $\sigz$ will {\it increase} for a given $\sddisk$ 
\citep{bottema1993}. This, again, results in estimates of $\mls$ that are upper limits.
The shape of the halo is unknown, but is here assumed to be spherical. There have been
studies showing that the dark-matter halo might be highly oblate or prolate
\citep[e.g.,][]{amorisco2010, vera-ciro2011}.
This would further increase $\sigz$.

\section{Summary and Conclusions}
\label{sec:Summary_DeComp}
We have presented rotation-curve mass decompositions of 30 spiral galaxies. From our
stellar-kinematic observations we measure vertical velocity dispersions of the stars in
the galaxy disks ($\sigz$). These are used to calculate the disk's dynamical mass surface
density ($\sddisk$) from the relation $\sddisk$$=$$\sigz^2/(\pi G k \hz)$, where we assume
a constant disk scale height ($\hz$) calculated from the disk scale length ($\hr$), and an
exponential vertical distribution of the disk mass and luminosity ($k$$=$1.5). Together
with measured atomic-gas mass surface densities from 21-cm radio synthesis observations,
and molecular-gas mass surface densities estimated from 24-$\mu$m {\it Spitzer}
observations, we derive the stellar mass surface density. Using near-infrared 2MASS
photometry, we calculate the mass-to-light ratio of the stellar disk ($\mls$), and with
the assumption of a constant $\mls$ with radius, we derive the stellar-bulge and 
stellar-disk mass surface densities from the observed surface brightness profiles.
The rotation curves of the baryonic components are calculated from their radial mass
surface density profiles. These are used together with \hone\ and \halp\ circular-speed
measurements to derive the structural parameters of the dark-matter halo, modeled as
either a pISO or NFW halo. We consider two different models for the mass content of the
baryonic components; Case~I, with mass surface densities dynamically derived directly from
our kinematic and photometric observations, and Case~II, with scaled-up stellar bulge and
disk masses which disregards our dynamical mass-density estimates and instead is designed
to contribute maximally to the total mass (a maximum-$\mls$ case).

Our derived gas-mass fractions have been compared to results in the recent literature.
These agree fairly well, except that we find systematically higher gas-mass fractions.
This is most likely due to different methods used to derive the stellar masses.

For individual galaxies, $\mls$ has systematic errors that are on average $\sim$70\%,
mainly due to the uncertainty in $\hz$ when converting $\sigz$ to $\sddisk$, and the
uncertainty in the subtraction of the molecular gas from $\sddisk$. All galaxies have
radially-averaged $K$-band $\overline{\mls}$ consistent with being equal. The 
error-weighted sample average and scatter is 
$\langle$$\overline{\mls}$$\rangle$$=$0.31$\pm$0.07. On average, we find a $\mls$ that is
a factor of 3.6 lower than required by the maximum-disk hypothesis. This factor ranges
between 1.6 to 8.1, with 20 of the galaxies having values in the range 2.1--4.6. In
general, for the galaxies in our sample, the dark-matter halo dominates the potential at
almost all radii.

We find the ratio of the baryonic to total rotation velocity, $\Fbary$$=$$\vbary/\vc$, to
be nearly constant between 1--6$\hr$ within a galaxy, ranging from $\sim$0.4--0.7 among
individual galaxies. This rather constant $\Fbary$ arises due to the atomic gas taking
over at larger radii where the stellar-disk contribution is declining. The result has a
consequence for measuring the maximality; for a comparison of the maximality between
different galaxies, it is not critical to measure it at exactly
$R_{\rm max}^{\rm disk}$$\approx$$2.2\hr$, and averaging over a band in radius, e.g.,
around $R$$=$(2.2$\pm$0.5)$\hr$, could be a way to obtain measurements with smaller errors.

All galaxies in the sample for which we have reliable measurements are submaximal.
On average, $\Fbary^{2.2\hr}$$=$0.57$\pm$0.07, with a weak trend of larger
$\Fbary^{2.2\hr}$ for more luminous galaxies, in agreement with our result in 
\citetalias{bershady2011}, but also with a trend of larger $\Fbary^{2.2\hr}$ for galaxies 
with higher central surface brightness.

The dark-matter rotation curves tend to be marginally better fit with a pISO than a NFW
halo; however, due to limitations in our data both in the central and outer regions of the
galaxies, this result has little statistical significance. The shapes of our inferred
dark-matter rotation curves in the nominal-$\mls$ case are well fit with parameters in
agreement with what has been found from dark-matter-only simulations. This result suggests
that the baryonic matter in our sample of galaxies has had only a minor effect on the
dark-matter distribution. However, when fitting a NFW halo in the maximum-$\mls$ case, we
find a concentration parameter that is too low compared to the simulations.

A number of improvements may be made to the analysis presented in this paper. With 
SparsePak, we have observed the stellar kinematics of 12 additional galaxies that, when 
combined with the PPak data, will increase our sample by more than a third and broaden our 
parameter space further. Inclusion of our deep optical ($UBVRI$) and near-infrared ($JHK$) 
data from the KPNO 2.1-m telescope will dramatically improve our surface photometry. Work 
is ongoing to determine the SVE shape in individual galaxies, with one of the goals being 
to reduce the systematic errors introduced from deprojecting $\slos$ to $\sigz$. Future 
analysis will benefit from an improved understanding of the relation between $\hr$ and 
$\hz$, taking advantage of ongoing surveys of edge-on galaxies. In future papers,
we will extend the analysis of the measured dark-matter distribution, e.g., to establish
any correlation with the \hone\ distribution.

In this paper, we have assumed Newtonian gravity. However, it would be interesting to 
investigate if the theory of modified Newtonian dynamics \citep[MOND;][]{milgrom1983} can 
produce convincing results using submaximal disks. Until then, we conclude that with our 
assumed Newtonian gravity, the results presented in this paper indicate that the dark 
matter in spiral galaxies is not only needed to explain the non-declining \hone\ rotation 
curves in the outer regions, but is also required to explain the rotation curves overall.

\acknowledgements
First, we wish to thank the referee for useful comments and suggestions.
T.P.K.M.\ and M.A.W.V.\ acknowledge financial support provided by NOVA, the Netherlands
Research School for Astronomy, and travel support from the Leids Kerkhoven-Bosscha Fonds.
Support for this work has also been provided by the National Science Foundation (NSF) via
grants AST-0307417 and AST-0607516 (M.A.B.\ and K.B.W.), OISE-0754437 (K.B.W.), and
AST-1009491 (M.A.B.). K.B.W. is also supported by grant 614.000.807 from the Netherlands
Organisation for Scientific Research (NWO). R.A.S.\ and M.A.B.\ acknowledge support from
NASA/Spitzer grant GO-30894.
This publication makes use of data products from the Two Micron All Sky Survey, which is a
joint project of the University of Massachusetts and the Infrared Processing and Analysis
Center/California Institute of Technology, funded by the National Aeronautics and Space
Administration and the National Science Foundation.
This work is based in part on observations made with the Spitzer Space Telescope, which is
operated by the Jet Propulsion Laboratory, California Institute of Technology under a
contract with NASA.

\bibliography{Martinsson_DMS_VII} 
\bibliographystyle{aa_DMS_VII}
\setlength{\bibsep}{1.3pt}

\appendix

\section{The Atlas}

{\small
\subsection{Content of the Atlas}
In this appendix we present data and results for individual galaxies. The next subsection
(Sect.~\ref{app:gal}) provides some notes on the individual galaxies. In
Sect.~\ref{sec:Atlas}, an Atlas page is provided for each galaxy. Each
page is divided into two columns, each with four panels. The left column
demonstrates the progression from the observed surface brightness ($\mu_{K}$) and vertical
velocity dispersions of the disk stars ($\sigz$) to mass surface densities ($\Sigma$) and
mass-to-light ratios ($\ml$). The right column provides four rotation-curve mass 
decompositions.

The top panel in the left column shows the surface brightness profile. Gray dots represent 
the observed $K$-band surface brightness derived in \citetalias{martinsson2012a}. The
dashed line shows the fitted bulge. This bulge has been subtracted from the total surface 
brightness to obtain the light profile of the disk (black errorbars).
The second panel shows the measured $\sigz$ (errorbars), azimuthally averaged in 
5$\arcsec$ wide rings. The dashed line indicates the exponential fit to the 
individual-fiber data \citepalias{martinsson2012a}.
The third panel shows the derived mass surface density of the disk ($\sddisk$) and its
three component; stars ($\sds$), atomic gas ($\sda$) and molecular gas ($\sdm$). The open
circles indicate $\sddisk$, calculated directly from the measured $\sigz$. The filled dots
with errorbars show the calculated $\sds$ (where the gas has been subtracted from
$\sddisk$). The dotted and dash-dotted lines show measured $\sda$ and $\sdm$, 
respectively. The black dashed line shows $\sds^{\rm disk}$, calculated from the surface
brightness (above) and the weighted average stellar mass-to-light ratio (below).
The dark gray dashed line, falling of most rapidly with radius, indicates
$\sds^{\rm bulge}$, calculated with the same mass-to-light ratio as the disk.
The bottom panel shows the calculated dynamical ($\mldyn$; open circles) and stellar 
($\mls$; filled circles with errorbars) mass-to-light ratios. The solid and dashed lines 
show the weighted average of $\mls$ and $\mldyn$, respectively. The arrow on the x-axis 
indicates $2.2\hr$.
In all figures, the light gray shaded region indicates the region excluded from any 
analysis, typically the ``bulge'' region, delimited by the radius $\rbulge$ at which the
light from the bulge contributes 10\% to the total light. We have excluded any points
within this region when calculating the average $\ml$. In all panels, the darker gray
areas around the measured points indicate systematic errors.

The right column shows four different rotation-curve decompositions. Note that the panels
in this column have different radial scaling than the panels in the left column. The
observed \halp\ and \hone\ rotation curves are indicated with filled and open dots,
respectively. The rotation curves of the stellar bulge (dark gray dashed line), stellar
disk (black dashed line), molecular gas (dash-dotted line) and atomic gas (dotted line)
have been calculated from the mass surface densities shown in the left column, as
described in Sect.~\ref{sec:RCmod}. The solid gray line indicates the fitted dark-matter
rotation curve. The solid black line is the quadrature sum of the various components. 
From top to bottom, the four panels in the right column show the rotation-curve mass
decomposition results when using the nominal $\mls$ with a fitted pISO dark-matter halo
(Case~Ia); the nominal $\mls$ with a fitted NFW dark-matter halo (Case~Ib); a maximum disk
(scaled-up $\mls$) with a fitted pISO dark-matter halo (Case~IIa); and a maximum disk with
a fitted NFW dark-matter halo (Case~IIb). The arrow in the bottom panel indicates
$2.2\hr$, the theoretical radius of maximum rotation speed of an infinitely-thin
exponential disk.

\subsection{Notes on individual galaxies}
\label{app:gal}
Here, we present a few notes on the individual galaxies.
See \citetalias{martinsson2012a} for more detailed notes on the data products
of the individual galaxies from PPak, for comments on \oiii\ emission, stellar and
\oiii\ kinematics, kinematic flaring, known supernovae within the galaxies, and for
notes on close field stars.
See \cite{martinsson2011} for more comments on the \hone\ observations of the individual
galaxies, and for notes on close companion galaxies (detected in \hone).

\begin{list}{}
  {
  \settowidth{\labelwidth}{\bf UGC 00000:}
  \setlength{\labelsep}{1em}
  \setlength{\itemsep}{0.2em}
  \setlength{\parskip}{0.2em}
  \setlength{\leftmargin}{\labelwidth}
  \setlength{\rightmargin}{0pt}
  }
  \item[{\bf UGC   448:}] IC 43.  High-quality kinematics, with regular stellar and gas
  kinematics. Significant bulge with second highest bulge-to-disk ratio in the sample
  ($B/D$$=$$0.32$). The rotation curve rises quite sharply. \hone\ rotation curve
  corrected for an inclination warp \citep{martinsson2011}. A small bar and significant 
  spiral structure are visible morphologically, but exhibit little kinematic influence.
  \item[{\bf UGC   463:}] NGC 234.  High-quality kinematics. PPak and SparsePak
  data studied in detail in \citetalias{westfall2011b}. Strong, three-arm
  spiral structure with minor streaming motions.
  \item[{\bf UGC  1081:}] NGC 575, IC 1710. Strongly barred galaxy. Bright
  field star within the PPak field-of-view. A Type-II break exists in $\mu_K(R)$
  at roughly $1\hr$ (approximately the same as the bar length).
  The mass surface density profile of the atomic gas is estimated using the results 
  in \cite{martinsson2011}.
  \item[{\bf UGC  1087:}] ``Ringing'' present in $\mu_K(R)$ associated with the
  azimuthal coherence of the tightly wound spiral arms.
  \hone\ rotation curve corrected for an inclination warp \citep{martinsson2011}.
  \item[{\bf UGC  1529:}] IC 193.  High-quality kinematics. Sc galaxy, rather
  typical of our sample, apart from the high inclination ($\itf$$=39\arcdeg$).
  The mass surface density profile of the atomic gas is estimated using the
  results in \cite{martinsson2011}.
  \item[{\bf UGC  1635:}] IC 208. A Type-II break in $\mu_K(R)$ occurs at $\sim$1$\hr$
  with a corresponding dip in the \oiii\ and \halp\ rotation curves; any dynamical
  association between these features is unknown. Gas poor.
  \item[{\bf UGC  1862:}] Unique among our sample: It has the lowest luminosity,
  $M_K$$=$$-21.0$, one magnitude fainter than the second least luminous galaxy (UGC~3701).
  Two Type-II breaks exist in $\mu_K(R)$ (at $\sim$23$\arcsec$ and $\sim$60$\arcsec$), but
  no indication of a bulge component; inner break is caused by the spiral arms. Appears to
  have a rather large bar. Data within $R$$=$$2\farcs5$ are excluded in the kinematic
  analysis, to avoid beam-smearing effects particularly strong in the center. The only
  galaxy in the sample for which the rotation curve does not reach $R$$=$$2.2\hr$.
  The mass surface density profile of the atomic gas is estimated using the results
  in \cite{martinsson2011}.
  \item[{\bf UGC  1908:}] NGC~927, Mrk~593. Barred galaxy with a weak Type-II break in
  $\mu_K(R)$ at $\sim$1$\hr$. Classified as a Starburst Nucleus Galaxy (SBNG), but its 
  nucleus has an ambiguous activity classification between $\rm H_2$ and LINER
  \citep{contini1998}.
  The mass surface density profile of the atomic gas is estimated using the results
  in \cite{martinsson2011}.
  \item[{\bf UGC  3091:}] Since the bulge/disk fitting routine resulted in a
  non-existing bulge, the excess light in the central region is interpreted as
  an inner disk. As with UGC~1862, $\slos$ and $\sigz$ data within $R$$=$$2\farcs5$
  are excluded from our analysis.
  The mass surface density profile of the atomic gas is estimated using the results in 
  \cite{martinsson2011}.
  \item[{\bf UGC  3140:}] NGC 1642. Very close to face on with $\itf$$=$$14\arcdeg$.
  Nicely defined spiral structure but slightly lopsided. The $\mu_K$ profile breaks
  to a {\it more extended} disk (larger scale length) at $R$$\sim$$16\arcsec$.
  There is a small offset between the \hone\ and \halp\ rotation curves
  (Section~\ref{sec:RC}), possibly due to \hone\ asymmetries.
  \hone\ rotation curve corrected for an inclination warp \citep{martinsson2011}.
  \item[{\bf UGC  3701:}] Second lowest disk surface brightness in our sample. Some
  ringing in the $\mu_K$ profile due to the spiral arms. Rotation curve rises slowly.
  \hone\ rotation curve corrected for an inclination warp \citep{martinsson2011}.
  \item[{\bf UGC  3997:}] Classified as Im by RC3 with low surface brightness.
  There is a small offset between the \hone\ and \halp\ rotation curves
  (Section~\ref{sec:RC}), maybe due to the warp in position angle, corrected for
  in the \hone\ rotation curve \citep{martinsson2011} but not in the \halp\
  rotation curve.
  \item[{\bf UGC  4036:}] NGC 2441.  Observations are dominated by ring-like structure,
  probably due to weak bar. Streaming motions likely affect the observed rotation curve.
  \item[{\bf UGC  4107:}] High-quality kinematics.  Well-defined three-arm
  spiral structure. Type-II break in $\mu_K(R)$ at $R$$\sim$$20\arcsec$.
  \item[{\bf UGC  4256:}] NGC 2532.  Two close companions $\sim$$4\arcmin$ to the north
  connected by an \hone\ bridge \citep{martinsson2011}. Interaction has likely produced
  the bright arm toward the east and the lopsidedness of the galaxy. High star-formation
  rate with very bright \oiii\ emission associated with visible star-formation regions
  \citepalias{martinsson2012a}. One of two galaxies in the sample with a molecular gas
  mass  which is larger than the stellar mass. Note that the bump feature in the modeled 
  molecular-gas rotation curve can also be seen in the observed rotation curve, and that
  the submaximal cases fit this feature better than the maximum-$\mls$ cases.
  There is a small offset between the \hone\ and \halp\ rotation curves 
  (Section~\ref{sec:RC}), maybe due to its kinematic lopsidedness.
  \item[{\bf UGC  4368:}] NGC 2575. Highest inclination in the sample
  ($\itf$$=$$45\arcdeg$). High-quality stellar kinematics. Type-II breaks in $\mu_K(R)$
  at $R$$\sim$$16\arcsec$ and $R$$\sim$$30\arcsec$.
  \item[{\bf UGC  4380:}]  Low-inclination galaxy ($\itf$$=$$13\fdg8$) with a  small
  apparent size and scale length. Stellar-kinematic data have limited radial extent.
  \hone\ rotation curve corrected for an inclination warp \citep{martinsson2011}.
  \item[{\bf UGC  4458:}]  NGC 2599, Mrk 389.  Earliest morphological type in our sample
  (Sa), with the largest bulge-to-disk ratio ($B/D$$=$0.72). Some spiral structure visible
  at large radii, but very smooth morphology otherwise. \hone\ rotation curve declines
  from 350 \kms\ to 250 \kms. The bulge dominates all stellar-kinematic data; 
  $\overline{\mls}$ derived from the last measured point only.
  \item[{\bf UGC  4555:}] NGC 2649.  Strong spiral structure affects $\mu_K(R)$.
  High-quality stellar-kinematic data.
  \item[{\bf UGC  4622:}] The most distant galaxy in the sample ($\vsys$$=$$12830$ \kms;
  $D$$=$$178$~Mpc). Stellar kinematics limited to $R$$<$$15\arcsec$. Type-II break in
  $\mu_K(R)$ at $R$$\sim$$15\arcsec$. Fourth highest bulge-to-disk ratio ($B/D$$=$0.16).
  \hone\ rotation curve corrected for an inclination warp \citep{martinsson2011}.
  \item[{\bf UGC  6903:}] Barred galaxy with rather low surface brightness. Poorest
  quality of stellar-kinematic data in our sample (one hour observation with PPak).
  Strong dip in $\mu_K(R)$ at $R$$\sim$$20\arcsec$.
  \item[{\bf UGC  6918:}] NGC 3982.  High-surface-brightness member of the Ursa
  Major cluster.  Very high-quality kinematic data. Classified as a Seyfert~1.9 
  \citep{veron2006}. Warped and lopsided extension to the \hone\ gas
  \citep{martinsson2011}; PPak kinematics are regular. Included in DMS pilot sample as
  presented in early publications \citep{verheyen2004,bershady2005,westfall2009}.
  Type-II break in $\mu_K(R)$ at $R$$\sim$$1\hr$; $\slos$ transitions to a shallower
  slope at this radius. The galaxy with the highest molecular gas-to-stellar mass ratio
  in the sample; together with UGC~4256 the only galaxy with higher molecular-gas mass
  than stellar mass. Like UGC~4256, this galaxy also shows a bump feature in the 
  calculated molecular gas rotation curve which can be seen in the observed rotation
  curve. \hone\ rotation curve corrected for an inclination warp \citep{martinsson2011}.
  \item[{\bf UGC  7244:}] NGC 4195. Barred galaxy, modeled assuming no bulge; inner excess
  in $\mu_K(R)$ ($R$$\leq$$8\arcsec$) interpreted as an inner disk. As with UGC~1862, we
  exclude $R$$<$$2\farcs5$ from our analysis. Stellar-kinematic measurements only reach
  $R$$\sim$$20\arcsec$.
  The receding part of the \hone\ rotation curve rises steeper than the approaching side
  \citep{martinsson2011}. This is not seen in the \halp\ rotation curve, and results in
  some offsets between the \hone\ and \halp\ rotation curves (Section~\ref{sec:RC}).
  In optical images \citepalias{bershady2010a}, this galaxy looks rather peculiar, with an
  offset bar and bent spiral arms. 
  \item[{\bf UGC  7917:}] NGC 4662.  High-quality stellar kinematics. Gas-poor in the
  center, resulting in no measured \halp\ and \hone\ kinematics in that region.
  Excluded kinematic data at $R<1\hr$ (shaded region in the Atlas) to remove
  bar-associated regions. Type-II break in $\mu_K(R)$ at $R$$\sim$$25\arcsec$.
  Fifth highest bulge-to-disk ratio ($B/D=0.14$).
  \item[{\bf UGC  8196:}]  NGC 4977. Early-type spiral (SAb). Third highest bulge-to-disk
  ratio in the sample ($B/D$$=$0.24).
  Gas-poor \citep[][\citetalias{martinsson2012a}]{martinsson2011}. \hone\ rotation curve
  corrected for an inclination warp \citep{martinsson2011}. Low-surface-brightness,
  extended disk with strong spiral structure not probed by our kinematics. Rather complex
  $\mu_K(R)$, transitions to a shallower slope at $R$$\sim$$10\arcsec$, which may be an
  extent of the bulge not accounted for in our bulge-disk decomposition. Has a
  non-physical measurement of the maximality and is generally excluded from the results
  in this paper.
  \item[{\bf UGC  9177:}]  Well-defined spiral structure affecting $\mu_K(R)$.
  High inclination ($\itf$$=$$40\arcdeg$) and high-quality rotation curves. 
  Measurements of $\slos$ limited to $R$$<$$25\arcsec$.
  \item[{\bf UGC  9837:}] Stellar data has a limited radial extent. Weak
   morphological bar does not affect the kinematics. Shallower slope in
   $\mu_K(R)$ beyond $\sim$$1\hr$. Regular \hone\ kinematics.
  \item[{\bf UGC  9965:}] IC 1132.  Very nearly face-on ($\itf$$=$$12\arcdeg$) with
  strong spiral structure visible in $\mu_K(R)$.  Bulgeless galaxy; as with
  UGC~1862, we exclude data within $R$$=$$2\farcs5$ in the kinematic analysis.
  \hone\ rotation curve corrected for an inclination warp \citep{martinsson2011}.
  \item[{\bf UGC 11318:}] NGC 6691.  Barred galaxy. Lowest inclination in the sample
  ($\itf$$=$$5\fdg8$), yielding a very low-amplitude projected rotation curve. Type-II
  break in $\mu_K(R)$ at $\sim$$1\hr$.
  \item[{\bf UGC 12391:}] NGC 7495. Type-II break in $\mu_K(R)$ at $R$$\sim$$25\arcsec$. 
  The mass surface density profile of the atomic gas is estimated using the 
  results in \cite{martinsson2011}.
\end{list}
%
}




\newpage
\onecolumn

\subsection{Atlas}
\label{sec:Atlas}

 \begin{figure}[!b]
 \centering
 \includegraphics[width=1.00\textwidth]{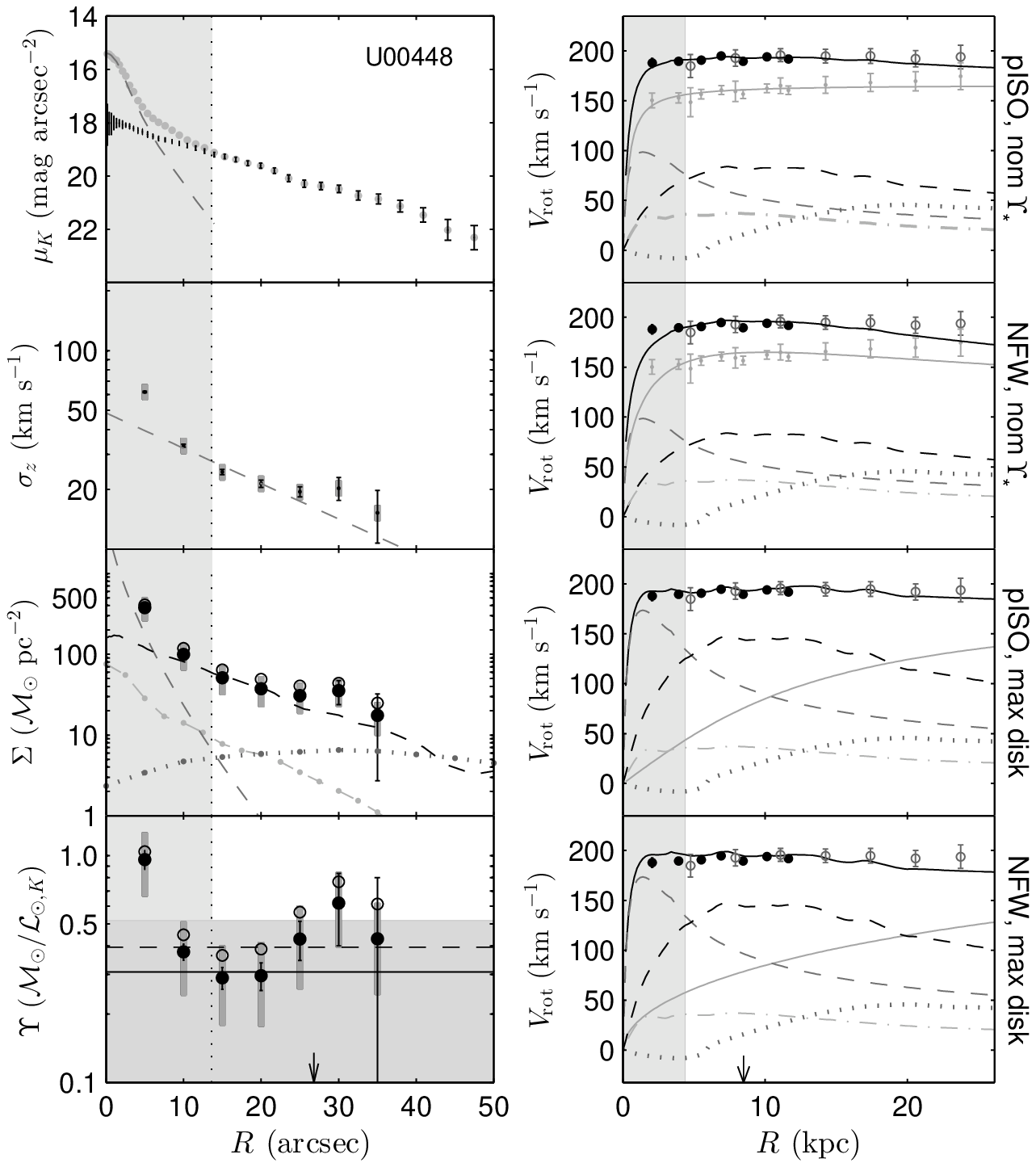}
 \end{figure}
 
 \begin{figure}
 \centering
 \includegraphics[width=1.00\textwidth]{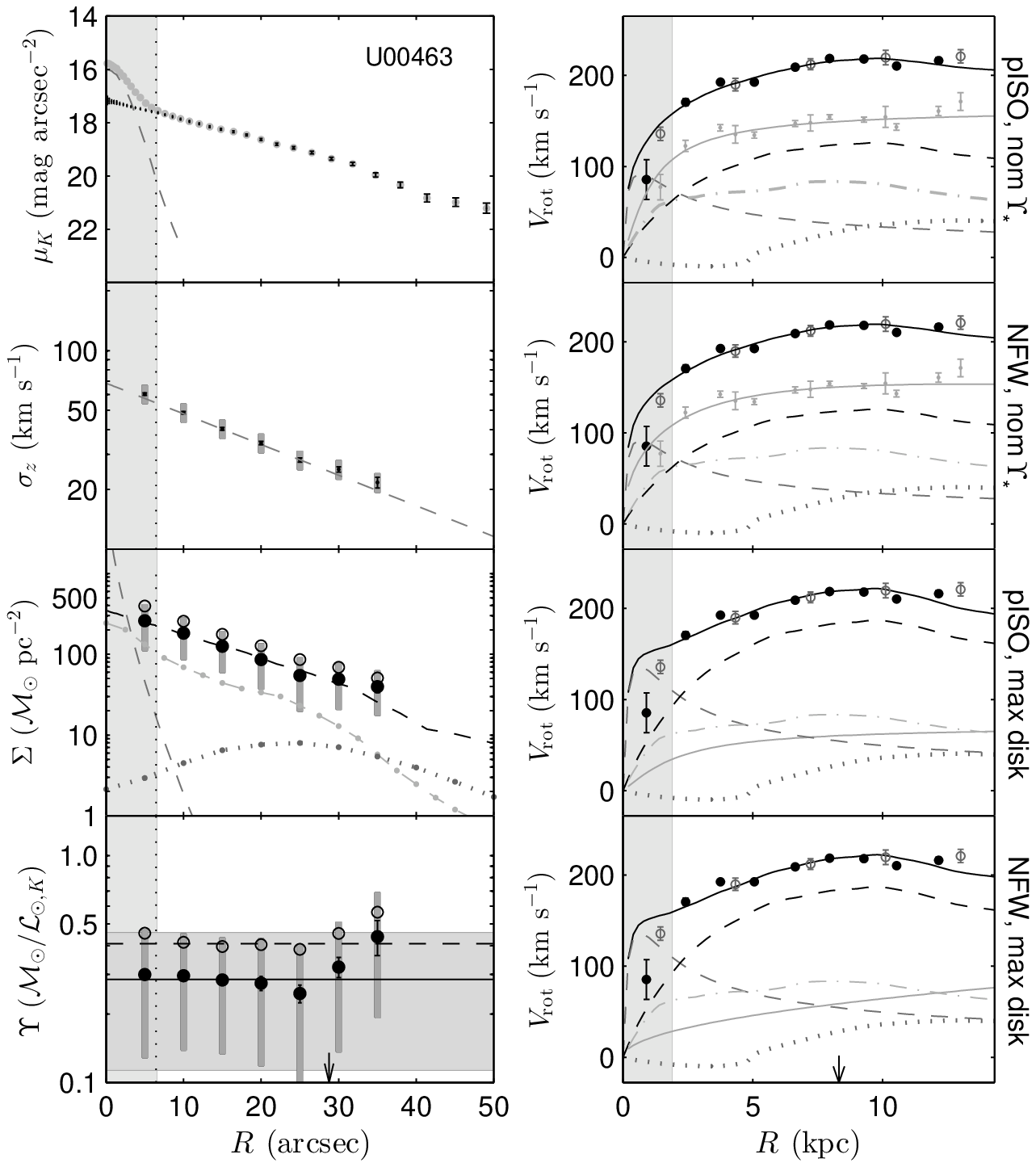}
 \end{figure}

 \begin{figure}
 \centering
 \includegraphics[width=1.00\textwidth]{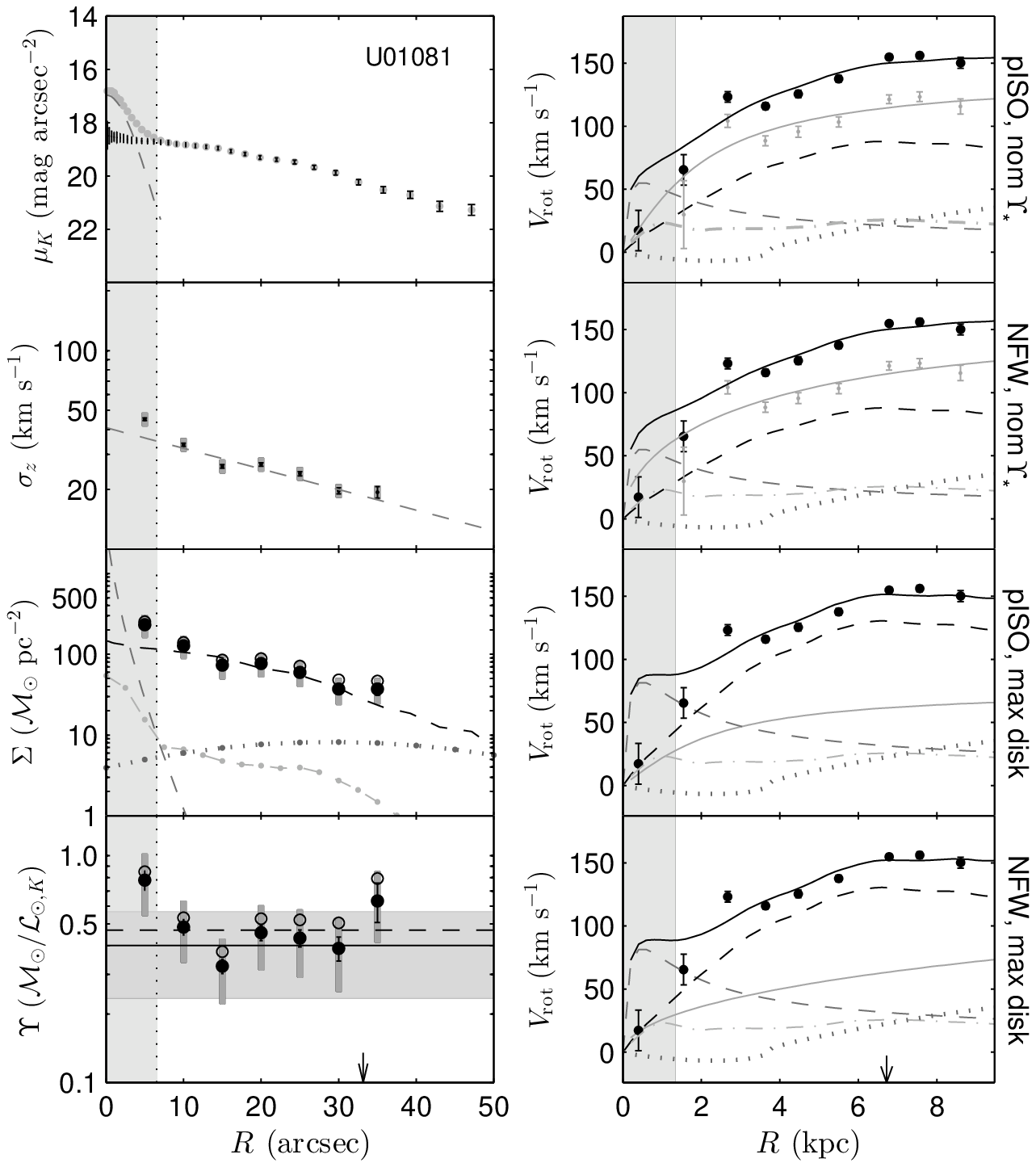}
 \end{figure}

 \begin{figure}
 \centering
 \includegraphics[width=1.00\textwidth]{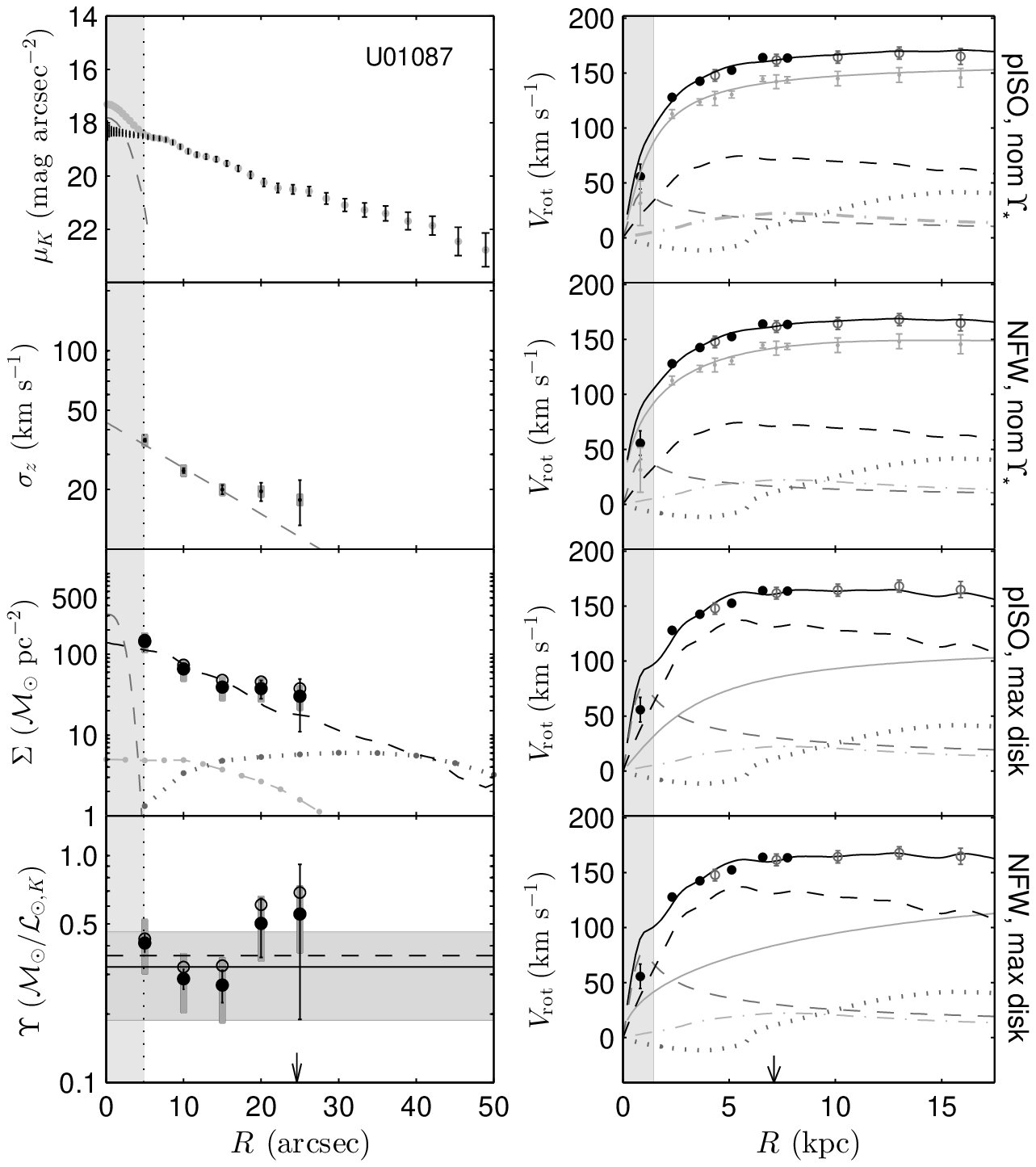}
 \end{figure}

 \begin{figure}
 \centering
 \includegraphics[width=1.00\textwidth]{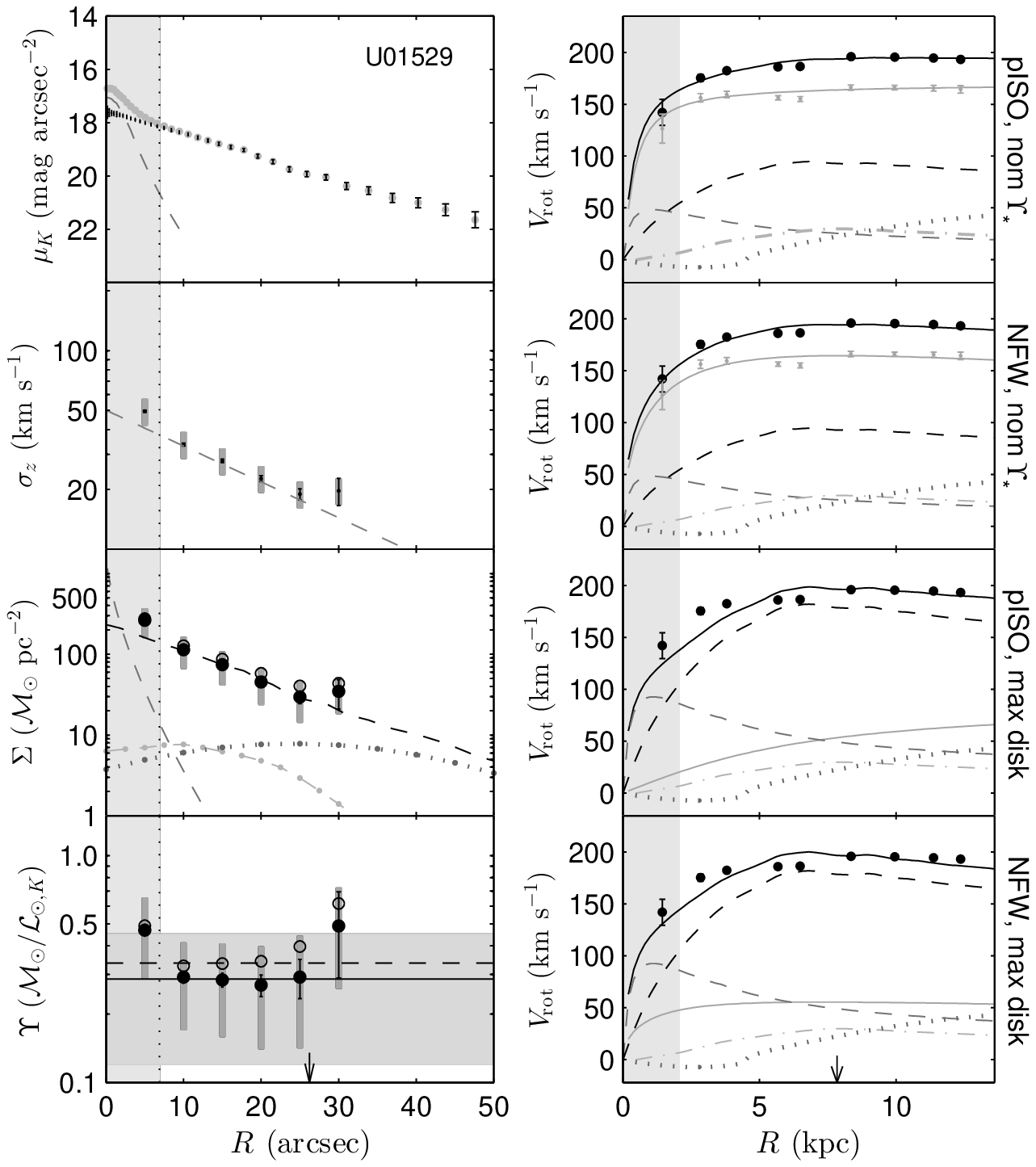}
 \end{figure}

 \begin{figure}
 \centering
 \includegraphics[width=1.00\textwidth]{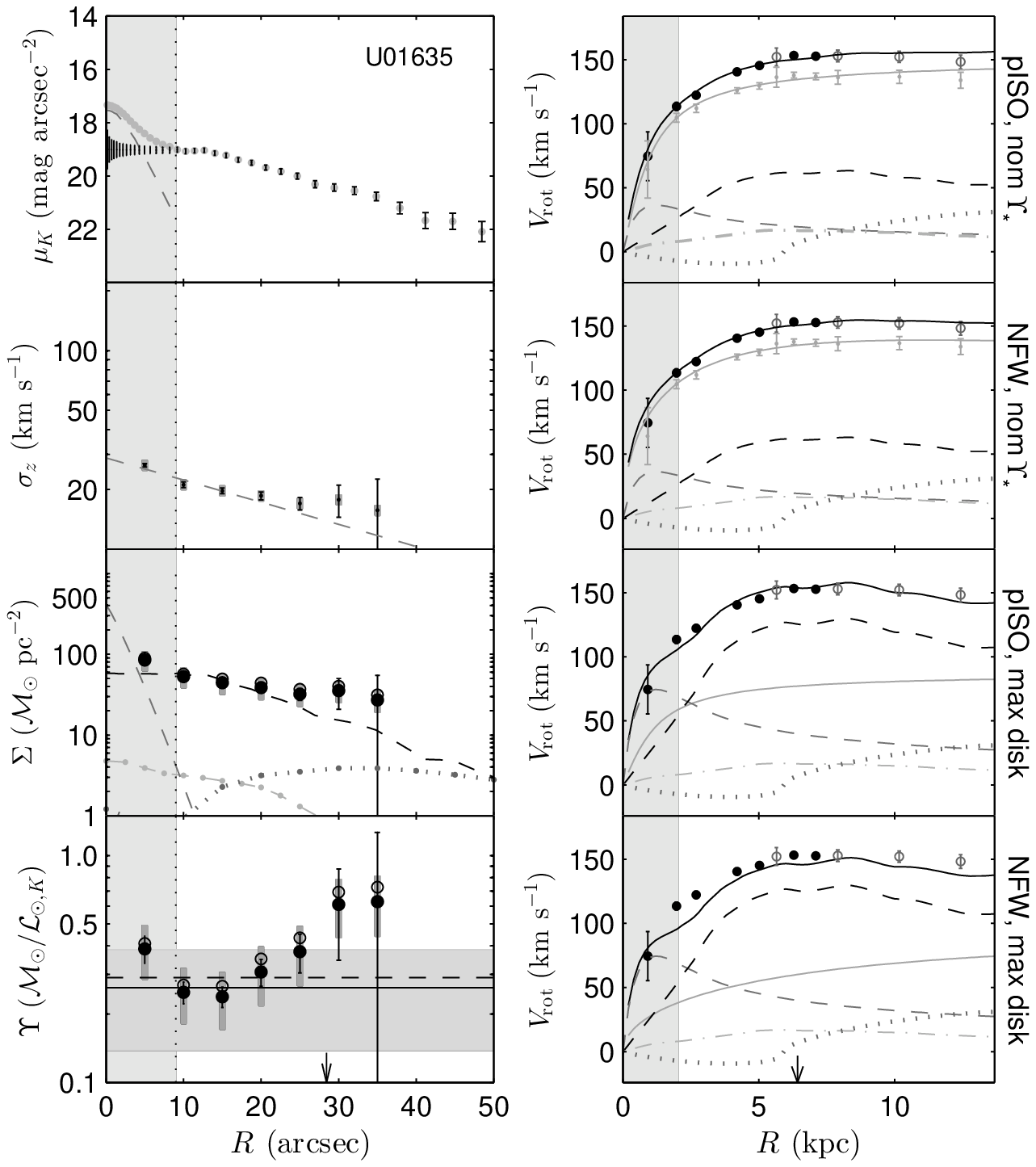}
 \end{figure}

 \begin{figure}
 \centering
 \includegraphics[width=1.00\textwidth]{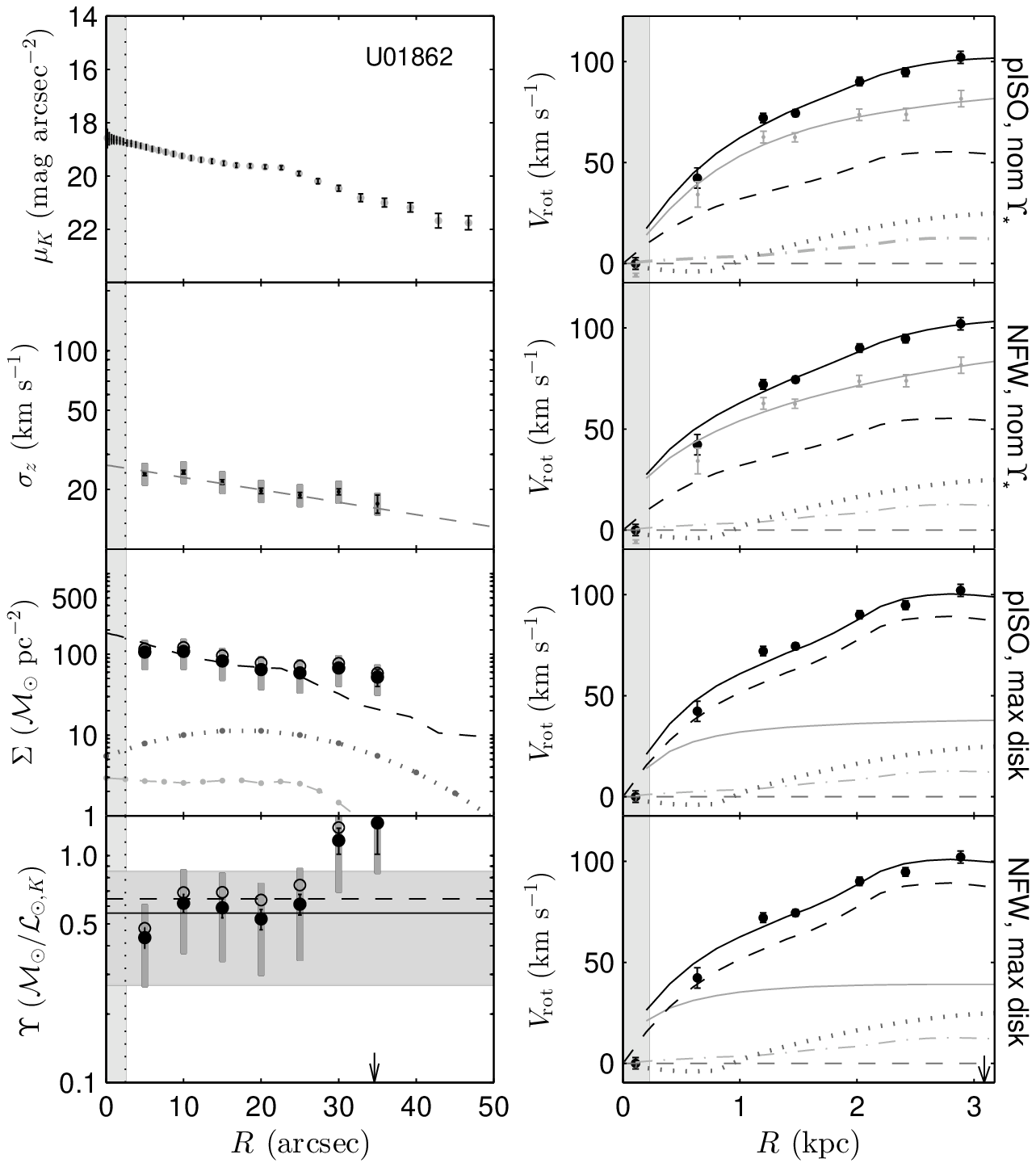}
 \end{figure}

 \begin{figure}
 \centering
 \includegraphics[width=1.00\textwidth]{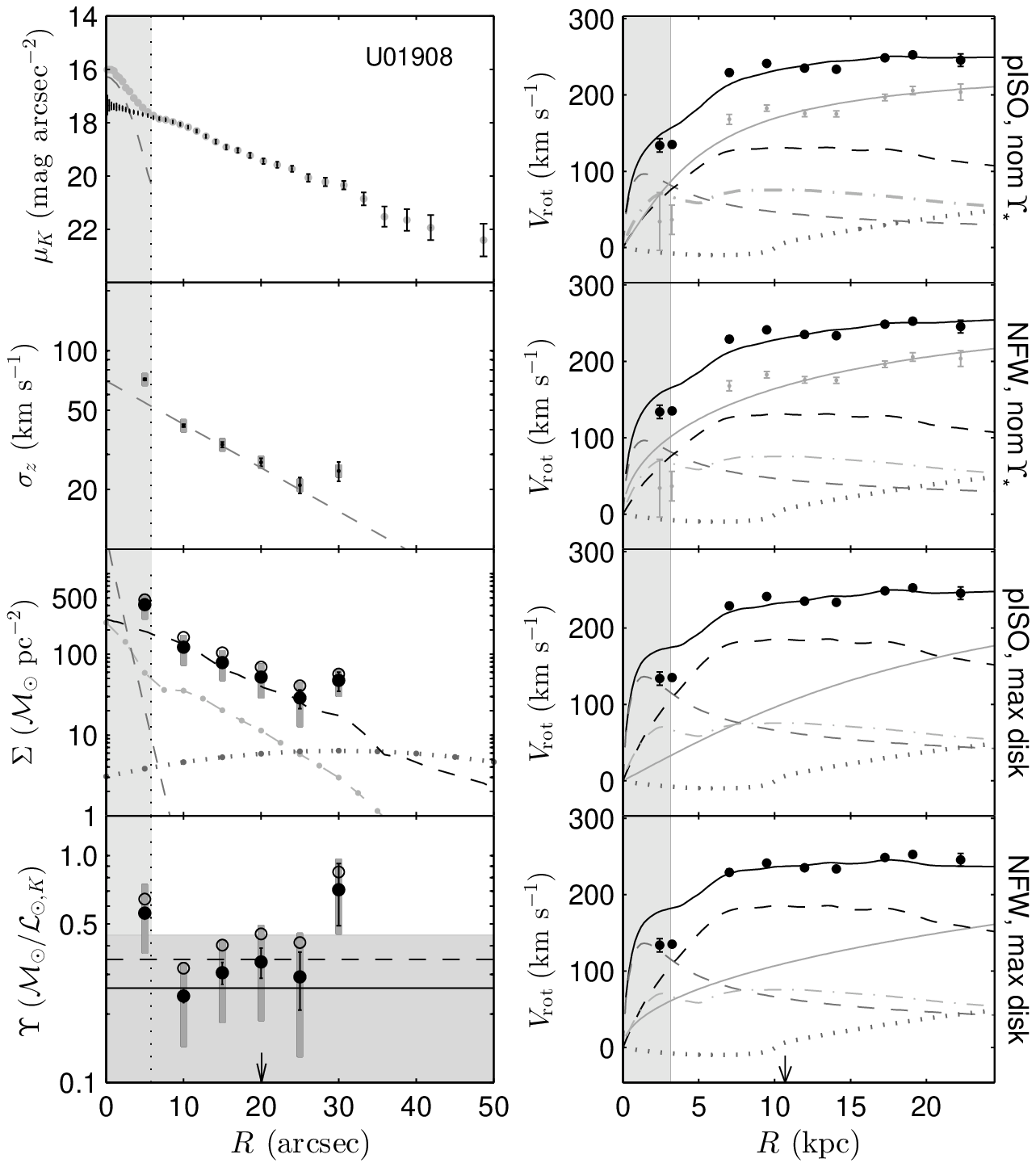}
 \end{figure}

 \begin{figure}
 \centering
 \includegraphics[width=1.00\textwidth]{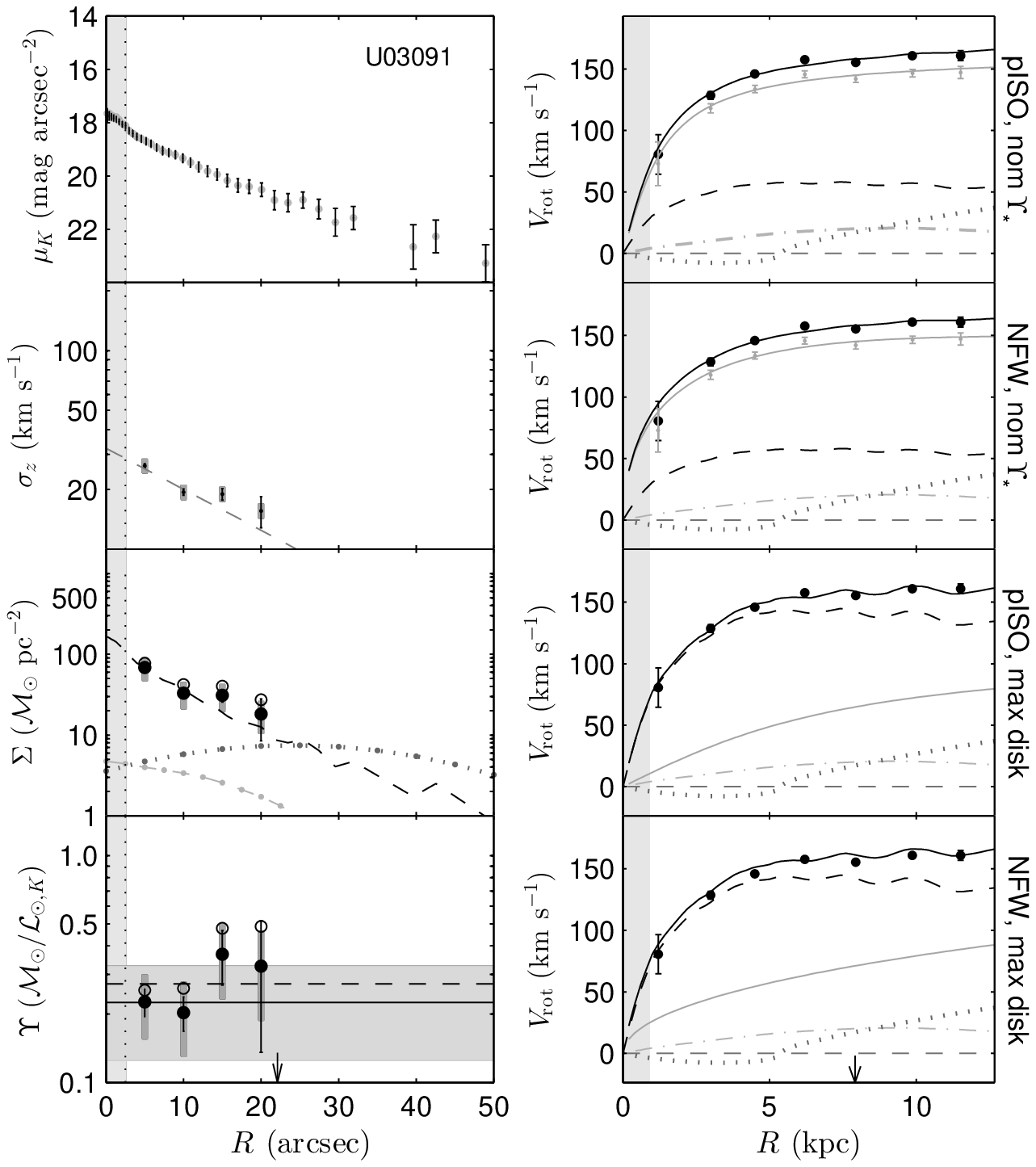}
 \end{figure}

 \begin{figure}
 \centering
 \includegraphics[width=1.00\textwidth]{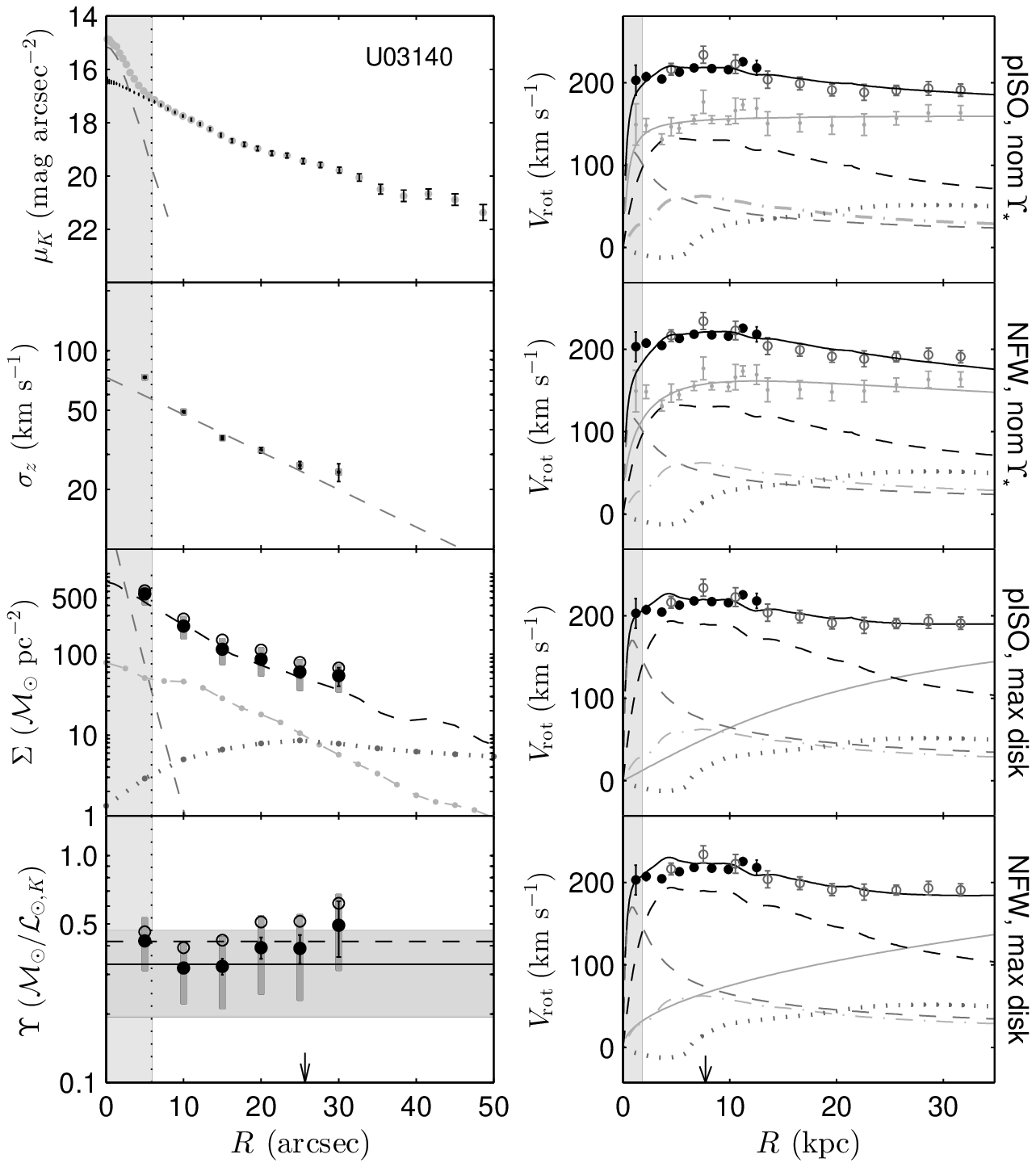}
 \end{figure}

 \begin{figure}
 \centering
 \includegraphics[width=1.00\textwidth]{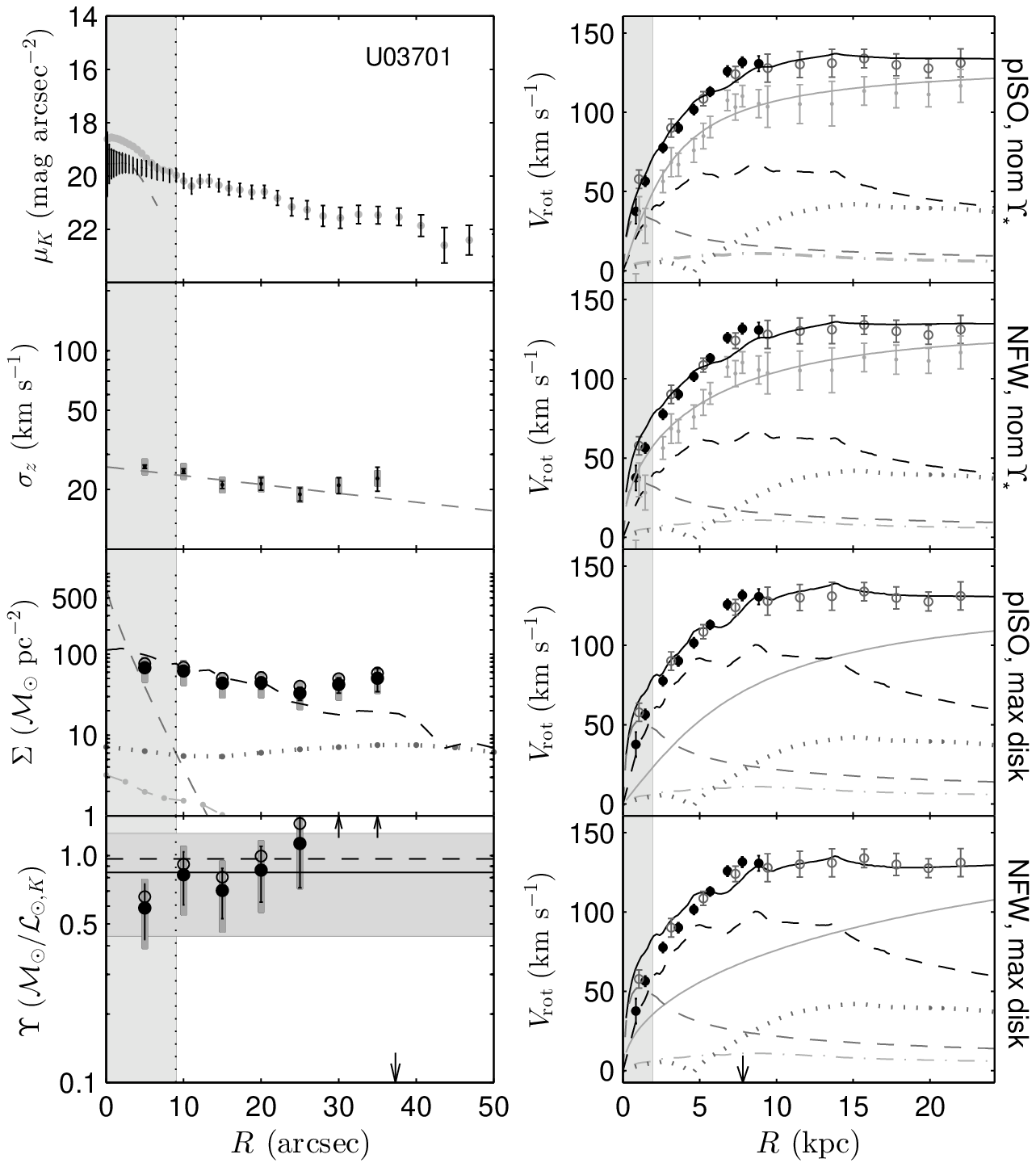}
 \end{figure}

 \begin{figure}
 \centering
 \includegraphics[width=1.00\textwidth]{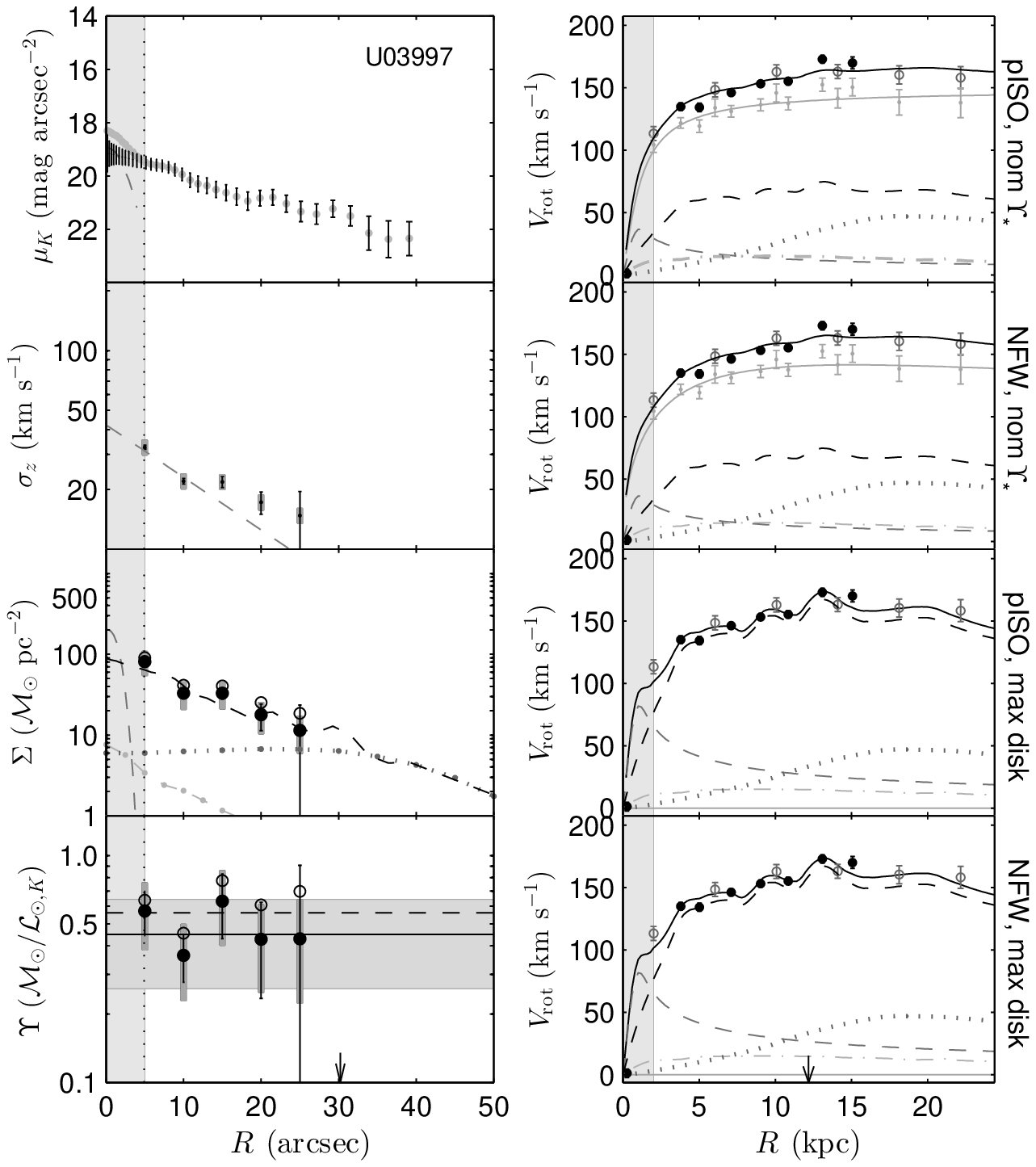}
 \end{figure}

\clearpage

 \begin{figure}
 \centering
 \includegraphics[width=1.00\textwidth]{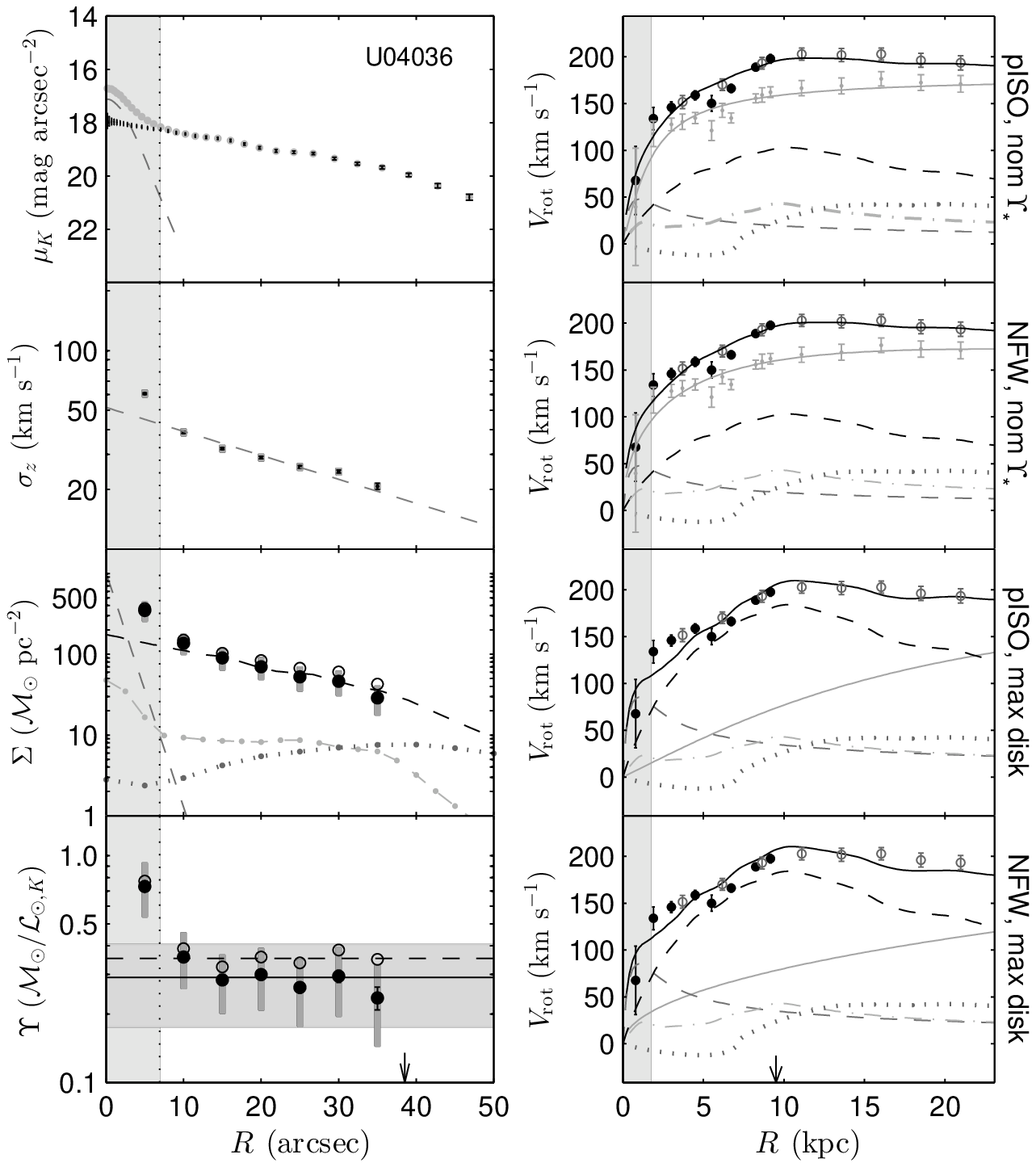}
 \end{figure}

 \begin{figure}
 \centering
 \includegraphics[width=1.00\textwidth]{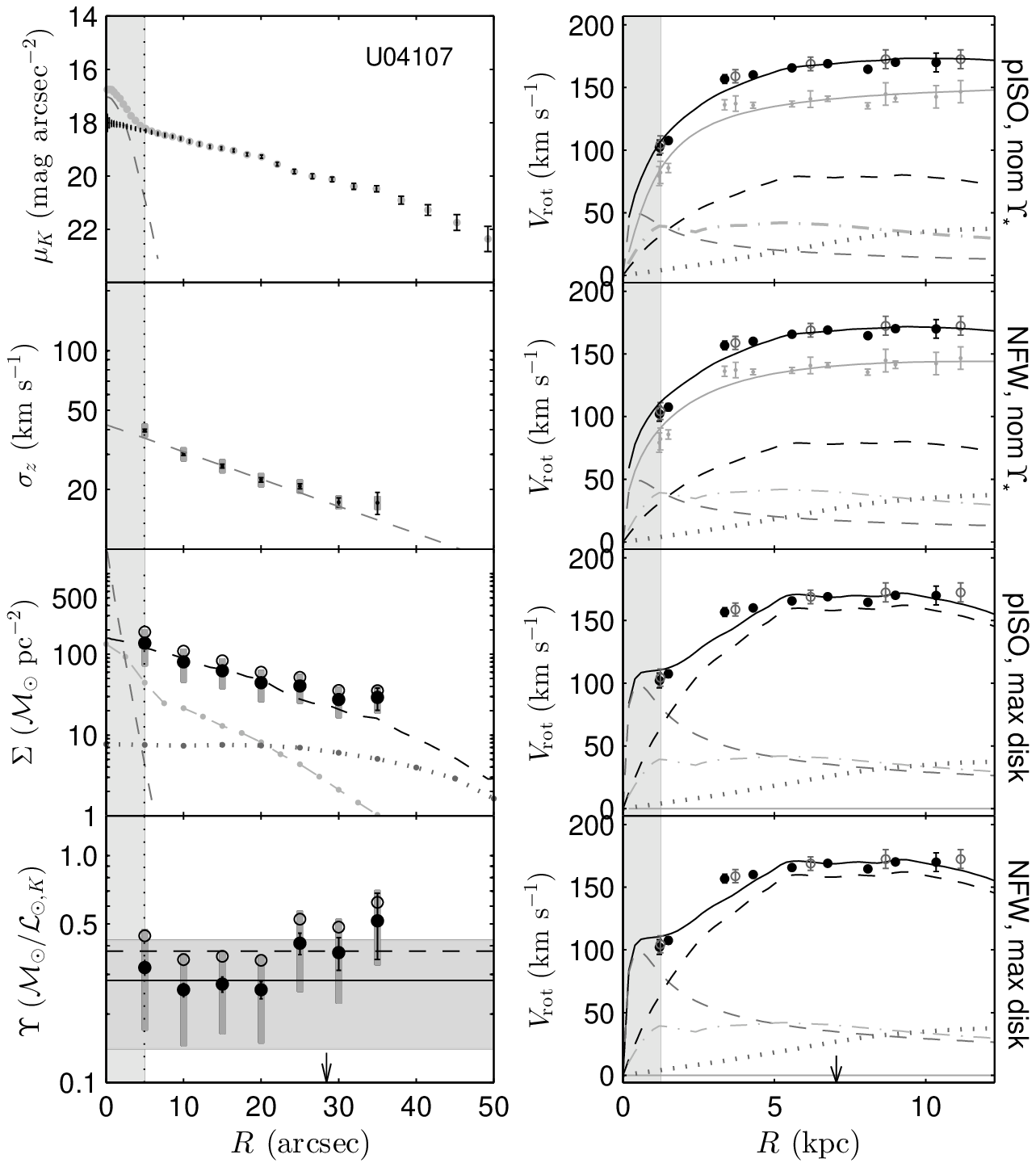}
 \end{figure}

 \begin{figure}
 \centering
 \includegraphics[width=1.00\textwidth]{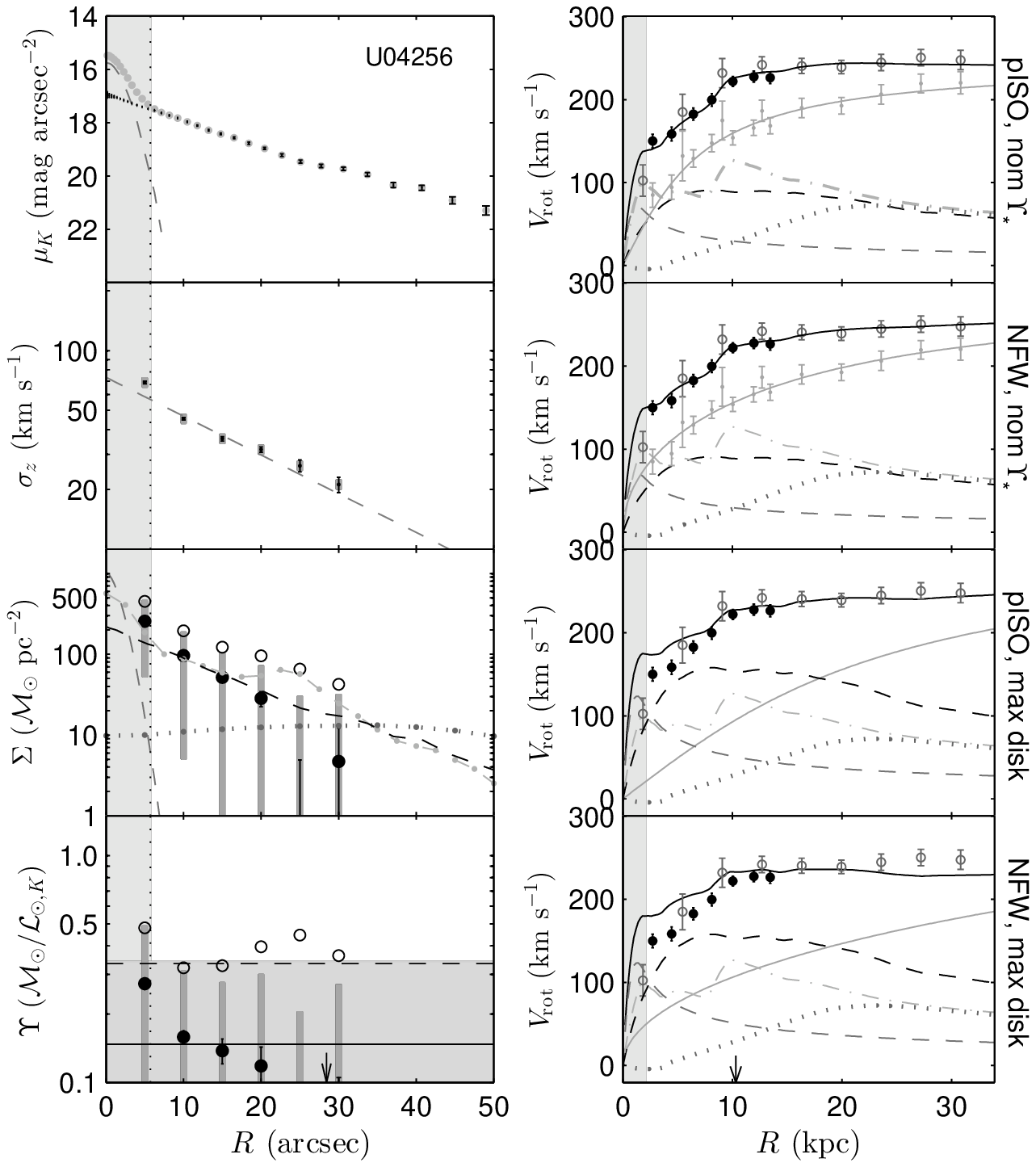}
 \end{figure}

 \begin{figure}
 \centering
 \includegraphics[width=1.00\textwidth]{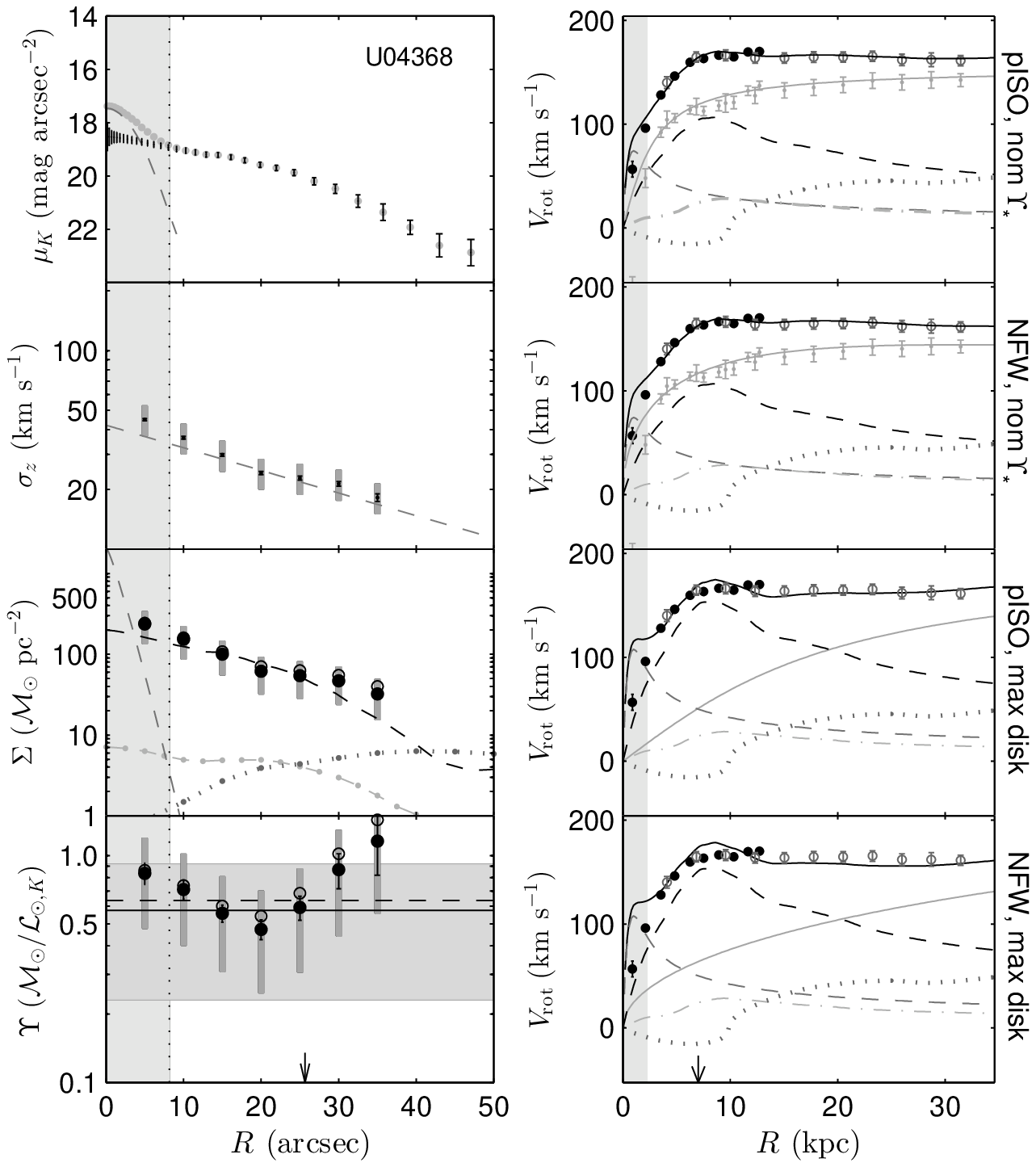}
 \end{figure}

 \begin{figure}
 \centering
 \includegraphics[width=1.00\textwidth]{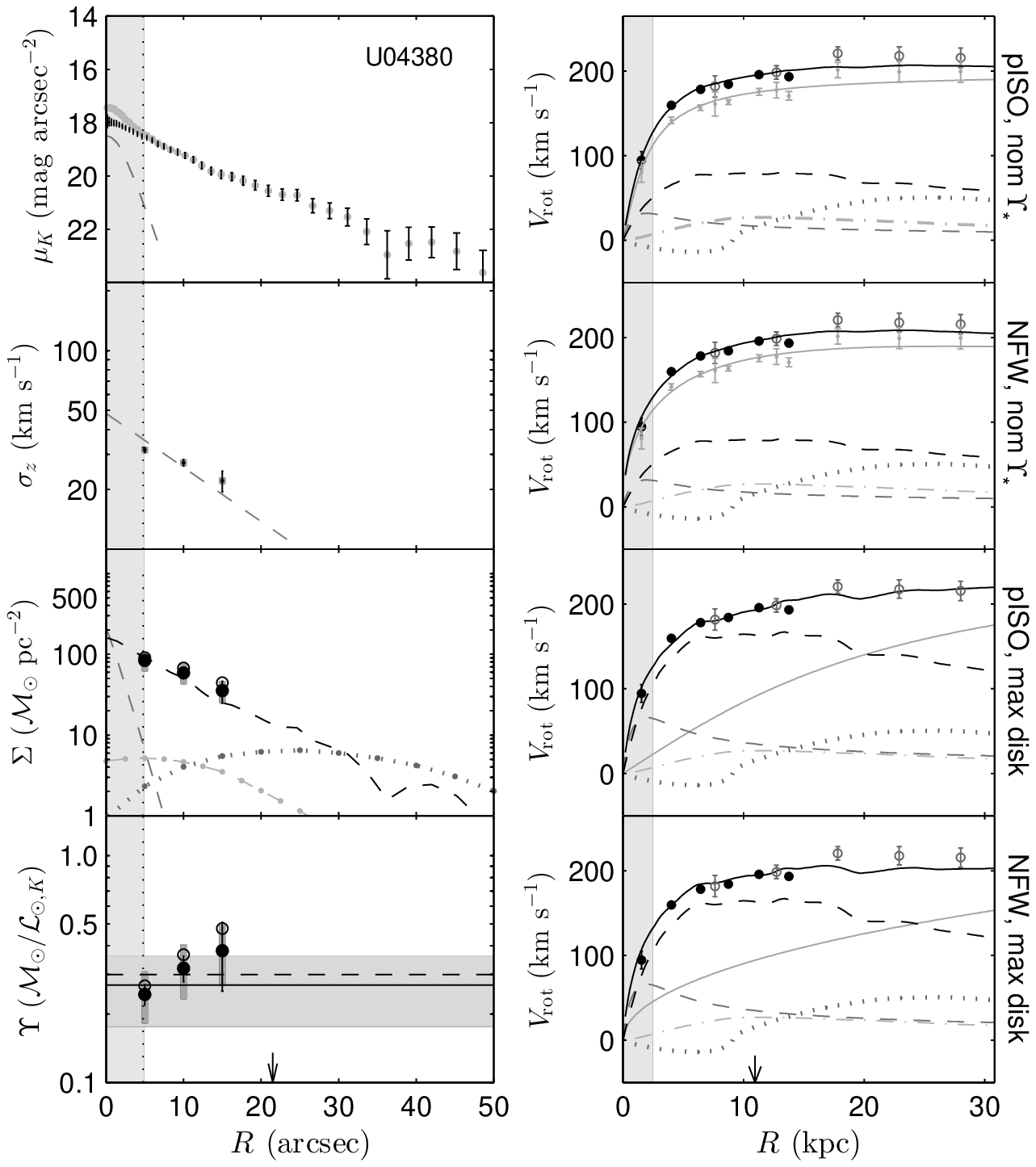}
 \end{figure}

 \begin{figure}
 \centering
 \includegraphics[width=1.00\textwidth]{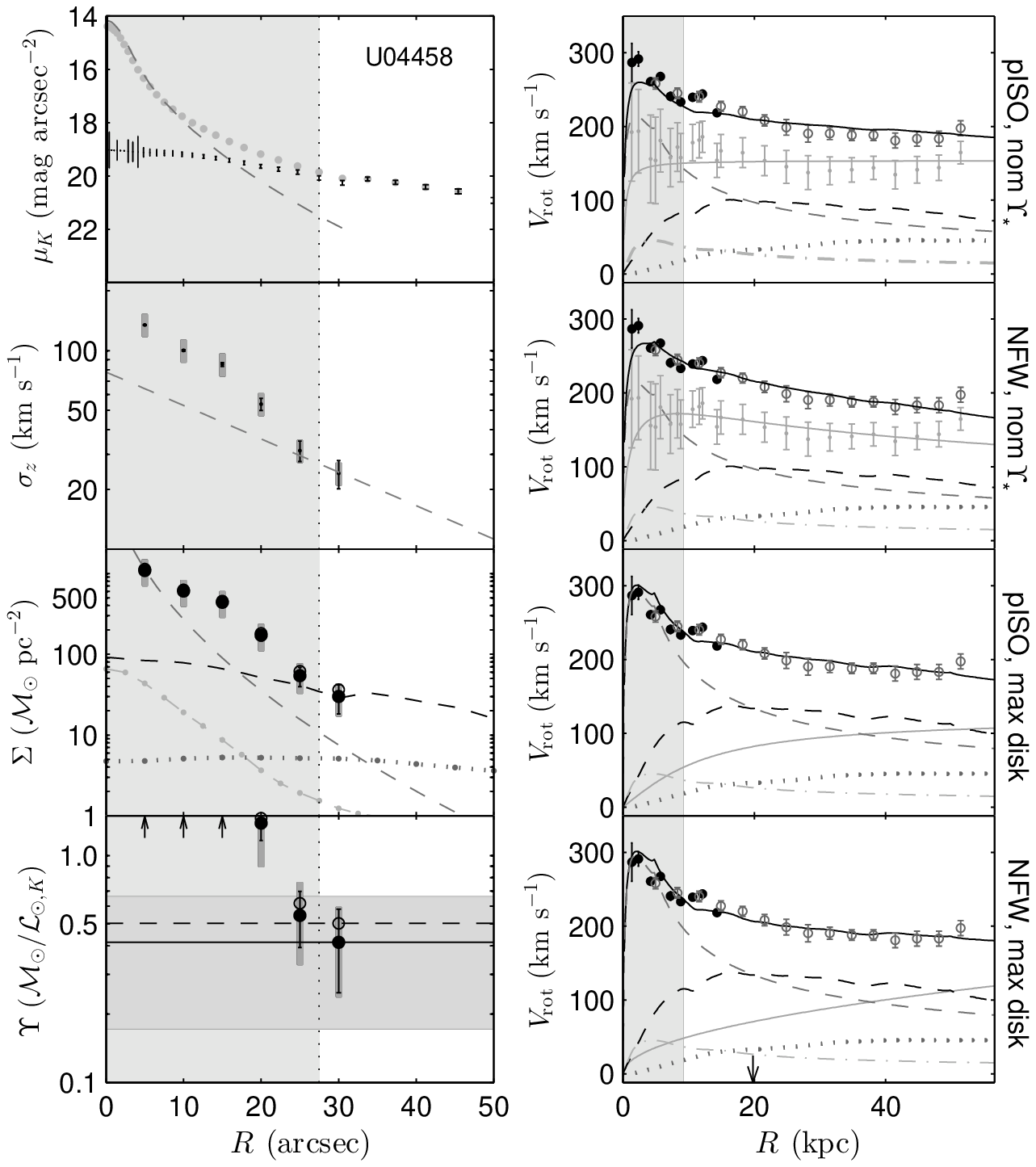}
 \end{figure}

\clearpage

 \begin{figure}
 \centering
 \includegraphics[width=1.00\textwidth]{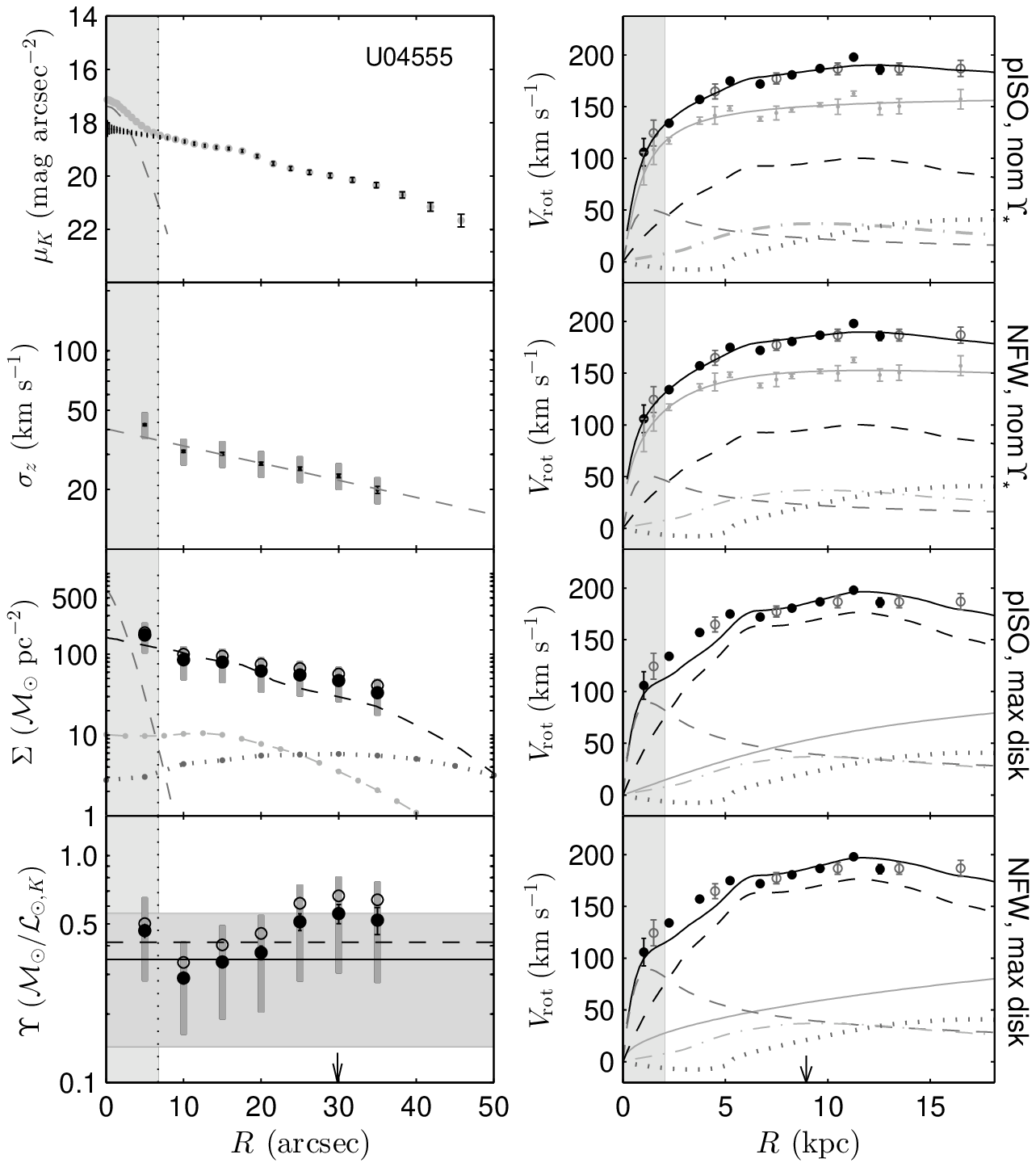}
 \end{figure}

 \begin{figure}
 \centering
 \includegraphics[width=1.00\textwidth]{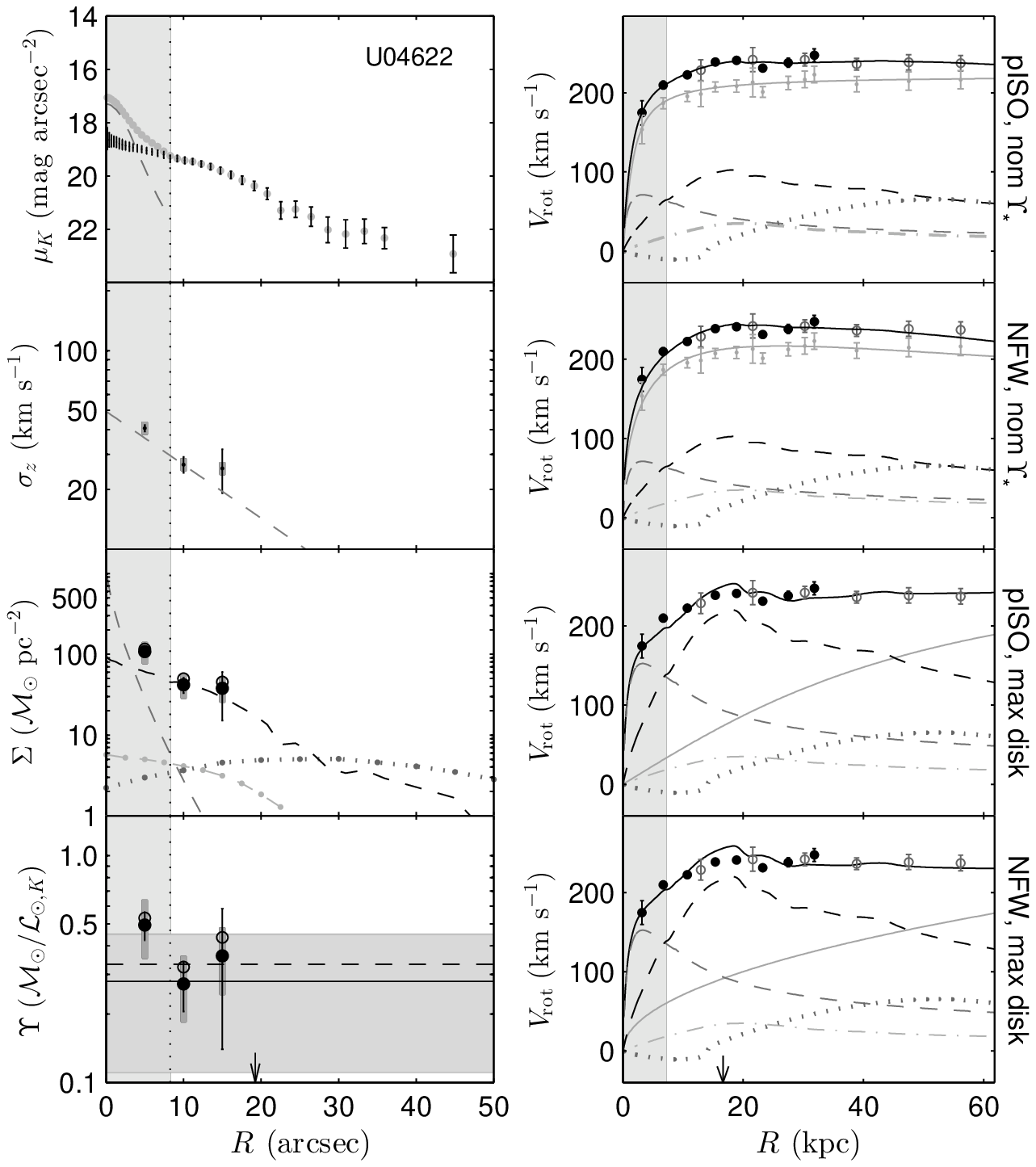}
 \end{figure}



 \begin{figure}
 \centering
 \includegraphics[width=1.00\textwidth]{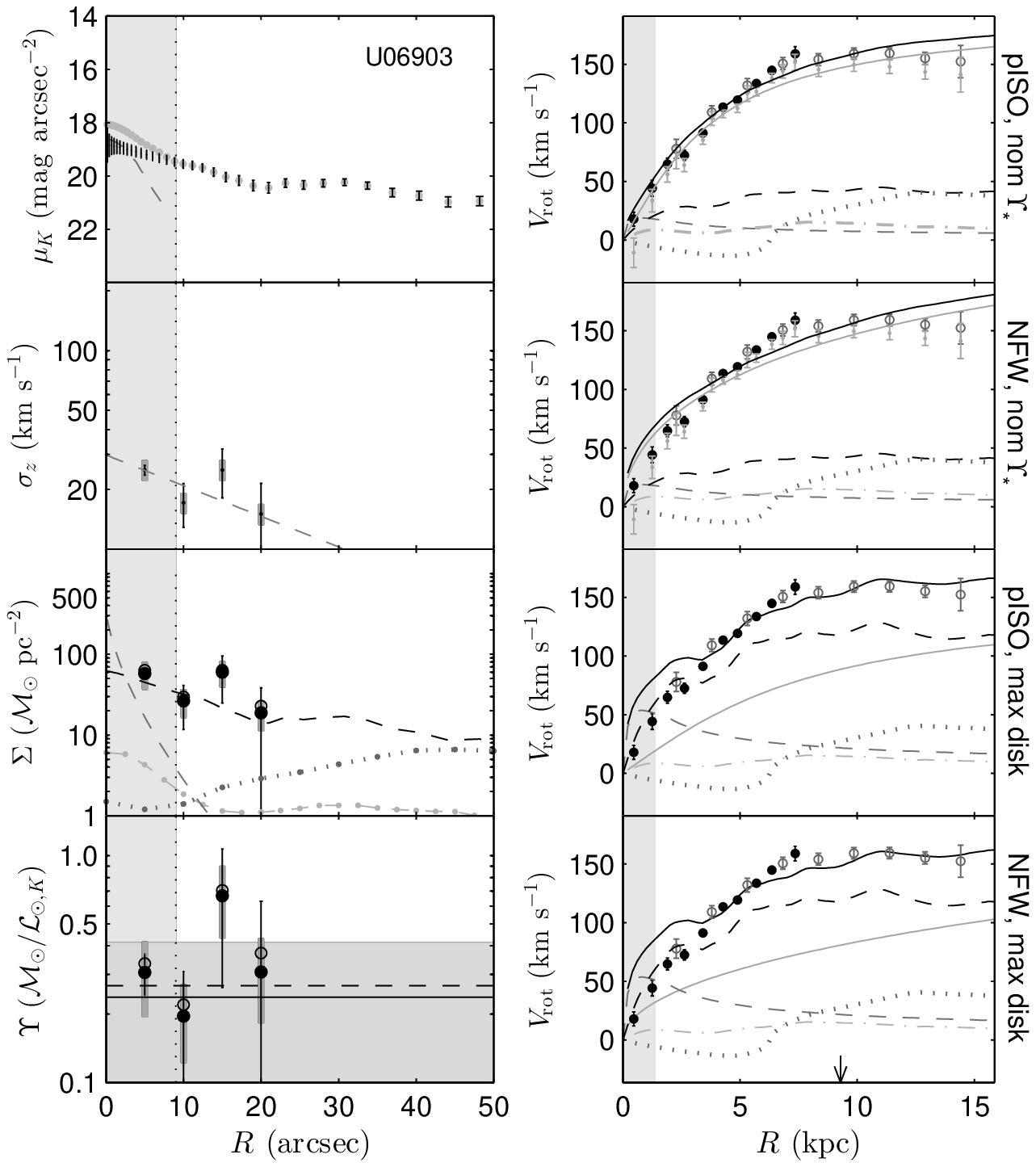}
 \end{figure}

 \begin{figure}
 \centering
 \includegraphics[width=1.00\textwidth]{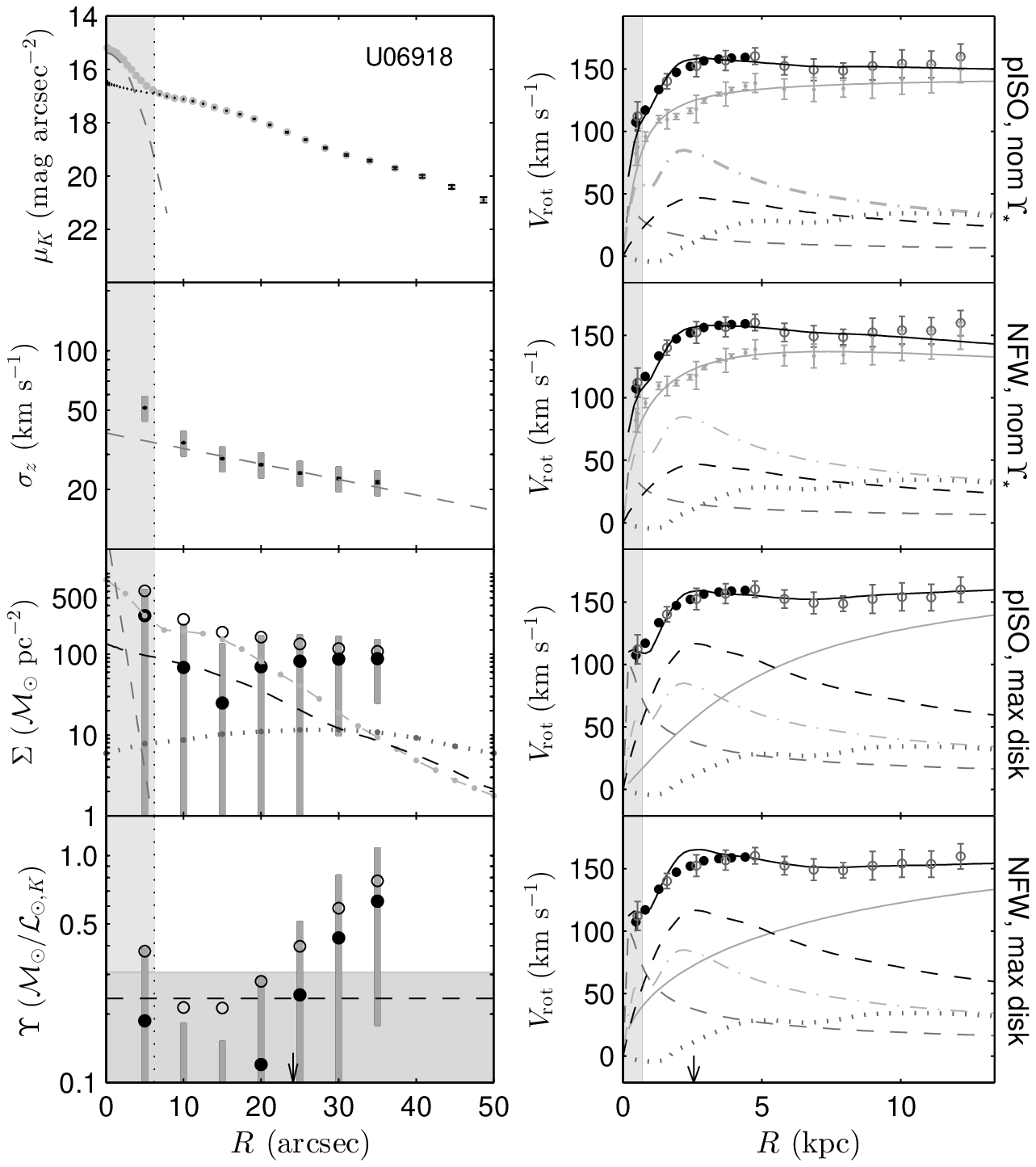}
 \end{figure}

\clearpage

 \begin{figure}
 \centering
 \includegraphics[width=1.00\textwidth]{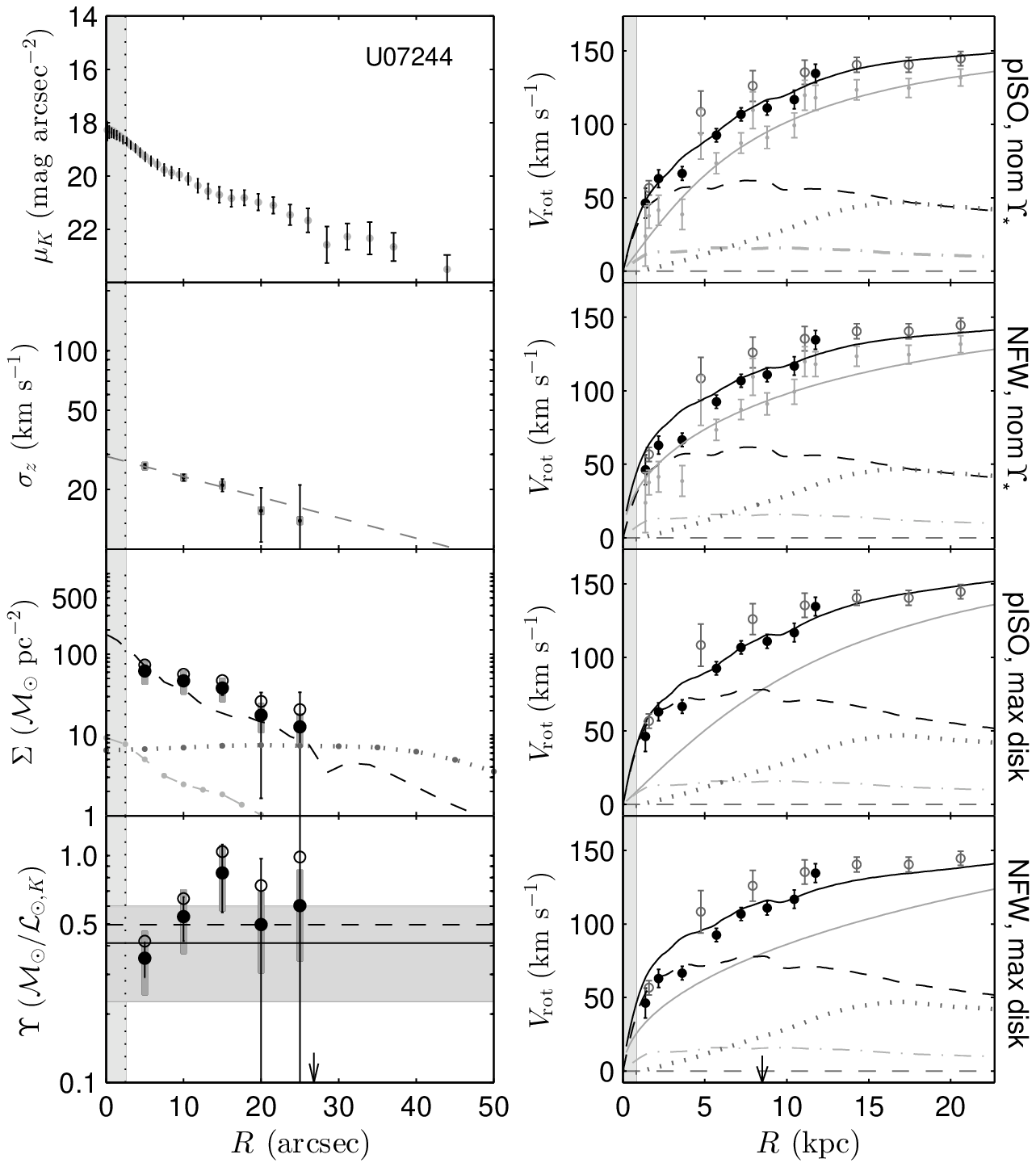}
 \end{figure}


 \begin{figure}
 \centering
 \includegraphics[width=1.00\textwidth]{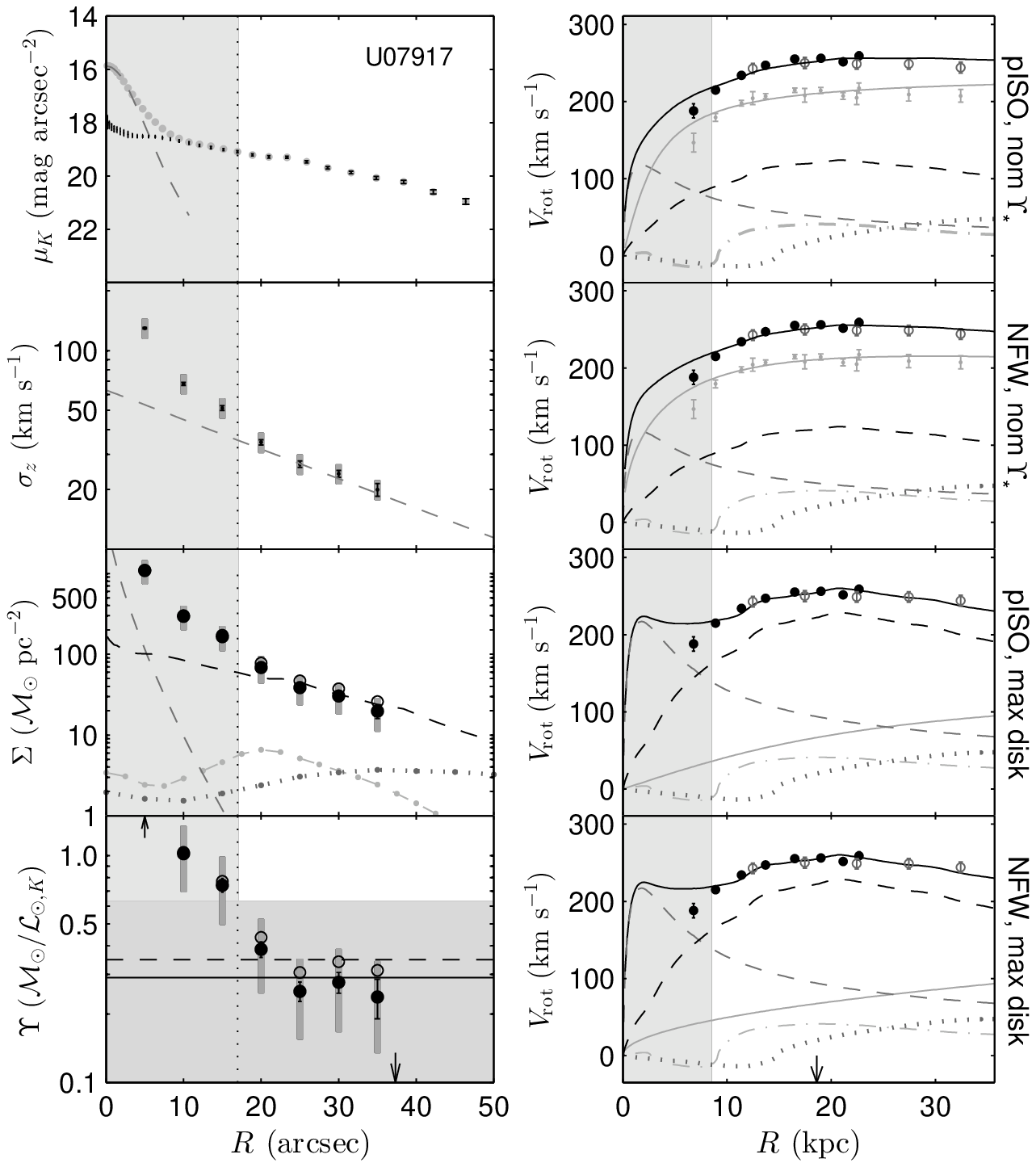}
 \end{figure}

 \begin{figure}
 \centering
 \includegraphics[width=1.00\textwidth]{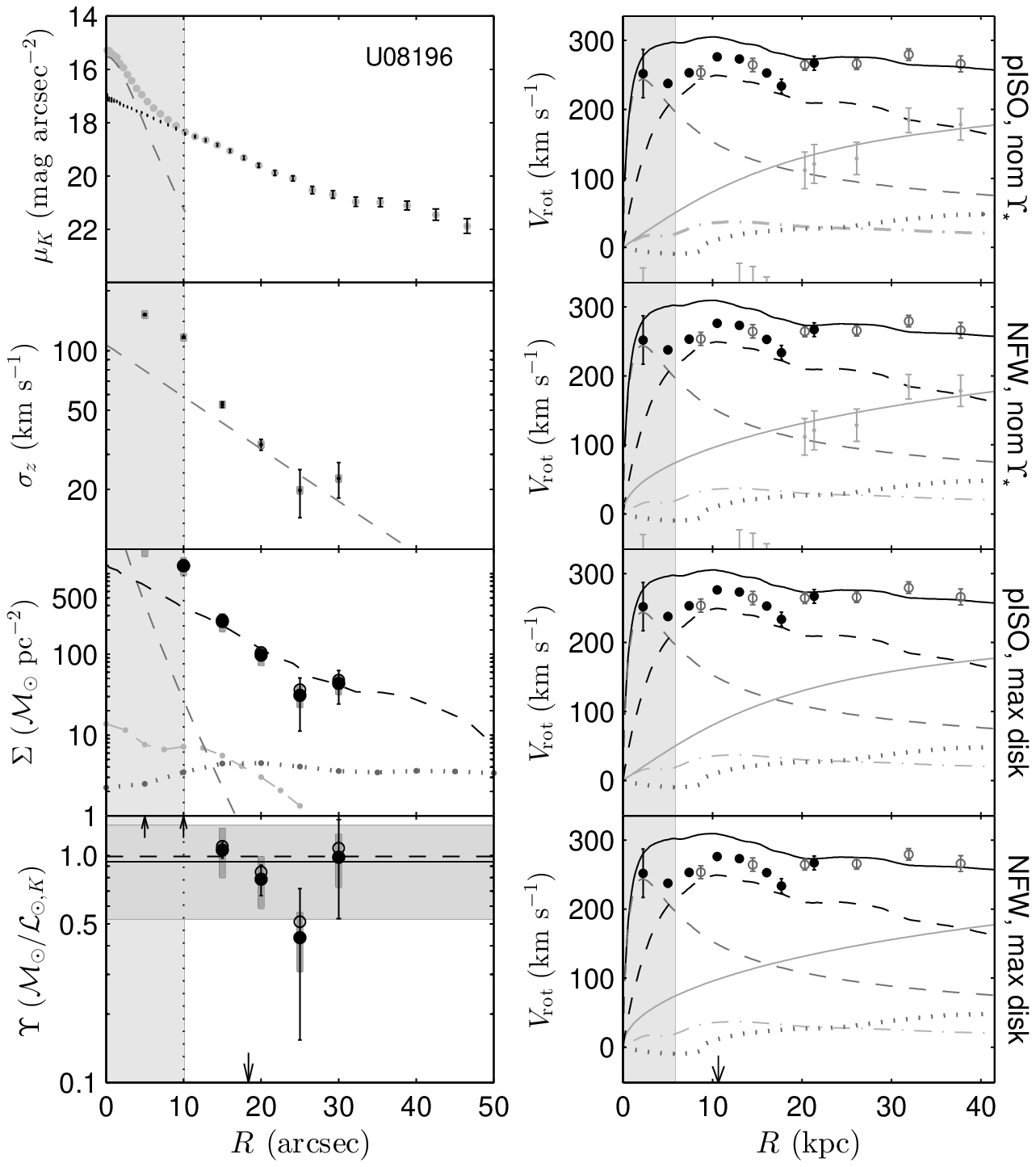}
 \end{figure}


 \begin{figure}
 \centering
 \includegraphics[width=1.00\textwidth]{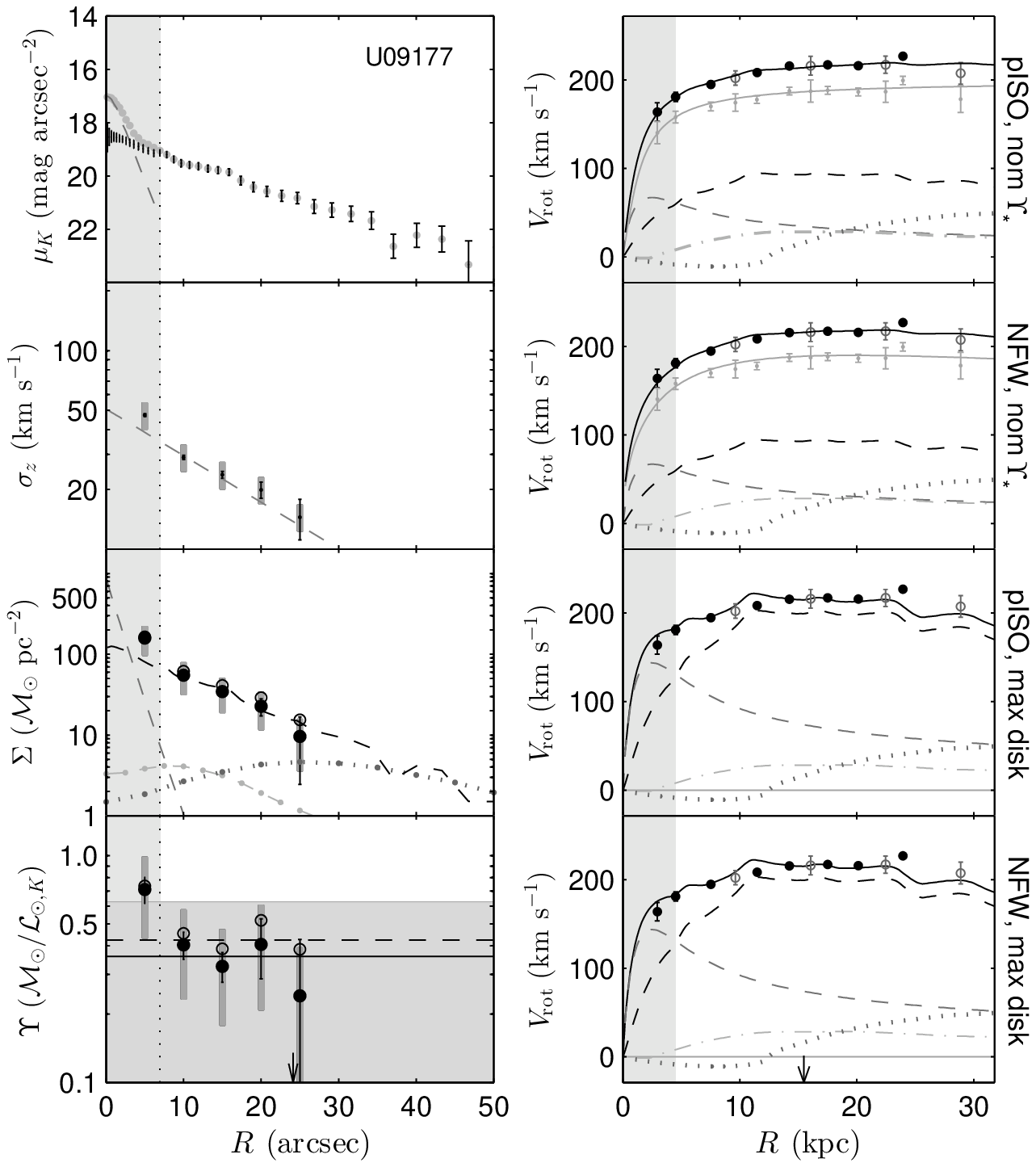}
 \end{figure}

 \begin{figure}
 \centering
 \includegraphics[width=1.00\textwidth]{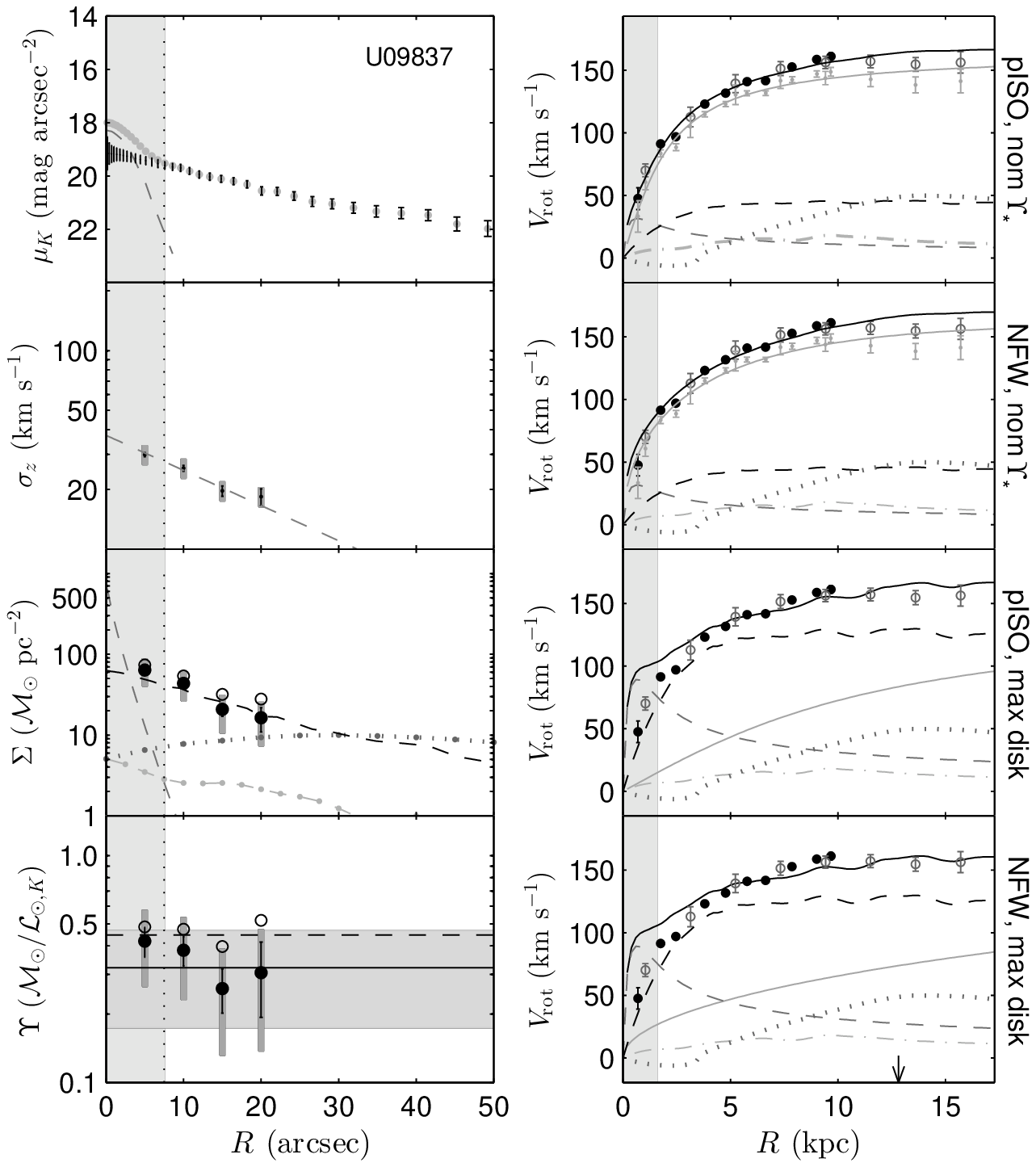}
 \end{figure}

 \begin{figure}
 \centering
 \includegraphics[width=1.00\textwidth]{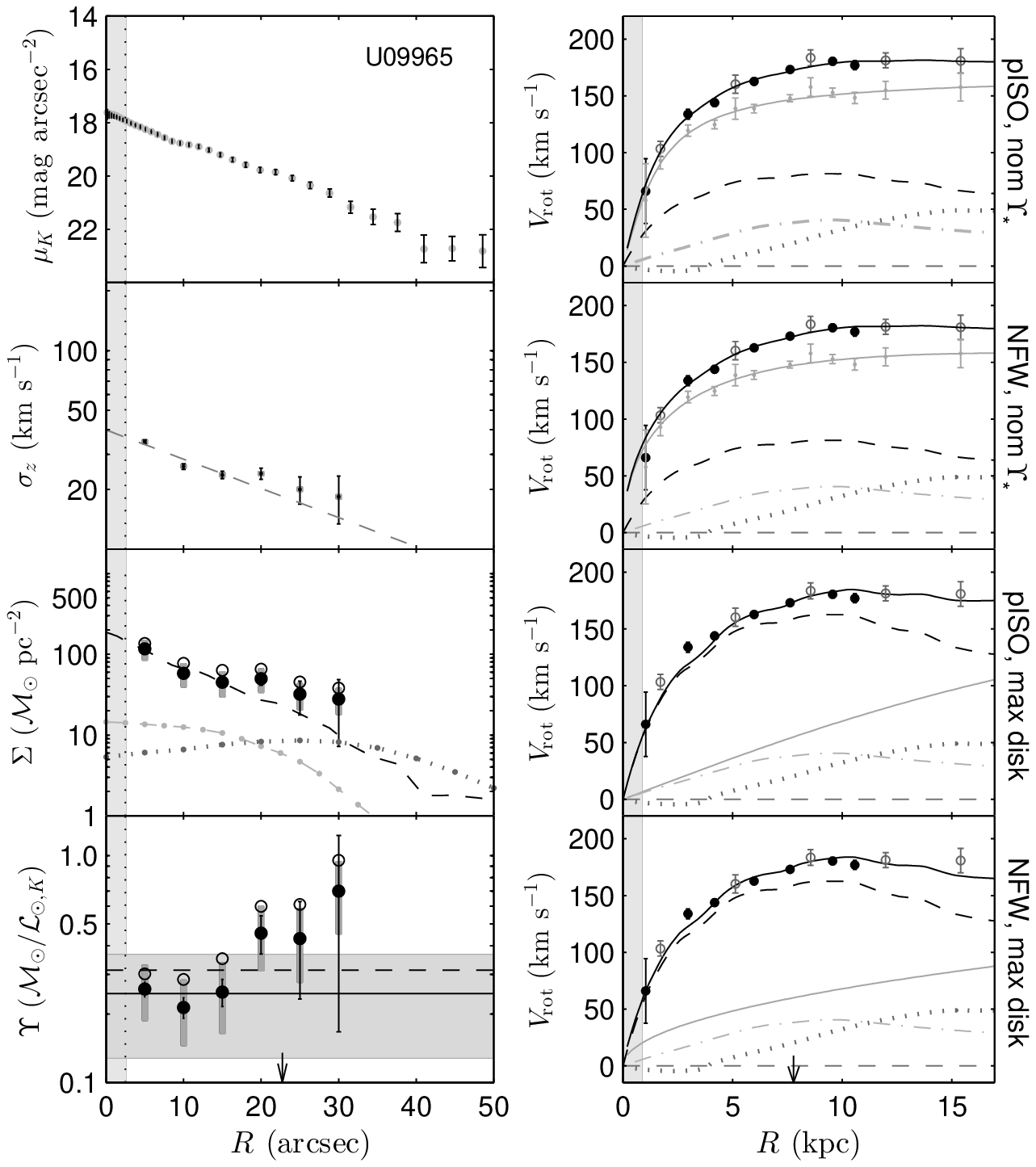}
 \end{figure}

 \begin{figure}
 \centering
 \includegraphics[width=1.00\textwidth]{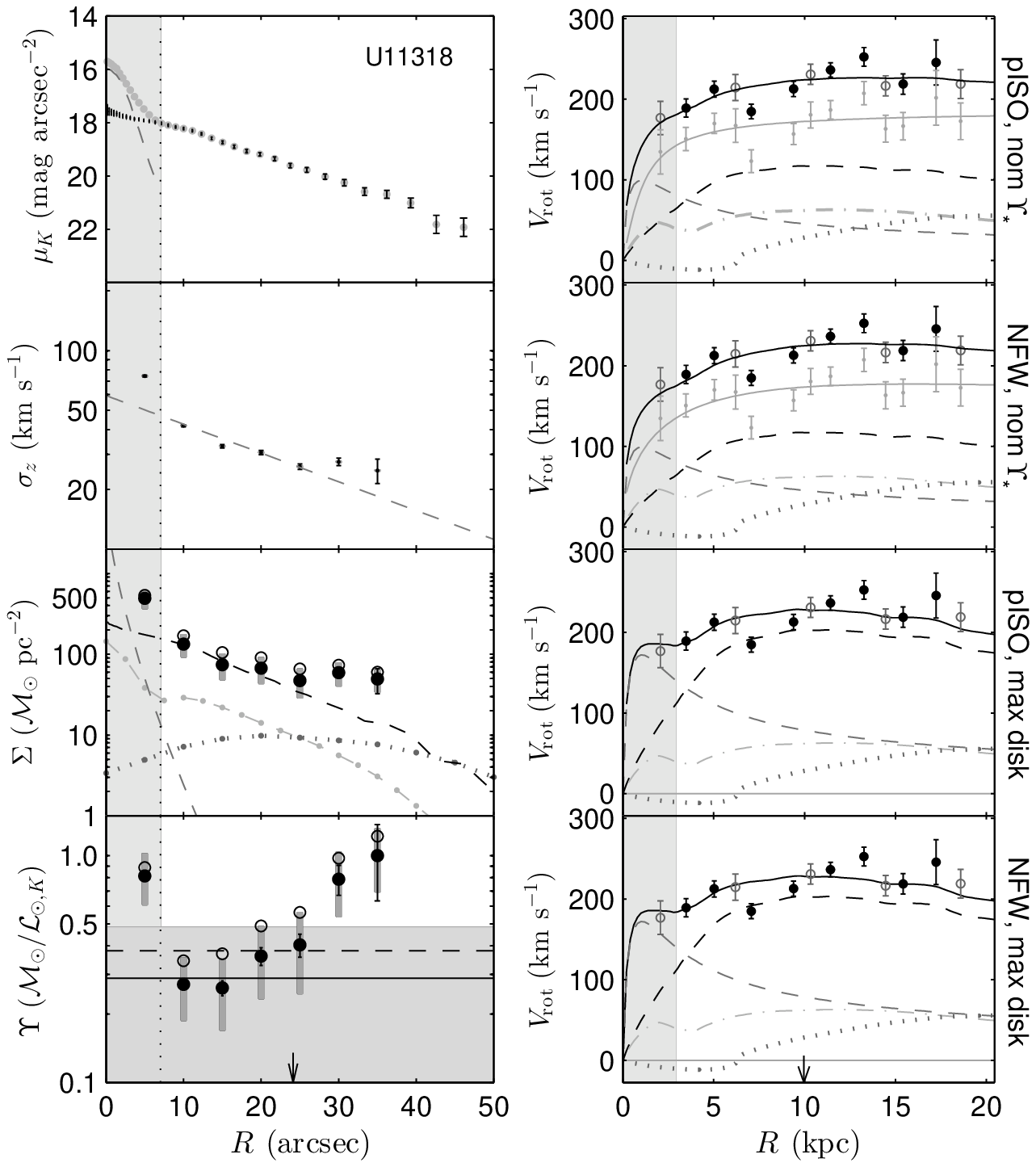}
 \end{figure}

 \begin{figure}
 \centering
 \includegraphics[width=1.00\textwidth]{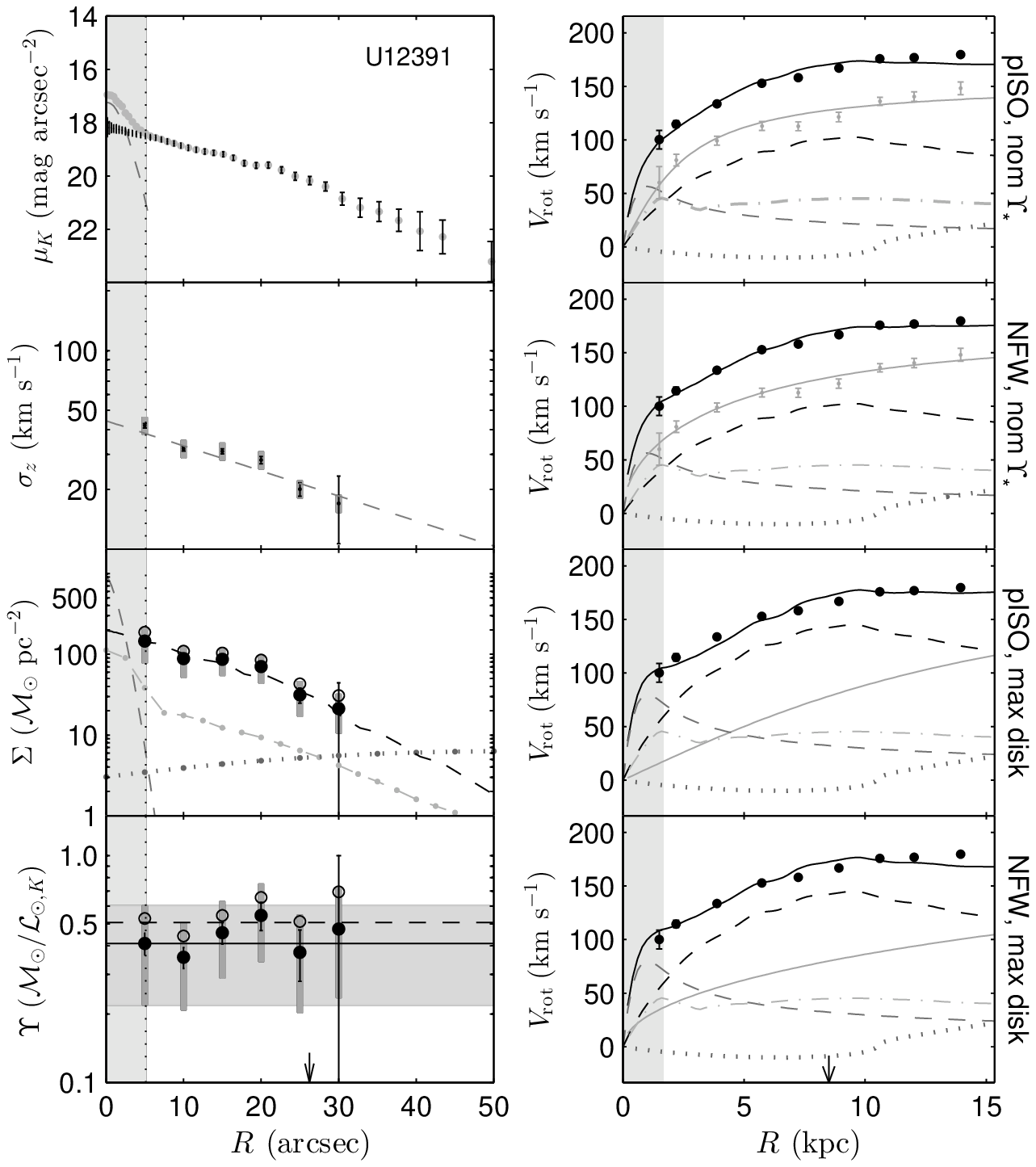}
 \end{figure}


\end{document}